\documentclass[pdftex,nofootinbib,superscriptaddress,aps,prfluids,notitlepage]{revtex4}
\usepackage{multirow,enumerate,epsfig,graphics,amssymb,amsmath,subeqnarray,mathrsfs,color,xcolor,epstopdf}
\usepackage{natbib}
\usepackage{color,epsfig,graphics,amssymb,amsmath,subeqnarray,graphicx,amsthm,subfigure,mathrsfs}
\usepackage{mathtools,bm} %for overbracket
\usepackage{rotating,mathrsfs}

\newcommand{\dd}{\mathrm{d}}

\newcommand{\Fb}{\mathbf{F}}

\newcommand{\ssb}{\mathbf{s}}
\newcommand{\nb}{\mathbf{n}}
\newcommand{\Rb}{\mathbf{R}}

\newcommand{\omegab}{\boldsymbol{\omega}}
\newcommand{\Omegab}{\boldsymbol{\Omega}}
\newcommand{\sigmab}{\boldsymbol{\sigma}}

\newcommand{\ub}{\mathbf{u}}
\newcommand{\eb}{\mathbf{e}}

\newcommand{\Ub}{\mathbf{U}}\newcommand{\xb}{\mathbf{x}}\newcommand{\rb}{\mathbf{r}}

\newcommand{\Pe}{\mbox{Pe}}

\let\grad\nabla
\let\grad\nabla

\newcommand{\Rey}{\mbox{Re}}

\newcommand{\Ck}[2]{\mathbf{C}^{#1}_{k,#2}}
\newcommand{\Cj}[2]{\mathbf{C}^{#1}_{j,#2}}

\newcommand{\vb}{\mathbf{v}}

\makeatletter
\def\sgn{\mathop{\operator@font sgn}}
\def\threevdots{\vbox{\baselineskip1\p@ \lineskiplimit\z@
  \kern6\p@\hbox{.}\hbox{.}\hbox{.}}}
\makeatother

\begin{document}
\title{Modeling chemo-hydrodynamic interactions of phoretic particles: a unified framework}
\author{Akhil Varma}
\email{akhil.varma@ladhyx.polytechnique.fr}
\affiliation{LadHyX -- D\'epartement de M\'ecanique, CNRS -- Ecole Polytechnique, Institut Polytechnique de Paris, 91128 Palaiseau Cedex, France}
\author{S\'ebastien Michelin}
\email{sebastien.michelin@ladhyx.polytechnique.fr}
\affiliation{LadHyX -- D\'epartement de M\'ecanique, CNRS -- Ecole Polytechnique, Institut Polytechnique de Paris, 91128 Palaiseau Cedex, France}

\begin{abstract}

Phoretic particles exploit local self-generated physico-chemical gradients to achieve self-propulsion at the micron scale. The collective dynamics of a large number of such particles is currently the focus of intense research efforts, both from a physical perspective to understand the precise mechanisms of the interactions and their respective roles, as well as from an experimental point of view to explain the observations of complex dynamics as well as formation of coherent large-scale structures. However, an exact modelling of such multi-particle problems is difficult and most efforts so far rely on the superposition of far-field approximations for each particle's signature, which are only valid asymptotically in the dilute suspension limit. A systematic and unified analytical framework based on the classical Method of Reflections (MoR) is developed here for both Laplace and Stokes' problems to obtain the higher-order interactions and the resulting velocities of multiple phoretic particles, up to any order of accuracy in the radius-to-distance ratio $\varepsilon$  of the particles. Beyond simple pairwise chemical or hydrodynamic interactions, this model allows us to account for the generic chemo-hydrodynamic couplings as well as $N$-particle interactions ($N\geq 3$). The $\varepsilon^5$-accurate interaction velocities are then explicitly obtained and the resulting implementation of this MoR model is discussed and validated quantitatively against exact solutions of a few canonical problems.

\end{abstract}
\maketitle

\section{Introduction}
Active matter comprises of a large collection of individually-active agents that continuously consume stored energy or energy from their surroundings to overcome external mechanical resistance and achieve self-propulsion. Being in a state of continuous non-equilibrium, they exhibit collective dynamics at a scale much larger than their size and a behaviour different from their individual dynamics \cite{Ramaswamy12,Marchetti13,Saintillan12,Zottl16,Elgeti15}, as observed, for example, in biological systems such as flocks of birds or bacterial colonies. In so-called dry active matter systems (e.g. vibrated granular matter, dry nematics~\cite{Aranson06,Deseigne10}), the influence of the surrounding medium has negligible influence on dynamics of individual agents, and their interactions are guided by short-range processes (e.g. steric repulsion) but are sufficient to create long-range order and phase transitions~\cite{Cates15,Fodor18}. In contrast, in wet active matter, interacting individuals are immersed in a fluid medium, thus allowing for medium-mediated couplings of their respective motions, such as  long-range hydrodynamic interactions. Examples of the complex ordering and/or dynamics at the collective levels include the turbulent nature of bacterial and algal suspensions \cite{Zhou17,Petroff15}, active polar gels  \cite{Julicher07,Sanchez12}, nematic liquid crystals \cite{Thampi13,deGennes95,Herminghaus14} and active liquid drop emulsions \cite{Herminghaus14,Thutupalli13}. 

From a physical standpoint, characterizing the development of such rich collective dynamics from simple individual behaviour is of particular interest
 \cite{Fodor18,Zottl16,Saintillan12,Bechinger16}. Historically, such efforts initially focused on phenomenological models based on short-range self-alignment rules between the individual agents~\cite{Viscek95,Toner95}. A clear advantage of such an approach is its generality which allowed these models to be applied across a wide variety of system sizes and physical nature. At  microscopic scales, similar ideas are at the root of  so-called Active Brownian Particle (ABP) and Run-and-Tumble Particle (RTP) models, which account for short-range steric interactions and also attempt to include long-range couplings through phenomenological interaction potentials~\cite{Zottl16,Cates15,Redner13,Tailleur08}. Despite attempts to include physical modeling of the coupling field~\cite{Liebchen17,Liebchen18}, such models fundamentally overlook major complexities of the long-range (e.g. hydrodynamic) coupling mechanisms, which motivates the development of another class of models based on a direct and detailed physical description of these interaction mechanisms and derived from first principles. Such modeling may thus fundamentally differ from one active system to another in order to account for the specific nature of the interaction routes between its agents.

Natural examples of microscopic active matter systems can be found in the behavior of the suspensions of swimming microorganisms (e.g. bacteria, algae). Yet, over the past decade, artificial systems have also gained much attention from physicists and engineers alike thanks to significant advances in controlled manufacturing~\cite{Fischer15,Walther13} and the successful parallel development of quantitative experimental measurements and adapted theoretical frameworks~\cite{Illien17,Moran17}. In such active systems, microscopic particles suspended in a fluid medium  are excited either through an externally-applied field  or exploiting direct interactions of individual particles with their physico-chemical environment.  In the former case, energy is supplied by an external directional field to individual agents (e.g. particles forced by rotating electromagnetic fields~\cite{Ghosh09,Bricard15,Driscoll16,Kaiser17}). The latter, which corresponds to so-called fuel-based systems where the source of energy is stored in the particle's immediate environment, includes autophoretic particles~\cite{Paxton04,Howse07,Theurkauff12}, active liquid drops \cite{Thutupalli11,Izri14,Kruger16} and bubble propelled micro-swimmers \cite{Ebbens16,Li16,Giacomo18}. In order to swim, all such systems exert a mechanical forcing on their surroundings and generate a displacement of their fluid environment. They further share two fundamental physico-chemical properties, namely their ability to act on a physico-chemical field by changing the local temperature, electric field or solute content of their fluid environment (\emph{Activity}) and their ability to convert inhomogeneities of this field into  phoretic flows and/or Marangoni stresses at their surface (\emph{Mobility}). Thanks to these generic common features, multiple interaction routes can be envisioned for this class of systems, either directly through the flow generated by the motion of one particle in the vicinity of its neighbors or through its physico-chemical signature and resulting gradients near other particles. How such long-range interaction routes compete and condition the collective dynamics of these systems as observed in experiments, and how this interplay is modified by the varying particle density or their environment are key questions, currently at the center of attention of the physical community~\cite{Thutupalli18,Kanso19,Liebchen19}. 

The goal of the present work is to provide a simple yet accurate and unified framework to analyze and model such interactions. In the following, we specifically focus on the particular case of self-diffusiophoretic particles to derive such models, keeping in mind that a similar formalism could be extended to other fuel-based systems sharing the fundamental properties outlined above. Such colloidal particles "swim" through self-generated gradients of a solute concentration using differences in the short-range interactions with their surface of solute and solvent molecules. To achieve this,  two distinct physico-chemical properties are necessary: (i) a surface activity ($\mathcal{A}$) that catalyses a chemical reaction which either produces or consumes the chemical solute and, (ii) a surface mobility ($\mathcal{M}$) which generates an effective hydrodynamic slip velocity along its surface in response to local concentration gradients \cite{anderson89,Golestanian07}. Together, these two properties create self-induced surface velocity which allows the particle to work against viscous forces in order to attain sustained motion.

Understanding completely their collective dynamics from a fundamental point of view  requires finding the joint solutions for the dynamic evolution of the flow field and chemical concentration of solute (driven by diffusion but also potentially by advection by the phoretic flows), under the influence of the chemical and resulting mechanical forcing of the (many) different  particles. In dense suspensions, i.e. for small inter-particle distances, this has to be done numerically, and a wide range of numerical methods are currently available or could be proposed based on state-of-the-art techniques for generic microswimmer suspensions. Theses include direct numerical solutions of the hydrodynamic and diffusion equations accounting for the detailed forcing of the particles (e.g. \emph{Boundary Element Methods}~\cite{Ishikawa06,TDMJ15,Uspal15} or \emph{Immersed Boundary Methods}~\cite{Lushi13,Lambert13}) or a reduced-order approximation of this forcing (e.g. \emph{Multipole Methods}~\cite{Delmotte15}), coarse-grained models such as \emph{Multiple Particle Collision Dynamics}~\cite{Yang14,Colberg17} or \emph{Lattice Boltzmann Methods}~\cite{Alarcon13}. {Direct computations of the particles' velocities can be performed using hybrid representations of their resistance/mobility matrices using \emph{Stokesian Dynamics}~\cite{Brady88,Ichiki01}. Recent advances have been made in developing schemes for efficiently determining these matrices numerically in suspensions of spheres \cite{Sierou01,Swan19,Yan16}}. A somewhat similar approach identifies \emph{Generalized Stokes laws} and solves numerically for the different irreducible resistance tensors linking particles velocities and surface tractions~\cite{Singh18,Singh19}. {Computational algorithms based on identification between coefficients of Lamb's solution to the known Green's function expansions for internal and external flows in a suspension of spheres have also been previously proposed and have been applied to solve certain problems in periodic domain such as Stokes flow through porous media \cite{Mo94,Sangani96}.} For large numbers of particles, direct or approximate simulations of the chemical and hydrodynamic problems can however become prohibitively expensive computationally, or impose drastic approximations in the representation of the particles' forcing on the fluid medium. As such obtaining an analytical or semi-analytical framework to compute the particles' velocity efficiently while retaining the underlying physical interaction mechanisms is of particular interest to understand the behaviour of phoretic suspensions.
 
Although full analytical solutions of the joint chemical and hydrodynamic problems can be obtained for  simple geometries (e.g. one or two spheres~\cite{Michelin14,Michelin15}), such exact derivations do not extend beyond two spherical particles. Yet, in the dilute suspension limit when the typical distance $d$ between particles is much greater than their typical radius $a$, a first physical insight on the particles' coupling can be obtained by evaluating the leading order correction to a particle's velocity in the slowest decaying components of the chemical and hydrodynamic fields of its neighbors.  In such far-field models, the drift velocities resulting from hydrodynamic and chemical coupling of the particles both scale as $(a/d)^2$. A fundamental assumption here is that interactions between particles can be analyzed pairwise, thus neglecting multi-particle interactions. Nonetheless, thanks to their simplicity of implementation, far-field models have significantly contributed to our understanding of dilute suspension dynamics~\cite{Kanso19,Spagnolie12,Saha14,Soto14} although they may be unable to capture even qualitatively several key features of the hydro-chemical coupling beyond the asymptotically-dilute limit~\cite{Michelin15}. 

The spirit of such models can however be extended to include both higher-order contributions to each particle's forcing on its surrounding environment and multi-particle interactions, still retaining the advantageous simplicity of solving the hydrodynamic and chemical problems only in the vicinity of a single particle, for which analytical solutions exist for simple geometries (e.g. spheres and spheroids),  but in a modified (non-uniform) environment. Using an iterative process where at each step the fields are corrected so as to satisfy the proper boundary conditions on each particle, a series solution for the particles' velocity can be obtained with increasing order of accuracy in $a/d$, and this iterative process effectively accounts for multi-particle interactions. This approach, wittingly termed as the \emph{Method of Reflections}, was initially introduced by Smoluchowski~\cite{Smoluchowski11} and has classically been implemented in hydrodynamics to analyse the collective sedimentation of interacting spheres~\cite{Kynch59,Wilson13} or the effect of a confining boundary~\cite{HappelBrenner}, and the conditions for its asymptotic convergence toward the exact solution have also been analyzed mathematically~\cite{Luke89,Traytak06}. Recently, this framework was applied to diffusion problems related to bubble dissolution~\cite{Michelin18} or phoretic propulsion and migration of homogeneous particles~\cite{Varma18,Rallabandi19}. 

In this work, we propose a systematic use of this approach to solve both for the chemical and hydrodynamic problems and obtain the interaction velocities of spherical phoretic particles of arbitrary surface properties as series expansion in the radius-to-distance ratio $a/d$. The result is a versatile and systematic framework to obtain the particles' velocity directly, which can be used to analyse suspension behavior in the not-so-dilute limit and provide significant improvement over simple far-field models. 

The rest of the paper is organized as follows. Section \ref{sec:isolated} briefly reviews the classical solutions of the chemical and hydrodynamic problems around a single Janus particle and its resulting swimming dynamics. Focusing on dilute systems, Section~\ref{sec:farfield} summarizes the derivation of far-field interactions where only the leading order chemical and hydrodynamic signatures of each particle are retained to obtain $2$-particle interaction velocities accurate till $O(a^2/d^2)$. The general framework at the center of present work is then presented in Section~\ref{sec:reflections} and builds upon the previous classical results using the method of reflections for  both the chemical (Laplace) and hydrodynamic (Stokes) problems, to obtain approximation of the interaction velocities up to a desired but arbitrary accuracy $O(a^n/d^n)$ with $n>2$. As a practical example, this framework is then used to obtain explicitly the particles' velocities to an $(a/d)^5$-accuracy in Section \ref{sec:interactions_ep5}. In Section \ref{sec:validations}, the predictions of this model are compared and validated against analytical solutions and/or direct numerical simulations for various configurations of multiple Janus particles thereby showing the significance of the proposed method in capturing crucial dynamics of the system. We finally draw conclusions and analyse future applications of this class of models in Section~\ref{sec:conclusions}.

\section{Single Janus particle}\label{sec:isolated}
In this Section, an active axisymmetric Janus colloid of radius $a$ is considered. Such polar phoretic particles, which generally consist of an inert rigid colloidal sphere coated on one half by an active catalyst or of two hemispheres of different chemical nature (e.g. bi-metallic swimmers), are commonly used in experiments~\cite{Duan15}. This axisymmetric particle is characterized by its position $\xb$ and a unit vector $\eb$ indicating the direction of its axis of symmetry, and along which self-propulsion occurs. $\rb$ denotes position with respect to the center of the particle. 

For simplicity of analysis, the activity of the particle is modeled as a spatially-dependent production (resp. consumption) of solute with a fixed rate $\mathcal{A}(\mu)>0$ (resp. $\mathcal{A}(\mu)<0$) which may vary along the surface; {here, we note $\mu = \eb \cdot \rb/r\; (=\cos \theta)$ with $r=|\rb|$}. Later on, we will focus on Janus particles with $\mathcal{A}(\mu)=\mathcal{A}$ uniform on one part of the surface (active site) and $\mathcal{A}=0$ on the passive part of the particle. The solute's diffusion within the solvent phase of viscosity $\eta$ and density $\rho$ is characterized by its molecular diffusivity $D$.  and the background (i.e. far-field) concentration of solute is   $C_{\infty}$. 

 Following the classical continuum framework \cite{Michelin14, Golestanian07}, the surface of the particle generates an effective slip velocity in response to local concentration gradients along the surface, $\tilde{\ub}= \mathcal{M}(\mu) \left.\nabla_\parallel C\right|_{r=a}$, as the result of an imbalance in osmotic pressure resulting from the differential interaction of solute and solvent molecules with the particle's surface. Here, $\mathcal{M}$ is the spatially-dependent surface mobility of the particle. In the following, we denote $\mathcal{A}^*$ and $\mathcal{M}^*$ the typical (positive) scales of the activity and mobility properties. The dimensionless activity and mobility are thus $A(\mu) = \mathcal{A}(\mu)/ \mathcal{A}^*$ and $M(\mu) = \mathcal{M}(\mu)/ \mathcal{M}^*$ respectively. The problem is made non-dimensional using the size of the particle, $a$, as reference length scale, $\mathcal{A}^*\mathcal{M}^*/D$ and $\mathcal{A}^* \mathcal{M}^*\eta/(aD)$ as characteristic velocity and pressure, respectively, while the dimensionless relative concentration field is defined as $c(\rb) = (C(\rb)-C_\infty )/(\mathcal{A}^*a/D)$. For large enough diffusivity (or small enough particles), the effect of solute advection by the fluid flow and fluid inertia are negligible (i.e. the characteristic Reynolds and P\'eclet numbers are negligibly small, $\Rey=\rho \mathcal{A}^* \mathcal{M}^*/(aD)\ll 1$ and $\Pe=\mathcal{A}^* \mathcal{M}^*a/D^2\ll 1$), and so is the transient redistribution  of solute molecules around the particle so that $c(\rb)$ satisfies a quasi-static Laplace problem around the particle:
\begin{equation}
\nabla^2 c =0,
\label{eq:laplace}
\end{equation} 
with boundary conditions in the far-field and on the particle's surface,
\begin{align}
c(r \to \infty,\mu)  =0 \qquad \mbox{and} \qquad
\nb \cdot \nabla c \biggr|_{r=a} = -A(\mu).
\label{eq:bc1}
\end{align}
Note that $a$ now denotes the non-dimensional particle radius (here $a=1$, trivially) and is retained for generality purpose so as to allow  later on the treatment of multiple particles of different radii. 
The general solution to the Laplace problem in \eqref{eq:laplace}--\eqref{eq:bc1} is obtained as an harmonic series \cite{Golestanian07,Michelin14,Varma18,Kanso19},
\begin{align}
 c(\rb) = \sum_{m=0}^\infty \frac{A_m}{m+1}\left(\frac{a}{r}\right)^{m+1} L_m(\mu) \qquad \mbox{with} \quad
A_m = \frac{2m+1}{2} \int_{-1}^1 A(\mu)\; L_m(\mu) d\mu, 
\label{eq:Am}
\end{align}
where $L_m(\mu)$ are the Legendre polynomials of order $m$. {The concentration field is thus decomposed into the superposition of an infinite number of polar modes of increasing order and spatial decay rate}:  $m=0$ represents a \emph{point source} $(\sim r^{-1})$, $m=1$ a \emph{source dipole} $(\sim r^{-2})$, $m=2$ a \emph{source quadrupole} $(\sim r^{-3})$ and so on. The strength of each mode, $A_m$, is obtained by a simple projection along $L_m(\mu)$ of the activity distribution, Eq.~\eqref{eq:Am}. For  a hemispheric Janus particle with $A(\mu)=1$ for $\mu \in [0,1]$ and $A(\mu)=0$ otherwise, the mode amplitudes $A_m$ can be obtained analytically as $A_0=1/2$, $A_1=3/4$, $A_2=0$, $A_3=-7/16$, etc...~\cite{Michelin14}

In response to the non-uniform distribution of solute at its surface, the particle generates a local phoretic slip $\widetilde\ub$,
\begin{equation}
\widetilde{\ub} = M(\nb)\; (\mathbf{I}-\nb \nb) \cdot \nabla c\biggr|_{r=a} =-M(\mu)\; \sqrt{1-\mu^2} \frac{\partial c}{\partial \mu}\biggr|_{r=a} \eb_{\theta},
\end{equation}
which in turns generates a flow around the particle and its locomotion. The flow velocity is obtained in the laboratory frame by solving Stokes' equations,
\begin{align}
\nabla^2 \ub = \nabla p , \qquad \nabla \cdot \ub = 0,
\label{eq:stokes}
\end{align}
around the particle, with boundary conditions  
\begin{equation}
\ub\biggr|_{r=a} = \Ub^{\mbox{\scriptsize self}}+ a \; \mathbf{\Omega}^{\mbox{\scriptsize self}} \times \nb + \widetilde{\ub} \qquad \mbox{and} \qquad \ub (\rb \to \infty) = 0, 
\label{eq:stokes_BC}
\end{equation}
where $\Ub^{\mbox{\scriptsize self}}$ and $\mathbf{\Omega}^{\mbox{\scriptsize self}}$ denote the particle's translational and rotational velocities respectively.

For force- and torque-free particles, the translational and rotational velocities can be obtained using the Lorentz Reciprocal Theorem applied to Stokes' flows~\cite{stone96},
\begin{equation}
\Ub^{\mbox{\scriptsize self}} = -\langle \widetilde{\ub} \rangle \qquad \mbox{and} \qquad \mathbf{\Omega}^{\mbox{\scriptsize self}} = \frac{3}{2a}\langle \widetilde{\ub} \times \nb \rangle,
\label{eq:self}
\end{equation}
where $\langle \; \rangle$ represents the averaging operator over the particle's surface. When the particle's mobility is uniform ($M(\nb)=M=1$), this simplifies as $\Ub^{\mbox{\scriptsize self}} =-(MA_1/3)\eb$ and $\mathbf{\Omega}^{\mbox{\scriptsize self}} =0$: the particle self-propels along its axis of symmetry with no rotation. Here $\eb$ is chosen to be directed from passive to active part. Note that the only chemical mode contributing to self-propulsion of the phoretic particle is a chemical source dipole ($m=1$). All the other modes of the concentration field generate only non-swimming flow fields. For a Janus particle with hemispherical active surface (i.e. $A(\nb)=1$ on the active half, and $A(\nb)=0$ otherwise), $\Ub^{\mbox{\scriptsize self}} = -\eb/4$. 

The complete axisymmetric hydrodynamic flow field is further obtained classically as a superposition of orthogonal squirming modes~ \cite{Blake71,Pak14,Michelin14}:
\begin{align}
\ub(\rb)=\frac{\alpha_1}{2r^3}\left(\frac{3\rb\rb}{r^2}-\mathbf{I}\right)\cdot\eb-\sum_{m\geq 2}\frac{(2m+1)\alpha_m}{2m(m+1)}\Bigg\{&\left(m(m+1)L_m(\mu)\left[\left(\frac{a}{r}\right)^{m+2}-\left(\frac{a}{r}\right)^{m}\right]\right)\frac{\rb}{r}\nonumber\\
&+L_m'(\mu)\left[(m-2)\left(\frac{a}{r}\right)^{m}-m\left(\frac{a}{r}\right)^{m+2}\right]\left(\mathbf{I}-\frac{\rb\rb}{r^2}\right)\cdot \eb\Bigg\}, \label{eq:flow_field}
\end{align}

and, for all $m\geq 1$,
\begin{align}
\alpha_m & = \frac{1}{2} \int_{-1}^1 \sqrt{1-\mu^2} \;L_m'(\mu)\; \widetilde{\ub} \cdot \eb_{\theta} \; d\mu.
\label{alpha}
\end{align}
In these notations, the swimming velocity is $\Ub^\textrm{self}=\alpha_1\eb$, and the successive squirming modes of the series in Eq.~\eqref{eq:flow_field} are associated with hydrodynamic singularities of increasing order: for example, $m=1$ includes a \emph{source dipole} ($\sim r^{-3}$), $m=2$ consists of a \emph{force dipole} ($\sim r^{-2}$) and a \emph{source quadrupole} ($\sim r^{-4}$) and so on with higher modes comprising of force and source multipoles. For an axisymmetric phoretic particle with uniform mobility $M$, a one-to-one relation between the coefficients of hydrodynamic and chemical modes can be further established \cite{Michelin14}.
\begin{equation}
\alpha_m = -\frac{mM A_m }{2m+1}\cdot
\label{alpha_A}
\end{equation}

\section{Far-field interactions}\label{sec:farfield}
The derivations of the previous section demonstrate that phoretic particles leave two types of imprints on their environment: a modified chemical field due to their activity and a hydrodynamic signature due to their swimming motion. Both of these modify the dynamics of their neighbours which will now evolve in a modified background environment. In this section we briefly review the associated resulting drifts, at the core of so-called far-field interaction models which are the lowest order of approximation for particles' interactions in the dilute limit. 

\subsection{Motion of particles in external chemical and hydrodynamic fields} 
Considering a single particle in externally-imposed non-uniform chemical and hydrodynamic fields, the concentration and velocity fields now satisfy the modified Laplace and Stokes problems:
\begin{eqnarray}
\nabla^2 c=0,\qquad &\nb\cdot\nabla c\biggr|_{r=a}=-A(\nb),\quad \mbox{and} \quad &c(r\gg a)\sim c_\infty(\rb),\label{eq:farfieldchem}\\
\nabla^2\ub=\grad p,\qquad \nabla\cdot\ub=0,\qquad &\left. \ub\right|_{r=a}=M(\nb)\;(\mathbf{I}-\nb\nb)\cdot \grad c \biggr|_{r=a}+\Ub+\Omegab\times\nb,\quad \mbox{and} \quad & \ub(r\gg a)\sim\ub_\infty(\rb).\label{eq:farfieldhydro}
\end{eqnarray}
together with the force- and torque-free condition on the particle.

Both problems are linear, therefore the self-propulsion velocities $(\Ub,\Omegab)$ can be decomposed as three independent problems, defined in response to the three forcings, namely the chemical activity of the particle and the background chemical and hydrodynamic fields 
\begin{enumerate}[(i)]
\item{$(\Ub^\textrm{self},\Omegab^\textrm{self})$: self-propulsion of the active particle with no background forcing ($A\neq 0$, $c_\infty=0$, $\ub_\infty=0$),}
\item{$(\Ub^\chi,\Omegab^\chi)$: drift of a passive particle in a background chemical field ($A=0$, $c_\infty \neq 0$, $\ub_\infty=0$),}
\item{$(\Ub^h,\Omegab^h)$: drift of a passive particle in a hydrodynamic background flow ($A=0$, $c_\infty=0$, $\ub_\infty \neq 0$).}
\end{enumerate}

  The self-propulsion problem (i) is the focus of the previous section. The drift in an external concentration field $c_\infty(\rb)$ is a classical problem discussed in~\cite{Kanso19,anderson89,MTC18}. For a particle with uniform mobility $M(\nb)=M$,
\begin{align}
\Ub^\chi = - M \; \nabla c_\infty \biggr|_{r=0} \quad \textrm{and,} \quad \Omegab^\chi = \mathbf{0}. \label{eq:chemdrift1}
\end{align}

The effect of an external disturbance flow $\ub_\infty$ is analyzed here by computing the hydrodynamic drift on a rigid particle exposed to a non-uniform background hydrodynamic field $\ub_\infty(\rb)$. This is a classical hydrodynamic problem, whose solution is given by the well-known Faxen's laws for a spherical particle~\cite{kimkarrila}
\begin{equation}
\Ub^{\scriptsize h} = \ub_\infty\biggr|_{r=0} + \frac{a^2}{6} \nabla^2\ub_\infty\biggr|_{r=0} \quad \mbox{and} \quad  \Omegab^{\scriptsize h} = \frac{1}{2}\nabla \times \ub_\infty\biggr|_{r=0}.
\label{eq:drift}
\end{equation}

\subsection{ Far-field interaction of active phoretic particles}
Combining these fundamental results, a first approximation to the collective dynamics of phoretic particles is then obtained in the dilute limit (i.e. when the particles are asymptotically far away from each other) by assuming that the background concentration and hydrodynamic fields experienced by a given particle $k$ result from the superposition of the chemical and hydrodynamic signatures of each of its neighbors (noted $j\neq k$) \emph{as if these particles were themselves isolated}. This assumption is critical as it amounts to neglecting the influence of surrounding particles (or boundaries) on the chemical and hydrodynamic fields they generate, thereby fundamentally restricting the order of accuracy of the approximation. Further, in the dilute limit, only the slowest decaying contribution to each signature needs to be retained to obtain the dominant chemical and hydrodynamic drifts. \\

\begin{figure}[t]
\includegraphics[width=0.55\textwidth]{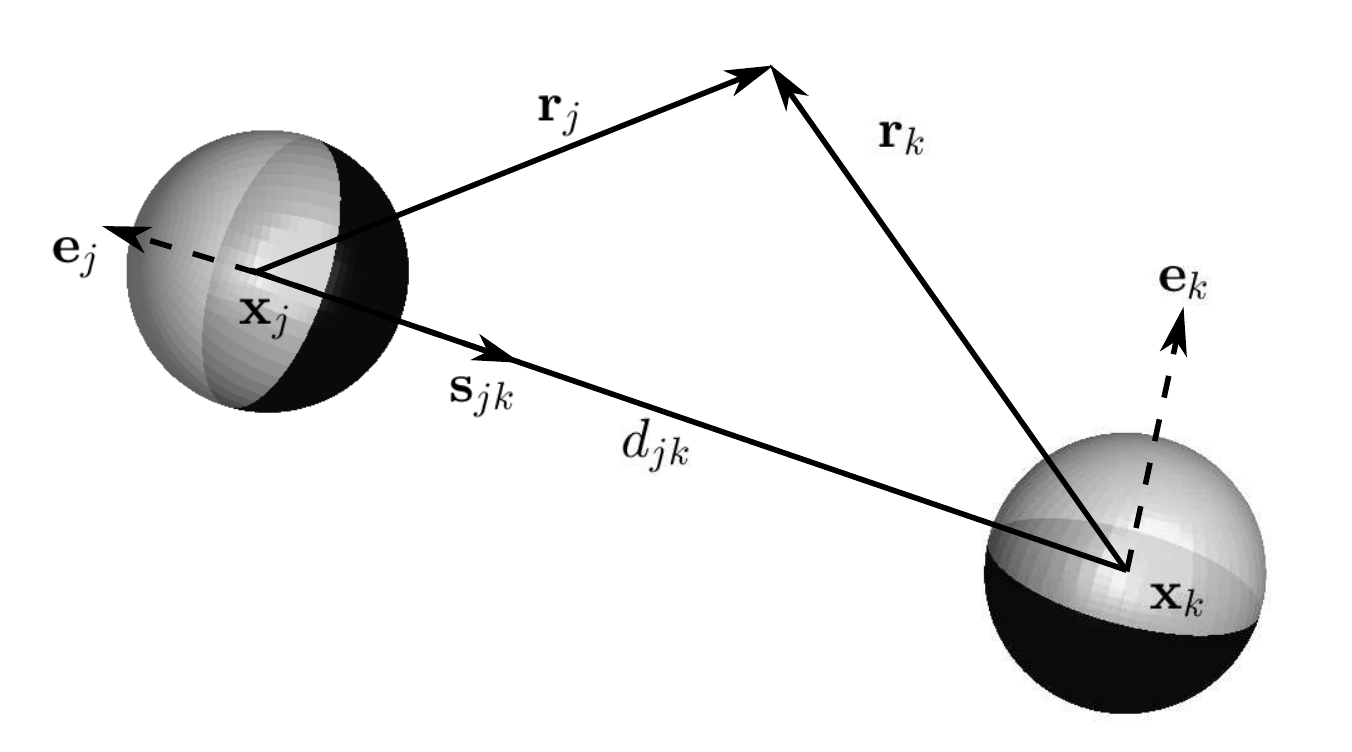}
\caption{{Notations used for geometric description of the arrangement of any two Janus particles $j$ and $k$. The Janus particles comprise of active (white) and inert (black) parts.}}
\label{fig:schematic}
\end{figure}

In the following, and in the rest of the manuscript, the position of particle $k$ is noted $\xb_k$, its radius is $a_k$, and its orientation is given by a unit vector $\eb_k$. For any two particles $j$ and $k$, $d_{jk}$ and $\ssb_{jk}$ respectively denote their center-to-center distance and the unit vector joining the centre of particles $j$ to $k$, i.e. $d_{jk}\ssb_{jk}=\xb_k-\xb_j$, as shown in figure \ref{fig:schematic}. We further denote $\rb_j$ the position vector measured with respect to particle $j$, i.e. $\rb_j=\rb-\xb_j$.

The concentration and hydrodynamic fields created by isolated particles (see Section~\ref{sec:isolated}) can be expanded as series of chemical and hydrodynamic singularities whose effect on neighboring particles scale like increasing powers of $\varepsilon=a/d$ (where $a$ and $d$ denote here the typical values of particle radius $a_j$ and interparticle distance $d_{jk}$, respectively). When sufficiently far apart (i.e. $\varepsilon\ll 1$), the phoretic particles behave, at the leading order, as the slowest decaying chemical and hydrodynamic singularities, i.e. a \emph{chemical point source} and a \emph{hydrodynamic force-dipole}.

Retaining only the dominant chemical signature of each particle, the external concentration field $c_{\infty,k}$ experienced by particle $k$ and its gradient at the particle's center ($r_k=0$) are obtained as 

\begin{equation}
c_{\infty,k}=\sum_{j\neq k}\frac{A_{j,0}a_j}{r_j}\quad \Rightarrow\quad \nabla c_{\infty,k} \biggr|_{r_k=0} = -\sum_{j\neq k}\frac{A_{j,0}\; a_j^2}{d_{jk}^2 }\ssb_{jk}.
\label{eq:gradcinf}
\end{equation}
The resulting chemical drift due to a point source is then obtained using Eqs.~\eqref{eq:chemdrift1},
\begin{equation}
\Ub_k^\chi = M_k \sum_{j\neq k} \frac{A_{j,0}\; a_j^2}{d_{jk}^2 }\ssb_{jk}, \quad \mbox{and} \quad \Omegab^\chi_k = \mathbf{0}.
\end{equation}
Each neighboring particle $j$ induces on particle $k$ a chemical drift  along their line of centers without any rotation (for uniform mobility).

Similarly, retaining only the leading-order flow field created by particle $j$ (i.e. that of a stresslet obtained for $m=2$ in Eq.~\eqref{eq:flow_field}), the background hydrodynamic field experienced by particle $k$ is given by
\begin{equation}
\ub_{\infty,k}(\rb)=\sum_{j\neq k}\frac{a_j^2 M_j A_{j,2}}{2} (3\eb_j\eb_j-\mathbf{I}):\left(\frac{\rb_j\rb_j\rb_j}{r_j^5}\right),
\end{equation}
and the hydrodynamic drifts are obtained from Eqs.~\eqref{eq:drift}, keeping only leading order contributions, as
\begin{align}
\Ub^{\scriptsize h}_k & = \sum_{j \neq k} \frac{M_j A_{j,2}}{2} \left(\frac{a_j}{d_{jk}}\right)^2 (3\eb_j\eb_j-\mathbf{I}):\ssb_{jk}\ssb_{jk}\ssb_{jk},
\label{eq:drift1}\\
\Omegab^{\scriptsize h}_k & = \sum_{j \neq k} \frac{3 a_j^2 M_j A_{j,2}}{2d_{jk}^3}(\eb_j \cdot \ssb_{jk}) (\ssb_{jk} \times \eb_j). \label{eq:drift2}
\end{align}

In non-dimensional units, the self-propulsion velocity of the particles is $O(1)$, Eq.~\eqref{eq:self}, while the chemical and hydrodynamic drifts introduced by the presence of other particles, Eqs~\eqref{eq:chemdrift1} and \eqref{eq:drift1}--\eqref{eq:drift2} are both of the same order, $O(\varepsilon^2)$. 

The resulting framework, termed \emph{far-field interaction model}, is fundamentally based on neglecting (i) higher order contributions to the chemical and hydrodynamic signatures of individual particles which would contribute to $O(\varepsilon^3)$ or smaller drift velocities and (ii) modification in the fields created by each particle due to the presence of others. The latter includes for example the drift on particle $k$ arising from the concentration induced by the activity of particle $j$ but in a finite domain due to the presence of particle $l$. Note that, while the former contributions could in principle be directly obtained from the results of Section~\ref{sec:isolated}, they must be discarded in order to remain consistent in the order of asymptotic approximation of the model with respect to approximation (ii). Similarly, far-field models must also ignore such higher order corrections as the Laplacian term in Faxen's law, Eq.~\eqref{eq:drift} which would contribute an $O(\varepsilon^4)$ to the particles' velocities. Such higher-order interactions however become increasingly significant as the separation between particles decreases. Therefore, obtaining a more accurate estimate of $\Ub_k$ and $\Omegab_k$ requires taking into account explicitly these faster-decaying terms. This idea of including multiple interdependent interactions between the particles is at the heart of the classical \emph{Method of Reflections} for both chemical and hydrodynamic problem, which we exploit in the following section to construct analytically consistent estimates of the velocities with increasing order of accuracy.

%%%%%%%
% MOR
%%%%%%%
\section{The method of reflections for phoretic problems}
\label{sec:reflections}
 To compute the multi-body dynamics of phoretic particles, we now derive a systematic framework to compute their propulsion velocities $(\Ub,\Omegab)$ explicitly to any degree of accuracy $O(\varepsilon^n)$, using the method of reflections to solve Laplace and Stokes equations around $N$ spherical particles. The linearity of both problems allows for a decomposition of both the chemical and hydrodynamic fields into truncated series expansions, matching piece-wise the chemical as well as the hydrodynamic boundary conditions.
Starting from a simple superposition of the chemical and hydrodynamic fields generated by each of the particles when it is isolated, this approach consists in eliminating at each stage, the spurious concentration flux (resp. disturbance flow) introduced on the surface of a given particle by the chemical (resp. hydrodynamic) field generated by all the other particles in the previous stage, thus introducing a new correction to the chemical (resp. hydrodynamic) field around each of the particles independently. 

In the following, the general framework is first presented for the chemical problem, generalizing the method proposed in~\cite{Varma18} for homogeneous particles to the general case of arbitrary surface activity, in order to determine the successive moments of the surface concentration on each particle as a result of their activity. In a second step, the corresponding method is presented for the hydrodynamic problem using the output of the chemical dynamics as a forcing and constructing the resulting particle velocities. Combining these two steps provide the particles' velocities as a function of their geometrical arrangement and orientations. This provides a systematic approach to construct the particles' velocities with a $O(\varepsilon^n)$ accuracy for any $n$. As a practical example, the application of this method is presented in Section~\ref{sec:interactions_ep5} to obtain the particles' velocities up to $O(\varepsilon^5)$, i.e. with leading order corrections scaling as $O(\varepsilon^6)$.

\subsection{Method of reflection for the chemical problem}
\label{sec:chem-ref}
The method is initiated by considering the superposition of the chemical fields created by isolated particles, noted $c_k^0$, which was obtained explicitly in Eq.~\eqref{eq:Am}. $c_k^0$ satisfies the correct boundary condition on particle $k$ only but introduces a spurious flux on the other particles.

At each subsequent stage ($r\geq 1$), known as a ``reflection", a correction $c_k^r$ to the concentration field created by a particle $k$ is introduced in order to correct the spurious normal flux introduced on the boundary of particle $k$ during the previous reflection at the other particles (e.g. $c_k^1$ must correct for the spurious flux introduced by $\displaystyle\sum_{j\neq k}c_j^0$). $c_k^r$ is therefore the unique solution to the following Laplace problem
\begin{align}
\nabla^2 c^r_k & = 0 \quad\textrm{ for   }r_k\geq a_k,\qquad \nb_k \cdot \nabla c^r_k\biggr|_{r_k=a_k}  = - \sum_{j\neq k} \nb_k \cdot \nabla c^{r-1}_j\biggr|_{r_k=a_k}, \qquad c_k^r(r_k\gg a_k)\longrightarrow 0.
\label{eq:conc_p}
\end{align}
and can be written as
\begin{equation}
c^r_k (\rb_k)= \sum_{q \geq 0} \frac{a_k^{q+1}}{r_k^{2q+1}} \Ck{r}{q} \overset{q}{\odot} [\rb_k \overset{q}{\otimes} \rb_k],
\label{eq:gen_cpk}
\end{equation} 
where $(\Ck{r}{q})_q$ is a unique set of $q^{\mbox{\scriptsize th}}$ order fully symmetric and deviatoric tensors. In the previous equation $\rb_k\overset{q}{\otimes}\rb_k$ denote the tensorial product of vector $\rb_k$ by itself repeated $q$ times, while $\mathbf{A}\overset{q}{\odot}\mathbf{B}$ denotes the $q$-fold contraction of tensors $\mathbf{A}$ and $\mathbf{B}$. Expanding  $c_j^{r-1}$ in Taylor series near the center of particle $k$, 
\begin{equation}
c^{r-1}_j (\rb_j)= \sum_{q\geq 0}\frac{1}{q!}\overset{q}{\grad}c^{r-1}_j \biggr|_{r_k=0} \overset{q}{\odot}[\rb_k\overset{q}{\otimes}\rb_k]=c^{r-1}_j \biggr|_{r_k=0} + \rb_k \cdot \nabla c^{r-1}_j \biggr|_{r_k=0} + \frac{\rb_k \rb_k}{2!} : \nabla \nabla c^{r-1}_j \biggr|_{r_k=0} + \hdots,
\label{eq:gen_cp_1j}
\end{equation}
the flux boundary condition in Eq.~\eqref{eq:conc_p} together with Eqs.~\eqref{eq:gen_cpk}--\eqref{eq:gen_cp_1j} imposes:
\begin{equation}
\Ck{r}{q} =\sum_{j \neq k} \frac{q a_k^q}{(q+1)!} \overset{q}{\grad} c^{r-1}_j \biggr|_{r_k=0}.
\label{eq:cpk}
\end{equation}
Substituting Eq.~\eqref{eq:gen_cpk} for particle $j$ at reflection $r-1$ into Eq.~\eqref{eq:cpk} provides the recursive relation
\begin{equation}
\Ck{r}{q} = \sum_{j \neq k} \sum_{s \geq 0} \mathbf{C}^{r-1}_{j,s} \overset{s}{\odot} \bm{\mathcal{F}^\chi}_{jk}(q,s),\qquad \textrm{with    }\bm{\mathcal{F}^\chi}_{jk}(q,s) =  \frac{q a_k^q a_j^{s+1}}{(q+1)!} \left[\overset{q}{\grad} \left(\frac{\rb_j \overset{s}{\otimes}\rb_j}{r_j^{2s+1}} \right)\right]_{r_k=0}=O(\varepsilon^{q+s+1}) . \label{eq:conc_ref_rec}
\end{equation}
Note that the formulation above corresponds to a \emph{parallel} form of the method of reflections (it relates the new concentration multipole on particle $k$ to that of other particles at the previous reflection). A \emph{sequential} approach of the method (i.e. obtaining $\Ck{r}{q}$ for each particle $k$ successively) would correspond to splitting the sum on $j$ in Eq.~\eqref{eq:conc_ref_rec} (respectively for $j<k$ and $j>k$) in order to exploit that for $j<k$, the new concentration multipole $\mathbf{C}_{j,s}^r$, being already available, would be used to compute $\Ck{r}{q}$.

Also, it should be noted that $\bm{\mathcal{F}^\chi}_{jk}(q=0,s)=0$ for all $s$ (reflections induce no net source) and that only the fully symmetric and deviatoric part of the transfer function $\bm{\mathcal{F}^\chi}_{jk}(q,s)$ with respect to its first $s$ indices contribute since $\mathbf{C}_{j,s}^{r-1}$ is fully symmetric and deviatoric. The lowest order transfer functions are thus obtained as: 
\begin{align}
\bm{\mathcal{F}^\chi}_{jk}(1,0)&=-\frac{a_ja_k\,\ssb_{jk}}{2d_{jk}^2}=O(\varepsilon^2), \\
 \bm{\mathcal{F}^\chi}_{jk}(2,0)&=\frac{a_k^2a_j(3\ssb_{jk}\ssb_{jk}-\mathbf{I})}{3d_{jk}^3}=-\frac{2a_k}{3a_j}\bm{\mathcal{F}^\chi}_{jk}(1,1)=O(\varepsilon^3),\\
\bm{\mathcal{F}^\chi}_{jk}(3,0)&=\frac{3a_k^3a_j}{8d_{jk}^4}\left(\mathbf{I}\ssb_{jk}+\ssb_{jk}\mathbf{I}+(\mathbf{I}\ssb_{jk})^{T_{23}}-5\ssb_{jk}\ssb_{jk}\ssb_{jk}\right)=-\frac{3a_k}{8a_j}\bm{\mathcal{F}^\chi}_{jk}(2,1)=\frac{3 a_k^2}{4a_j^2}\bm{\mathcal{F}^\chi}_{jk}(1,2)=O(\varepsilon^4).
\end{align}

{Here $\mathbf{A}^{T_{ab}}$ represents the transpose of the tensor matrix $\mathbf{A}$ with respect to its $a^{\mathrm{th}}$ and $b^{\mathrm{th}}$ indices}. The recursive relation in Eq.~\eqref{eq:conc_ref_rec} is initiated by noting that the tensors $\Ck{0}{q}$ are obtained from the activity distribution coefficients $A_{k,n}$ of the individual particle as 
\begin{equation}
C^0_{k,0} = A_{k,0},\qquad \Ck01= \frac{A_{k,1}\; \eb_k}{2},\qquad\Ck{0}{q\geq 2}=\frac{(2q-1)!\,A_{k,q}}{2^{q-1}(q-1)!\times(q+1)!}\overbracket{\eb_k\overset{q}{\otimes}\eb_k},\label{eq:conc_ref_0}
\end{equation}
with $\overbracket{\mathbf{B}}$ denoting the fully symmetric and deviatoric part of any given tensor $\mathbf{B}$~\cite{nasouri18}. It should be stressed here that the method is presented for axisymmetric particles (i.e. the successive moments $\Ck{0}{q}$ are function of the axis of the particle $\eb_k$ only), yet could easily be extended to particles of arbitrary coverage~\cite{lisicki18} by modifying Eq.~\eqref{eq:conc_ref_0} accordingly.

The change in surface concentration of particle $k$ introduced at reflection $r\geq 1$, noted $\tilde{c}_k^r$, is obtained within this framework as the sum of $c_k^r$ and of the contributions $c_j^{r-1}$ of all the other particles ($j\neq k$) evaluated at $r_k=a_k$: 
\begin{equation}
\widetilde{c}^r_k = c^r_k\biggr|_{r_k=a_k} + \sum_{j \neq k} c^{r-1}_j\biggr|_{r_k=a_k} = \sum_{q \geq 1} \frac{2q+1}{q} \Ck{r}{q} \overset{q}{\odot} [\nb_k\overset{q}{\otimes} \nb_k].
\label{eq:cpsurface}
\end{equation}
For $r=0$, the surface concentration is similarly obtained as
\begin{equation}
\tilde{c}_k^0=c_k^0\biggr|_{r_k=a_k}=\sum_{q\geq 0}\Ck{0}{q}\overset{q}{\odot} [\nb_k\overset{q}{\otimes} \nb_k].
\end{equation}
Equation~\eqref{eq:cpsurface}  provides an interpretation of the tensorial coefficients $\Ck{r}{q}$ as the fully symmetric and deviatoric moment of order $q$ of the surface concentration introduced at reflection $r$,
\begin{equation}
\Ck{r}{q} = \frac{q}{2q+1} \langle \widetilde{c}^r_k \overbracket{\nb_k \overset{q}{\otimes} \nb_k} \rangle.
\end{equation}

Finally, after all the desired reflections have been performed, the surface concentration $\tilde{c}_k$ of particle $k$ is obtained by superimposing all the different contributions $\tilde{c}_k^r$, 
\begin{equation}
\tilde{c}_k=C^0_{k,0} +\sum_{q\geq 1}\left[\Ck{0}{q} + \sum_{r\geq 1} \frac{(2q+1)}{q}\Ck{r}{q}\right]\overset{q}{\odot}[\nb_k\overset{q}{\otimes} \nb_k].\label{eq:c_surf}
\end{equation}

\subsection{Method of reflections for the hydrodynamic problem}
\label{sec:hydro-ref}

A similar framework can be formulated for the hydrodynamic problem. At each stage $p$, for a given particle $k$, we seek the unique solution of Stokes equation around particle $k$ that decays in the far-field,
\begin{equation}
\nabla^2\ub_k^p=\grad p_k^p,\qquad \nabla\cdot\ub_k^p=0,\qquad \ub_k^p(r_k\gg a_k)\rightarrow 0,\label{eq:stokes_k}
\end{equation} 
and further satisfies the following Dirichlet condition on the particle's surface
\begin{align}
\ub_k^p\biggr|_{r_k=a_k}=\vb_k^p+\Ub_k^p+\Omegab_k^p\times\nb_k,
\end{align}
where $\Ub_k^p$ and $\Omegab_k^p$ are the translation and rotation velocity corrections for particle $k$ at reflection $p$ (determined by enforcing the linear and angular momentum balances on particle $k$), and $\vb_k^0$ (initialization) corresponds to the phoretic slip resulting from the concentration distribution at the particle's surface, while $\vb_k^p$ with $p\geq 1$ (subsequent reflections) balances the spurious flow created at stage $p-1$ by all the other particles.

\subsubsection{General solution of the hydrodynamic problem}
The general solution to Eq.~\eqref{eq:stokes_k} is obtained classically from three sets of spherical harmonics~\cite{kimkarrila,Lamb},
\begin{equation}
\ub^{p}_k=\sum_{q=1}^\infty\left[\grad\phi^p_{k,q}+\grad\times\left(\chi^p_{k,q}\rb_k\right)+\frac{2(q+1)p^p_{k,q}\rb_k-(q-2)r_k^2\grad p^p_{k,q}}{2q(2q-1)}\right],\quad \textrm{with  }\left[\begin{array}{c}\phi^p_{k,q}(\rb_k)\\ p^p_{k,q}(\rb_k)\\ \chi^p_{k,q}(\rb_k) \end{array}\right]=\left[\begin{array}{c}\boldsymbol\Phi^p_{k,q}\\ \mathbf{P}^p_{k,q}\\ \mathbf{X}^p_{k,q} \end{array}\right]\overset{q}{\odot}\left(\frac{\rb_k\overset{q}{\otimes} \rb_k}{r_k^{2q+1}}\right),\label{eq:sol_ukp}
\end{equation}
and $(\boldsymbol\Phi^p_{k,q})_q$, $(\mathbf{P}^p_{k,q})_q$ and $(\mathbf{X}^p_{k,q})_q$ are three sets of fully-symmetric and deviatoric tensors of order $q$:
\begin{align}
\boldsymbol\Phi^p_{k,q}&=\frac{a_k^{q+2}}{2 (q+1)}\Big(q\mathcal{P}^k_q[\nb_k\cdot\vb_k^p]+\mathcal{P}^k_q[-a_k\grad_s\cdot\vb_k^p]\Big)+\frac{\delta_{q,1}a_k^3}{4}\Ub_k^p,\label{eq:Phi_kq_def}\\
\mathbf{P}^p_{k,q}&=\frac{(2q-1)a_k^{q}}{q+1}\Big((q+2)\mathcal{P}_q^k[\nb_k\cdot\vb_k^p]+\mathcal{P}_q^k[-a_k\grad_s\cdot\vb_k^p]\Big)+\frac{3\delta_{q,1}a_k}{2}\Ub_k^p,\label{eq:P_kq_def}\\
\mathbf{X}^p_{k,q}&=\frac{a_k^{q+1}}{ q(q+1)}\mathcal{P}_q^k[a_k\nb_k\cdot(\grad_s\times\vb_k^p)]+\delta_{q,1}a_k^3\Omegab_k^p,\label{eq:X_kq_def}
\end{align}
where $[\mathcal{P}_q^k(f)]_q$ is the unique set of fully symmetric and deviatoric tensors of order $q$ such that the expansion of a scalar field $f(\xb)$ into spherical harmonics at the surface of particle $k$ writes as  
\begin{equation}
f(\xb)\Big|_{r_k=a_k}=\sum_{q\geq 0} \mathcal{P}_q^k[f]\overset{q}{\odot}[\nb_k\overset{q}{\odot}\nb_k].\label{eq:sphere_harm_dec}
\end{equation}

{In contrast with the spherical harmonic decompostion of the chemical field which includes a single set of tensor coefficients ($\mathbf{C}^p_{k,q}$), the hydrodynamic field includes three such sets, ($\mathbf{P}^p_{k,q},\;\boldsymbol{\Phi}^p_{k,q},\; \mathbf{X}^p_{k,q}$).} In Eq.~\eqref{eq:sol_ukp}, each term corresponds to flow singularities of increasing order~\cite{Blake71,kimkarrila,HappelBrenner}, namely (i) source/potential multipoles, $(\boldsymbol\Phi^p_{k,q})_q$, with a flow field decaying as $1/r^{q+2}$, (ii) symmetric force multipoles $(\mathbf{P}^p_{k,q})_q$, with a flow field decaying as $1/r^{q}$ and (iii) rotlet (torque) multipoles $(\mathbf{X}^p_{k,q})_q$, with a flow field decaying as $1/r^{q+1}$. For instance, $\boldsymbol\Phi_{k,1}^p$ corresponds to a source dipole of intensity $-4\pi \boldsymbol\Phi_{k,1}^p$ while $\mathbf{P}_{k,2}^p$ corresponds to a stresslet of intensity $-4\pi\mathbf{P}_{k,2}^p/3$.

The conservation of linear and angular momentum for each particle  imposes two further conditions that uniquely determine $\Ub_k^p$ and $\Omegab_k^p$. For example, for force- and torque-free particles, $\mathbf{X}_{k,1}^p=\mathbf{P}_{k,1}^p=0$ (there is no rotlet or stokeslet contribution to particle $k$'s hydrodynamic signature). 

\subsubsection{Recursive relations for the hydrodynamic singularities ($p\geq 1$)}

When $p\geq 1$, $\vb_k^p$ must exactly cancel the flow introduced at the surface of particle $k$ by the previous reflection at all the other particles $j\neq k$; using a Taylor series expansion of those flow fields near the center of particle $k$,
\begin{align}
\vb_k^p=-\sum_{j\neq k}\ub_j^{p-1}\biggr|_{r_k=a_k}=-\sum_{q\geq 1}\left[\sum_{j\neq k}\frac{a_k^{q-1}}{(q-1)!}\overset{q-1}{\grad}\ub_j^{p-1}\biggr|_{r_k=0}\right]\overset{q-1}{\odot}[\nb_k\overset{q-1}{\otimes}\nb_k],\label{eq:surf_vel_exp}
\end{align}

{The normal velocity, surface divergence as well as surface vorticity on particle $k$ are then obtained from Eq.~\eqref{eq:surf_vel_exp} in terms of the reflected velocities at its center (see Appendix~\ref{sec:app_hydro} and Eqs.~\eqref{eq:dec_vn}--\eqref{eq:dec_rotsurf}). Furthermore, by taking the required gradients at the center of particle $k$, linear recursive definitions are obtained for the flow singularities intensity $(\boldsymbol\Phi_{k,q}^p,\mathbf{P}_{k,q}^p,\mathbf{X}_{k,q}^p)_q$ and particle velocities $(\Ub_k^p,\Omegab_k^p)$ at reflection $p$ in terms of their counterparts at the previous reflection; these take the form of transfer functions that  are independent of $p$ and solely depend on the particles' arrangement (see Appendix~\ref{sec:app_hydro})}. For force- and torque-free particles, these write:
\begin{align}
\Ub_k^p&=\sum_{j\neq k}\sum_{s\geq 1}\left[\boldsymbol\Phi_{j,s}^{p-1}\overset{s}{\odot}\boldsymbol{\mathcal{F}^1}_{jk}(1,s)-\mathbf{X}_{j,s}^{p-1}\overset{s}{\odot}\boldsymbol{\mathcal{F}^2}_{jk}(1,s)+\mathbf{P}_{j,s}^{p-1}\overset{s}{\odot}\left(\boldsymbol{\mathcal{F}^3}_{jk}(1,s)+\frac{a_k^2}{6}\boldsymbol{\mathcal{F}^1}_{jk}(1,s)\right)\right],\label{eq:U_rec}\\
\Omegab_k^p&=-\frac{1}{2}\sum_{j\neq k}\sum_{s\geq 1}\left[\mathbf{P}_{j,s}^{p-1}\overset{s}{\odot}\boldsymbol{\mathcal{F}^2}_{jk}(1,s)+s\mathbf{X}_{j,s}^{p-1}\overset{s}{\odot}\boldsymbol{\mathcal{F}^1}_{jk}(1,s)\right],\label{eq:Omega_rec}\\
\boldsymbol\Phi_{j,1}^{p}&=-\frac{a_k^5}{30}\sum_{j\neq k}\sum_{s\geq 1}\left[\mathbf{P}_{j,s}^{p-1}\overset{s}{\odot}\boldsymbol{\mathcal{F}^1}_{jk}(1,s)\right],\label{eq:Phi1_rec}
\end{align}
and for $q\geq 2$:
\begin{align}
\boldsymbol\Phi^p_{k,q}&=\sum_{j\neq k}\sum_{s\geq 1}\left[\boldsymbol\Phi^{p-1}_{j,s}\overset{s}{\odot}\boldsymbol{\mathcal{F}^{\Phi\rightarrow\Phi}}_{jk}(q,s)+\mathbf{P}^{p-1}_{j,s}\overset{s}{\odot}\boldsymbol{\mathcal{F}^{P\rightarrow\Phi }}_{jk}(q,s)+\mathbf{X}^{p-1}_{j,s}\overset{s}{\odot}\boldsymbol{\mathcal{F}^{X\rightarrow\Phi}}_{jk}(q,s)\right],\label{eq:Phip_rec}\\
\mathbf{P}^p_{k,q}&=\sum_{j\neq k}\sum_{s\geq 1}\left[\boldsymbol\Phi^{p-1}_{j,s}\overset{s}{\odot}\boldsymbol{\mathcal{F}^{\Phi\rightarrow P}}_{jk}(q,s)+\mathbf{P}^{p-1}_{j,s}\overset{s}{\odot}\boldsymbol{\mathcal{F}^{P\rightarrow P}}_{jk}(q,s)+\mathbf{X}^{p-1}_{j,s}\overset{s}{\odot}\boldsymbol{\mathcal{F}^{X\rightarrow P}}_{jk}(q,s)\right],\label{eq:P_rec}\\
\mathbf{X}^p_{k,q}&=\sum_{j\neq k}\sum_{s\geq 1}\left[\boldsymbol\Phi^{p-1}_{j,s}\overset{s}{\odot}\boldsymbol{\mathcal{F}^{\Phi\rightarrow X}}_{jk}(q,s)+\mathbf{P}^{p-1}_{j,s}\overset{s}{\odot}\boldsymbol{\mathcal{F}^{P\rightarrow X}}_{jk}(q,s)+\mathbf{X}^{p-1}_{j,s}\overset{s}{\odot}\boldsymbol{\mathcal{F}^{X\rightarrow X}}_{jk}(q,s)\right],\label{eq:X_rec}
\end{align}
{with the transfer functions above defined in Eqs.~\eqref{eq:transfer1}--\eqref{eq:transfer7}}. Of utmost importance to truncate the reflection process at a fixed order in $\varepsilon$ consistently, {their respective scalings are given in Table~\ref{hydro_interactions}.}

\begin{table}[h]
\begin{tabular}{|c|c|c|c|}
\hline
$\qquad$ & $\mathbf{P}^{p}_{k,q}$ & $\boldsymbol{\Phi}^{p}_{k,q}$ & $\mathbf{X}^{p}_{k,q}$  \\
\hline
$\mathbf{P}^{p-1}_{j,s}$ & $\varepsilon^{s+q-1},\; \varepsilon^{s+q+1}$ & $\varepsilon^{s+q-1},\; \varepsilon^{s+q+1}$ & $\varepsilon^{s+q}$ \\
\hline
$\boldsymbol{\Phi}^{p-1}_{j,s}$ & $\varepsilon^{s+q+1}$ & $\varepsilon^{s+q+1}$ & $0$ \\
\hline
$\mathbf{X}^{p-1}_{j,s}$ & $\varepsilon^{s+q}$ & $\varepsilon^{s+q}$ & $\varepsilon^{s+q}$\\
\hline
\end{tabular}
\caption{{Scaling of the transfer functions between the spherical harmonic coefficients ($q \geq 2$) after each reflection. Note that reflection of potential flows do not yield any torque multipoles i.e. $\boldsymbol{\mathcal{F}^{\Phi\rightarrow X}}_{jk}(q,s)=0$}}
\label{hydro_interactions}
\end{table}

As an example, using the results of Appendix~\ref{sec:app_hydro}, 
\begin{align}
 \boldsymbol{\mathcal{F}^{P\rightarrow P}}_{jk}(2,2) = -\frac{5 a_k^3}{12d^3_{jk}} \biggl[&\frac{(\mathbf{I}\ssb_{jk}\ssb_{jk})^{T_{24}}+(\mathbf{I}\ssb_{jk}\ssb_{jk})^{T_{23}}
 +(\ssb_{jk}\mathbf{I}\ssb_{jk})^{T_{34}}+\ssb_{jk}\mathbf{I}\ssb_{jk}}{2}+\ssb_{jk}\ssb_{jk}\mathbf{I} - 5\ssb_{jk}\ssb_{jk}\ssb_{jk}\ssb_{jk} \biggr]\nonumber\\ 
  -\frac{a_k^5}{12d_{jk}^5} \biggl[&\frac{(\mathbf{I}\mathbf{I})^{T_{23}}+(\mathbf{I}\mathbf{I})^{T_{24}}+\mathbf{II}}{5} -\mathbf{I}\ssb_{jk}\ssb_{jk} -(\mathbf{I}\ssb_{jk}\ssb_{jk})^{T_{23}}-(\mathbf{I}\ssb_{jk}\ssb_{jk})^{T_{24}}-(\ssb_{jk}\mathbf{I}\ssb_{jk})^{T_{34}}-\ssb_{jk}\mathbf{I}\ssb_{jk}
\nonumber \\ &-\ssb_{jk}\ssb_{jk}\mathbf{I}+7\ssb_{jk}\ssb_{jk}\ssb_{jk}\ssb_{jk}\biggr]
\end{align}
and the stresslet induced during reflection $p$ on particle $k$ by the stresslet signature of all the other particles at the previous reflection is:
\begin{align}
\mathbf{P}^p_{k,2} = - \sum_{j \neq k} \frac{5 a_j^3}{12 d_{jk}^3}\biggr[ (\mathbf{P}^{p-1}_{j,2}\cdot \ssb_{jk})\ssb_{jk} + \ssb_{jk}(\mathbf{P}^{p-1}_{j,2}\cdot \ssb_{jk})+ (\mathbf{P}^{p-1}_{j,2}:\ssb_{jk}\ssb_{jk})(\mathbf{I}-5\ssb_{jk}\ssb_{jk})\biggr] +O(\varepsilon^5\,\mathbf{P}^{p-1}_{j,2}).\label{eq:Pp_1}
\end{align}

The results above provide an explicit approach to obtain the successive reflections for the hydrodynamic flow field and to truncate them to a required degree of approximation in $\varepsilon$. Note that the method is completely general and could be applied formally to any low-$\mbox{Re}$ problem involving a suspension of spherical particles. 

\subsubsection{Initialization from the phoretic slip distribution ($p=0$)}
In the context of the present work, i.e. the collective dynamics of phoretic particles, the hydrodynamic problem is initiated by considering the flow field generated by a single isolated particle ($p=0$) with a phoretic slip distribution $\vb_k^0$ at its surface. By definition, $\vb_k^0=M(\nb_k)\grad_s C$ is purely tangential. Also, $a_k\nb_k\cdot(\grad_s\times\vb_k^p)=a_k\nb_k\cdot(\grad_sM\times\grad_s\tilde{c}_k)$ which is strictly zero for particles of uniform mobility. Finally, the surface divergence of $\vb_k^0$ is obtained from the spherical harmonic decomposition of the surface concentration on that particle, Eq.~\eqref{eq:c_surf}. For particles of uniform mobility $M_k$, we finally obtain
\begin{equation}
\nb_k\cdot\vb_k^0=0,\qquad -a_k\grad_s\cdot\vb_k^0=M_k\sum_{q\geq 1}(q+1)\left[q\Ck{0}{q} + \sum_{r\geq 1} (2q+1)\Ck{r}{q}\right]\overset{q}{\odot}[\nb_k\overset{q}{\otimes}\nb_k],\qquad a_k\nb_k\cdot(\grad_s\times\vb_k^p)=0.\label{eq:surf_vel_exp0}
\end{equation}
The last equation above imposes that $\mathbf{X}_{k,q}^0=0$ for all $q$, so that there is no self-rotation associated with phoretic slip for torque-free particles of uniform mobility. 
For force- and torque-free particles of uniform mobility, we finally obtain
\begin{align}
\Ub_k^0&=-\frac{2M_k}{3}\left[\Ck{0}{1} + \sum_{r\geq 1} 3\Ck{r}{1}\right],\qquad \Omegab_k^0=0,\qquad
\boldsymbol\Phi^0_{k,1}=-\frac{a_k^3 \Ub_k^0}{2}, \label{eq:U_zero} \\
\boldsymbol\Phi^0_{k,q\geq 2}&=\frac{a_k^2 \mathbf{P}^0_{k,q}}{2(2q-1)}=\frac{a_k^{q+2} M_k}{2}\left[q\Ck{0}{q} + \sum_{r\geq 1} (2q+1)\Ck{r}{q}\right],\qquad
\mathbf{X}^0_{k,q\geq 1}=0.
\label{eq:P_zero}
\end{align}

{From the above equations, note that the phoretic propulsion velocity, $\Ub_k^0$ for $r \geq 1$, arises only from the first mode ($\Ck{r}{1}$, the source dipole) of the reflected concentration field, which fundamentally corresponds to the gradient of external concentration field at the center of the particle. This implies that the propulsion velocity from the first reflection ($r=1$) is simply the drift created by the superimposed chemical fields of isolated particles.  }

{It may be noted from Eqs.~\eqref{eq:U_zero}--\eqref{eq:P_zero} the existence, when $M$ is uniform, of a direct one-on-one relation between the chemical and hydrodynamic coefficients, which was also observed in Eq.~\eqref{alpha_A}}. It should also be emphasized that the mobility distribution at the surface of the particles only impacts the initialization of the hydrodynamic problem ($p=0$), and not the recursive relations for $p\geq 1$ which are completely general. Although Eqs.~\eqref{eq:U_zero}--\eqref{eq:P_zero} are only valid for particles of uniform mobility, they can be generalized straightforwardly to particles of non-uniform mobility (e.g. Janus particles with different activities and mobilities on both hemispheres), by performing a tensor reduction process to rewrite the modified Eqs.~\eqref{eq:surf_vel_exp0} in terms of fully-symmetric and deviatoric tensors (see Appendix~\ref{sec:app_hydro} for an example of such reduction). This would potentially introduce a non-zero surface vorticity in Eq.~\eqref{eq:surf_vel_exp0}. 

%%%%%%%%%%%%%%%%%%%%%%

\subsection{Chemical vs. hydrodynamic vs. chemo-hydrodynamic interactions}
Performing successive reflections as described in the previous sections then provides a systematic framework to obtain the velocity and rotation rate $(\Ub_k,\Omegab_k)$ in terms of the position and orientation of the different particles $(\Rb_k,\eb_k)$ in the form of a series of terms in increasing powers of $O(\varepsilon)$. Truncating to a particular degree of accuracy  provides a computationally-efficient and asymptotically-consistent approach to determine the collective dynamics of $N$ particles.

This convenient framework also provides a clear understanding of the different interactions routes between the particles, and an explicit way to analyse only certain components of the coupling. Formally, we show below that the particles' velocity  includes four different contributions~\cite{Singh19}:
\begin{enumerate}
\item{\emph{Self-propulsion velocity}: velocity of the isolated particle in an unbounded fluid (no chemical and no hydrodynamic reflections).}
\item{\emph{Chemical interactions}: modification of the particle velocity resulting from the perturbation of its own surface chemical concentration by the presence of the other particles (i.e. chemical reflections with $r\geq 1$ in Sec.~\ref{sec:chem-ref}) but solving for its swimming velocity as if it was hydrodynamically-isolated (i.e. no hydrodynamic reflections).}
\item{\emph{Hydrodynamic interactions}: modification of the particle velocity resulting from the hydrodynamic influence of the other particles (i.e. performing hydrodynamic reflections with $p\geq 1$ in Sec.~\ref{sec:hydro-ref}) but neglecting any chemical influence of the other particles (i.e. no chemical reflections).}
\item{\emph{Chemo-hydrodynamic interactions}: modification to the particle velocity resulting from the hydrodynamic influence of the particles (hydrodynamic reflections with $p\geq 1$) and forced by the modification in surface concentration distribution due to the presence of other particles (chemical reflections with $r\geq 1$).}
\end{enumerate}
In the present framework, it is therefore particularly easy to analyse the effect of one interaction route over another, by simply including or not any chemical and/or hydrodynamic reflections of order $r,p\geq 1$.

It should also be noted that the classical view on phoretic particles' interactions is that of two distinct and independent routes, namely chemical and hydrodynamic couplings. While this dichotomy may be relevant for far-field (dilute) interactions which essentially are limited to two-particle interactions (i.e. the chemical or hydrodynamic influence of particle $i$ on particle $j$'s velocity), the present results emphasize that this does not hold in general and instead reveal the more intricate nature of the particles' coupling: in fact, a third coupling occurs as a result of the dual influence of the chemical and hydrodynamic of particles on each other. This third route, termed here ``chemo-hydrodynamic'' interactions, is fundamentally a three-particle coupling as its simplest occurence involves the chemical influence of particle $i$ on particle $j$'s surface concentration, resulting in a modified flow field near particle $k$ (note that particles $i$ and $k$ may be identical). As a result such interactions only arise at higher order of accuracy and are therefore subdominant in the far-field limit. 
%%%%%%

\section{An $\varepsilon^5$-accurate framework for phoretic particle interactions}
\label{sec:interactions_ep5}

In this section, we apply the previous formalism explicitly and systematically determine the particles' velocity and rotation rate resulting from the different interaction routes described in the previous section, up to an order of accuracy of $\varepsilon^5$, i.e. with the largest asymptotic errors for large distances scaling  as $O(\varepsilon^6)$. This choice of truncature order is motivated by the inclusion at that order of the dominant $3$-particle interactions (i.e. the interaction between two particles due to the presence of a third one) and chemo-hydrodynamic coupling. In principle however, the framework of Section~\ref{sec:reflections} can be repeated to any number of reflections and hence, achieve any stated degree of accuracy.

\subsection{Self-propulsion ($p=0$)}
The leading order contribution  to the particles' velocities corresponds to the self-generated concentration gradients at its surface (i.e. self-propulsion). It is obtained by neglecting any chemical or hydrodynamic interaction with other particles. Hence, no reflection should be performed and using the results of Eqs.~\eqref{eq:U_zero} is obtained as
\begin{equation}
\Ub^{sp}_k = -\frac{2 M_k}{3}\Ck{0}{1} = -\frac{A_1 M_k}{3} \eb_k \qquad \mbox{and} \quad \mathbf{\Omega}^{0}_k=0.\label{eq:selfprop_final}
\end{equation}

\subsection{Chemical interactions between particles}
As for self-propulsion, the hydrodynamic effect of other particles is neglected, hence no hydrodynamic reflections are performed. The chemical interactions correspond to the contributions in the surface concentration moments $\Ck{r}{1}$ with $r\geq 1$:
\begin{equation}
\Ub_k^{\chi}=-2M_k\sum_{r\geq 1}\Ck{r}{1},\qquad \Omegab_k^\chi=0,
\label{eq:u_chem}
\end{equation}
and $\Ck{1}{r}$ with $r\geq 1$ are obtained using the recursive relations, Eq.~\eqref{eq:conc_ref_rec}. Chemical reflections with $r\geq 3$ (i.e. $4$-particle interactions) do not contribute to the $O(\varepsilon^5)$ approximation of the velocity and are therefore ignored. The contribution to the chemical interaction velocity $\Ub_k^\chi$ can therefore be decomposed into two main groups whether (i) they involve the gradient of the concentration field near a given particle and created individually by all its neighbours ($2$-particle interactions, $r=1$) or (ii) they involve the  gradient near the particle of interest of the correction to the concentration field introduced by a second particle due to the presence of a third one ($3$-particle interactions, $r=2$).

\subsubsection{$2$-particle chemical interactions} 

We focus first on the contribution of $r=1$ to Eq.~\eqref{eq:u_chem}, i.e. the concentration gradient created directly by other particles, which is obtained from Eq.~\eqref{eq:conc_ref_rec}. The induced velocity $\Ub^{\chi,r=1}_{k}$ is of order $O(\varepsilon^{s+2})$ where, $s\geq 0$ represents the $s^{\textrm{th}}$ chemical mode. Hence, truncating terms smaller than $\varepsilon^5$,
\begin{equation}
\Ck{1}{1} =  \sum_{j \neq k}\left[C^{0}_{j,0} \bm{\mathcal{F}^\chi}_{jk}(1,0)+\mathbf{C}^{0}_{j,1} \cdot \bm{\mathcal{F}^\chi}_{jk}(1,1)+\mathbf{C}^{0}_{j,2} : \bm{\mathcal{F}^\chi}_{jk}(1,2)+\mathbf{C}^{0}_{j,3} \overset{3}{\odot} \bm{\mathcal{F}^\chi}_{jk}(1,3)\right].
\end{equation}
Using the expression for the transfer function $\bm{\mathcal{F}^\chi}_{jk}(q,s)$ provided in Eq.~\eqref{eq:conc_ref_rec}, the resulting chemical drift velocity is
\begin{align}
\Ub^{\chi,r=1}_{k}=M_k\sum_{j\neq k}&\left[\frac{a_ja_k C_{j,0}^0\ssb_{jk}}{d_{jk}^2}+\frac{a_ka_j^2(3\ssb_{jk}\ssb_{jk}-\mathbf{I})\cdot\mathbf{C}_{j,1}^0}{d_{jk}^3}+\frac{a_ka_j^3}{d_{jk}^4}(\mathbf{C}_{j,2}^0\cdot\ssb_{jk})\cdot(5\ssb_{jk}\ssb_{jk}-2\mathbf{I})\right.\nonumber\\
&\qquad+\left.\frac{a_ka_j^4}{d_{jk}^5}[\mathbf{C}_{j,3}^0:(\ssb_{jk}\ssb_{jk})]\cdot(-3\mathbf{I}+7\ssb_{jk}\ssb_{jk})\right],
\label{eq:chem1}
\end{align}

with $\mathbf{C}_{j,s}^0$ given in terms of the particles' orientation $\eb_j$ in Eq.~\eqref{eq:conc_ref_0}. One recognizes the successive contribution of the first four chemical singularities contributing to the signature of particle $j$ (monopole $C_{j,0}^0$, dipole $\mathbf{C}_{j,1}^0$, quadrupole $\mathbf{C}_{j,2}^0$ and octopole $\mathbf{C}_{j,3}^0$) to the concentration gradient near particle $k$ and its resulting chemical drift. Also note that the leading order term proportional to $C_{j,0}^0$ is  the velocity obtained from the far-field model (Section~\ref{sec:farfield}).

\subsubsection{$3$-particle chemical interactions} 
Proceeding now with the second reflection ($3$-particle interactions), we note that the concentration moments satisfy $\mathbf{C}_{j,q}^1=O(\varepsilon^{q+1})$ (i.e. the velocity induced by 3-particle chemical interactions are $O(\varepsilon^{2q+s+3})$ with $q\geq1$ and $s \geq 0$). Using the expression for the transfer function $\bm{\mathcal{F}^\chi}_{jk}(q,s)$ given in Eq.~\eqref{eq:conc_ref_rec}, the gradient of concentration $\mathbf{C}_{k,1}^2$ near particle $k$ responsible for its chemical drift includes a single $O(\varepsilon^5)$-contribution, namely
\begin{equation}
\mathbf{C}_{k,1}^2=\sum_{j\neq k}\mathbf{C}_{j,1}^1\cdot\bm{\mathcal{F}^\chi}_{jk}(1,1)=\sum_{j\neq k}\sum_{l\neq j}{C}_{l,0}^0\bm{\mathcal{F}^\chi}_{lj}(1,0)\cdot\bm{\mathcal{F}^\chi}_{jk}(1,1),
\end{equation}
and the resulting $3$-particle chemical interaction drift velocity of particle $k$ is obtained as
\begin{equation}
\Ub^{\chi,r=2}_{k}=-M_k\sum_{l}\sum_{j\neq (k,l)}{C}_{l,0}^0\frac{a_ka_j^3a_l(3\ssb_{jk}\ssb_{jk}-\mathbf{I})\cdot\ssb_{lj}}{2 d_{jl}^2d_{jk}^3}\cdot
\label{eq:chem2}
\end{equation}
\begin{figure}[t]
\begin{minipage}{0.45\textwidth}
\centering
\includegraphics[scale=0.6]{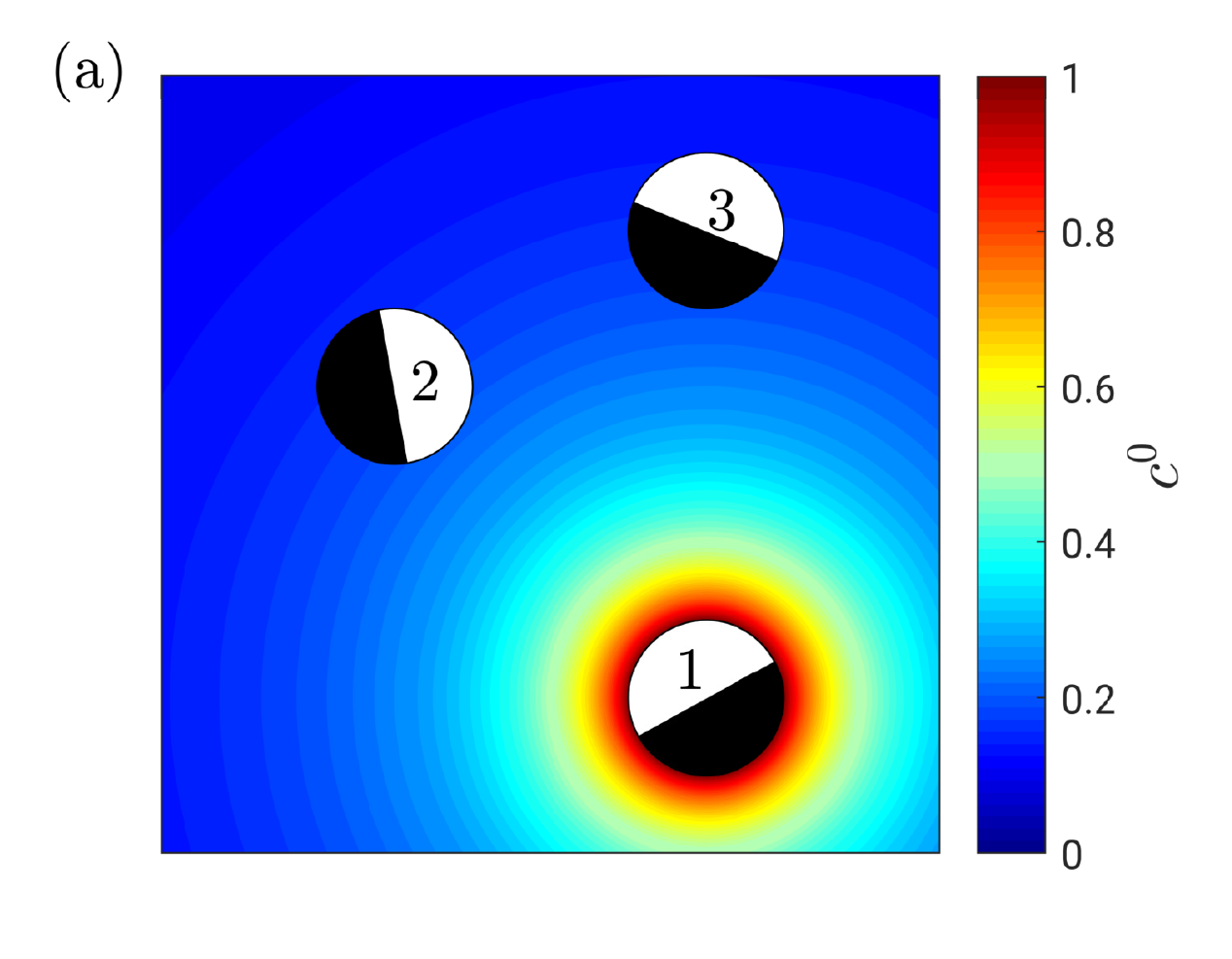}
\end{minipage}
\hspace{2em}
\begin{minipage}{0.45\textwidth}
\centering
\includegraphics[scale=0.6]{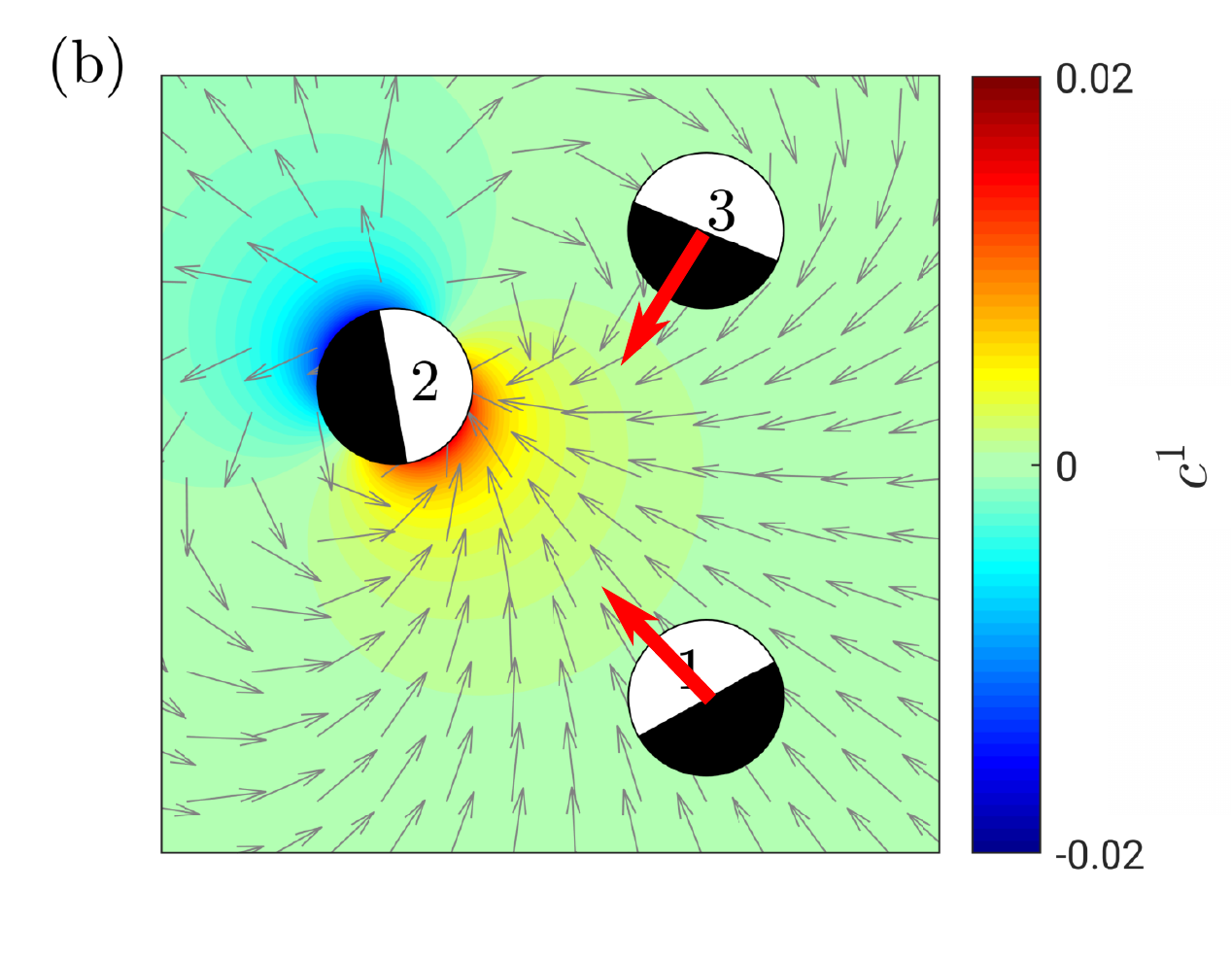}
\end{minipage}
\caption{Illustration of 3-particle chemical interactions arising from a single reflection of the concentration field. The chemical source from particle 1 (left) induces a chemical dipole (right) at the surface of particle 2. In turn, this corrected field and its gradient (arrows) induce a drift of particles 1 and 3. 
}\label{fig:chem3}
\end{figure}

Note that in the previous equation $l=k$ is possible, i.e. this also provides the interaction of particle $k$ with itself due to the presence of a second particle $j$. The sole contribution to the $3$-particle chemical interaction drift is therefore the gradient of concentration generated near particle $k$ by the dipolar correction near particle $j$ due to the monopolar  (Figure~\ref{fig:chem3}). The total velocity induced through purely chemical reflections is hence obtained from Eq. \eqref{eq:chem1} and Eq. \eqref{eq:chem2}.
\begin{equation}
\Ub^\chi_{k} = \Ub^{\chi,r=1}_{k} + \Ub^{\chi,r=2}_{k}.
\end{equation}

We further note from the considerations above that the leading $4$-particle interactions ($r=3$) would be at most $O(\varepsilon^8)$ and all such $4$-particle interactions are therefore ignored here.

\subsection{Drift from purely hydrodynamic interactions}
We turn now to the hydrodynamic drift of particles arising from the flow fields created by their neighbours. For purely hydrodynamic interactions, the flow forcing applied by each particle on the surrounding fluid is that resulting from its own chemical signature (i.e. no chemical reflections): hydrodynamic reflections are thus initiated with Eqs.~\eqref{eq:U_zero}--\eqref{eq:P_zero} using $\Ck{q}{0}$ defined in Eqs.~\eqref{eq:conc_ref_0}, and recursive relations in Eqs.~\eqref{eq:U_rec}--\eqref{eq:X_rec} are used to obtain the hydrodynamic drifts $\Ub_k^h$ and $\Omegab_k^h$:
\begin{align}
\Ub_k^h=\sum_{p\geq 1}\Ub^{p}_k \quad \textrm{and,} \quad \Omegab_k^h=\sum_{p\geq 1}\Omegab^{p}_k
\label{eq:upk}
\end{align}
where $\Ub^p_k$ an $\Omegab^p_k$ are defined in Eqs.~\eqref{eq:U_rec}--\eqref{eq:Omega_rec}; the transfer functions are given in Appendix \ref{sec:app_hydro}.

\subsubsection{$2$-particle hydrodynamic interactions}

For 2-particle interactions ($p=1$), the correction to propulsion velocity induced by a force multipole $\mathbf{P}_{j,s}^0$ of order $s$ is $O(\varepsilon^{s})$ for the translational velocity and of $O(\varepsilon^{s+1})$ for the angular velocity (with $s \geq 2$ in both cases). Similarly, the correction to propulsion velocity from a potential multipole $\boldsymbol\Phi_{j,s}^0$ of order $s$ is $O(\varepsilon^{s+2})$ (with $s \geq 1$), and there are no rotlet multipoles in the signature of an isolated phoretic particle of uniform mobility ($\mathbf{X}_{j,s}^0=0$). $\boldsymbol\Phi_{j,s}^0$ and $\mathbf{P}_{j,s}^0$ are $O(1)$ quantities, and using Table~\ref{hydro_interactions}, the drifts with $p=1$ in Eq.~\eqref{eq:upk} are obtained by retaining terms that are $O(\varepsilon^5)$ or larger:
\begin{align}
\Ub^{h,p=1}_k = \sum_{j\neq k}\biggl[ \boldsymbol\Phi_{j,1}^0 \cdot \boldsymbol{\mathcal{F}^1}_{jk}(1,1)  +& \boldsymbol\Phi_{j,2}^0 : \boldsymbol{\mathcal{F}^1}_{jk}(1,2) + \mathbf{P}_{j,2}^0:\left(\boldsymbol{\mathcal{F}^3}_{jk}(1,2)+ \frac{a_k^2}{6}\boldsymbol{\mathcal{F}^1}_{jk}(1,2)\right) + \boldsymbol\Phi_{j,3}^0 \overset{3}{\odot} \boldsymbol{\mathcal{F}^1}_{jk}(1,3) \nonumber \\ +& \mathbf{P}_{j,3}^0\overset{3}{\odot}\left(\boldsymbol{\mathcal{F}^3}_{jk}(1,3)+\frac{a_k^2}{6}\boldsymbol{\mathcal{F}^1}_{jk}(1,3)\right)+\mathbf{P}_{j,4}^0\overset{4}{\odot}\boldsymbol{\mathcal{F}^3}_{jk}(1,4)+\mathbf{P}_{j,5}^0\overset{s}{\odot}\boldsymbol{\mathcal{F}^3}_{jk}(1,5)\biggr],
\label{eq:hydro_Urefln1}\\
\Omegab^{h,p=1}_k=-\frac{1}{2}\sum_{j\neq k}\biggl[\mathbf{P}_{j,2}^0:\boldsymbol{\mathcal{F}^2}_{jk}(1,2) &+ \mathbf{P}_{j,3}^0 \overset{3}{\odot} \boldsymbol{\mathcal{F}^2}_{jk}(1,3) + \mathbf{P}_{j,4}^0 \overset{3}{\odot} \boldsymbol{\mathcal{F}^2}_{jk}(1,4)  \biggr].
\label{eq:hydro_Omrefln1}
\end{align} 
As expected, only force multipoles contribute to the rotation of the particles (potential flows do not create any vorticity). The strength of the different multipoles in the previous equations are directly related to the multipoles of concentration using Eqs.~\eqref{eq:P_zero} (e.g. $\Phi_{j,1}^0=a_j^3 M_j \mathbf{C}^0_{j,1}/3$,   $\mathbf{P}_{j,2}^0= 6 a_j^2 M_j \mathbf{C}^0_{j,2}$ and so on). Using the definition of the transfer functions provided in Appendix~\ref{sec:app_hydro}, the $\varepsilon^5$-accurate $2$-particle hydrodynamic interaction velocities are finally obtained as 

\begin{align}
\Ub^{h,p=1}_k = \sum_{j\neq k} & \biggl[\frac{M_j a_j^3}{3 d_{jk}^3} \Cj01 \cdot (\mathbf{I}-3\ssb_{jk}\ssb_{jk}) + M_j(\mathbf{C}_{j,2}^0\cdot\ssb_{jk})\cdot\biggl(\frac{3a_j^2}{d_{jk}^2}\ssb_{jk}\ssb_{jk}+\frac{a_j^2(a_j^2+a_k^2)}{d_{jk}^4}\left(2\mathbf{I}-5\ssb_{jk}\ssb_{jk}\right)\biggr) \nonumber \\ & +M_j(\mathbf{C}_{j,3}^0:\ssb_{jk}\ssb_{jk})\cdot\biggl(-\frac{3a_j^3}{2d_{jk}^3}\left(\mathbf{I}-5\ssb_{jk}\ssb_{jk}\right)+\frac{a_j^3(3a_j^2+5a_k^2)}{d_{jk}^5}\left(3\mathbf{I}-7\ssb_{jk}\ssb_{jk}\right)\biggr) \nonumber \\  & - \frac{3 a_j^4 M_j}{d_{jk}^4}(\mathbf{C}^0_{j,4}\overset{3}{\odot} \ssb_{jk}\ssb_{jk} \ssb_{jk}) \cdot (\mathbf{I} +4 \ssb_{jk}\ssb_{jk}) - \frac{15 a_j^5 M_j}{2 d_{jk}^5}(\mathbf{C}^0_{j,5}\overset{4}{\odot} \ssb_{jk}\ssb_{jk} \ssb_{jk}\ssb_{jk}) \cdot (\mathbf{I} -3 \ssb_{jk}\ssb_{jk})\biggr],\label{eq:hydro1a}\\
\Omegab^{h,p=1}_k  = \sum_{j \neq k} &\biggl[\frac{3 a_j^2 M_j}{d_{jk}^3}\Big(\ssb_{jk}\times [\Cj02 \cdot \ssb_{jk}]\Big) - \sum_{j \neq k} \frac{15 a_j^3 M_j}{2d_{jk}^4}\Big(\ssb_{jk}\times [\Cj03 : \ssb_{jk} \ssb_{jk}]\Big)  - \frac{14 a_j^4 M_j}{d_{jk}^5}\Big(\ssb_{jk}\times [\Cj04 \overset{3}{\odot} \ssb_{jk} \ssb_{jk} \ssb_{jk}]\Big)\biggr].\label{eq:hydro1b}
\end{align}

 %%%%%%%%%%%

\subsubsection{$3$-particle hydrodynamic interactions}

\begin{figure}[t]
\begin{minipage}{0.45\textwidth}
\centering
\includegraphics[scale=0.6]{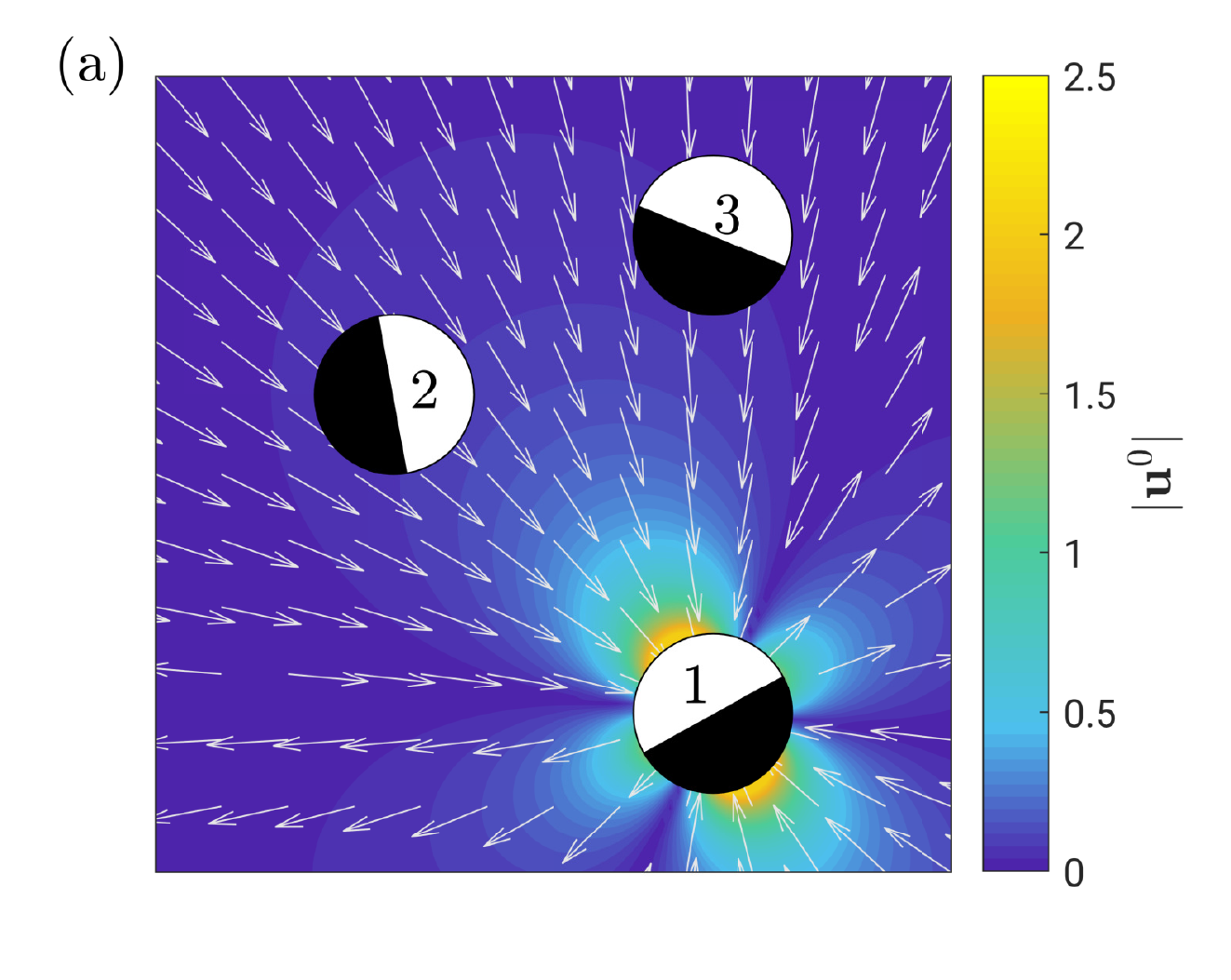}
\end{minipage}
\hspace{2em}
\begin{minipage}{0.45\textwidth}
\centering
\includegraphics[scale=0.6]{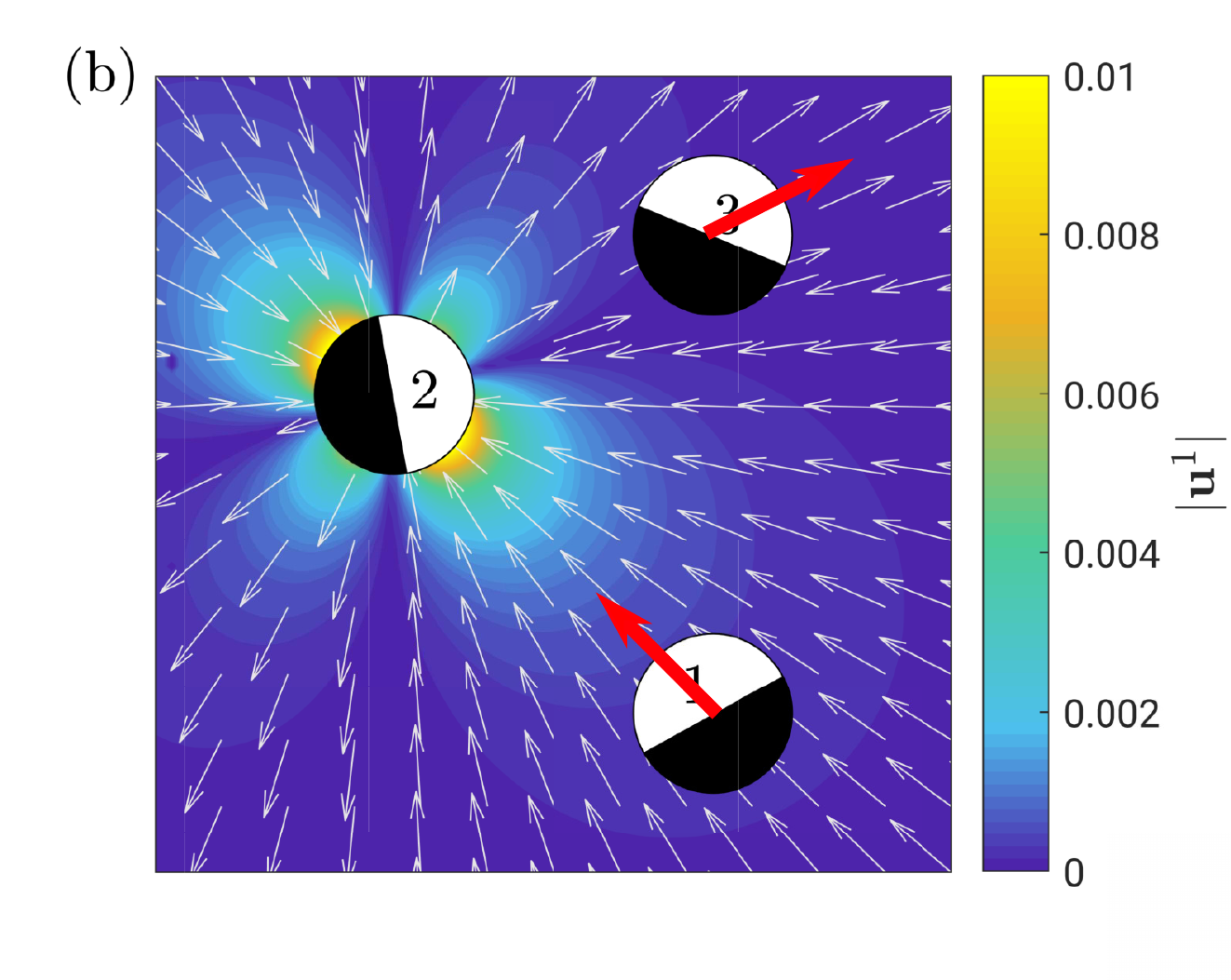}
\end{minipage}
\caption{Illustration of 3-particle hydrodynamic interactions resulting from a single hydrodynamic reflection: The stresslet induced by the self-propulsion of particle 1 (a) induces a reflected stresslet at particle 2 (b). In turn, this modifies the hydrodynamic environment of particle 1 and 3 and induces their hydrodynamic drift (red arrow). The velocity magnitude (color) and direction (white arrow) are shown. Note that a rotation is also induced but scales as $O(\varepsilon^6)$ and is neglected here.
}\label{fig:hydro3}

\end{figure}

The slowest decaying transfer function listed in Table~\ref{hydro_interactions} corresponds to the stresslet induced by a stresslet on another particle at the previous reflection and scales as $\varepsilon^3$. The slowest-decaying $3$-particle interaction therefore corresponds to the hydrodynamic drift of particle $k$ associated with the stresslet induced by particle $j$ after reflection of the flow field generated by the stresslet of particle $l$ (Figure~\ref{fig:hydro3}), and its dominant contribution scales as $\varepsilon^5$:

\begin{align}
\Ub^{h,p=2}_k & = \mathbf{P}^1_{j,2} : \boldsymbol{\mathcal{F}^3}_{jk}(1,2) = \sum_{l \neq j}(\mathbf{P}^0_{l,2} : \boldsymbol{\mathcal{F}^{P\rightarrow P}}_{lj}(2,2)) : \boldsymbol{\mathcal{F}^3}_{jk}(1,2), 
\end{align}

Knowing $\boldsymbol{\mathcal{F}^{P\rightarrow P}}_{lj}(2,2)$ from equation \ref{eq:Pp_1} and remembering $\mathbf{P}_{l,2}^0= 6 a_l^2 M_l \mathbf{C}^0_{l,2}$
\begin{align}
\Ub^{h,p=2}_k =  - \sum_{l}\sum_{j \neq (k,l)} \frac{5 a_j^3a_l^2M_l}{2 d_{lj}^3 d_{jk}^2}\biggr[ 2(\ssb_{jk}\cdot\mathbf{C}^0_{l,2}\cdot \ssb_{lj})(\ssb_{jk}\cdot\ssb_{lj})+ (\ssb_{lj}\cdot\mathbf{C}^0_{l,2}\cdot\ssb_{lj})(1-5(\ssb_{lj}\cdot\ssb_{jk})^2)\biggr]\ssb_{jk}.\label{eq:hydro2}
\end{align}
An illustration of the drift created by this 3 particle hydrodynamic interaction is shown in figure \ref{fig:hydro3}. The induced rotation from $3$-particle hydrodynamic interactions scales as $O(\varepsilon^6)$ and is therefore ignored here.
Indeed, rotational effects of the stresslet $\mathbf{P}_{j,2}^1$ considered above is $O(\varepsilon^6)$. The only other singularity that can contribute to $\Omega_k^2$, namely the rotlet dipole $\mathbf{X}_{j,2}^1$, has an $O(\varepsilon^3)$ intensity (see Table~\ref{hydro_interactions}) and the associated rotation rate is therefore $O(\varepsilon^7)$.

\subsection{Drift from chemo-hydrodynamic interactions}

A third type of interactions arise when accounting for reflections both in the hydrodynamic and chemical problems between at least 3 particles. These are \emph{chemo-hydrodynamic interactions}, which are the hydrodynamic drifts generated by a given particle on its neighbors as a {result of their reflected chemical signature}. Such {multi-body} interactions are completely absent in the far-field model (Section~\ref{sec:farfield}) as these frameworks solely focused on pairwise and direct interactions of particles. They also correspond to higher-order corrections of the particles' velocity and therefore become particularly important in not-so-dilute regimes. In the following, we show that the leading-order chemo-hydrodynamic interactions is $O(\varepsilon^5)$. 

\begin{figure}[t]
\begin{center}
\begin{tabular}{ccc}
\includegraphics[width=.33\textwidth]{Fig2a.pdf}& 
\includegraphics[width=.33\textwidth]{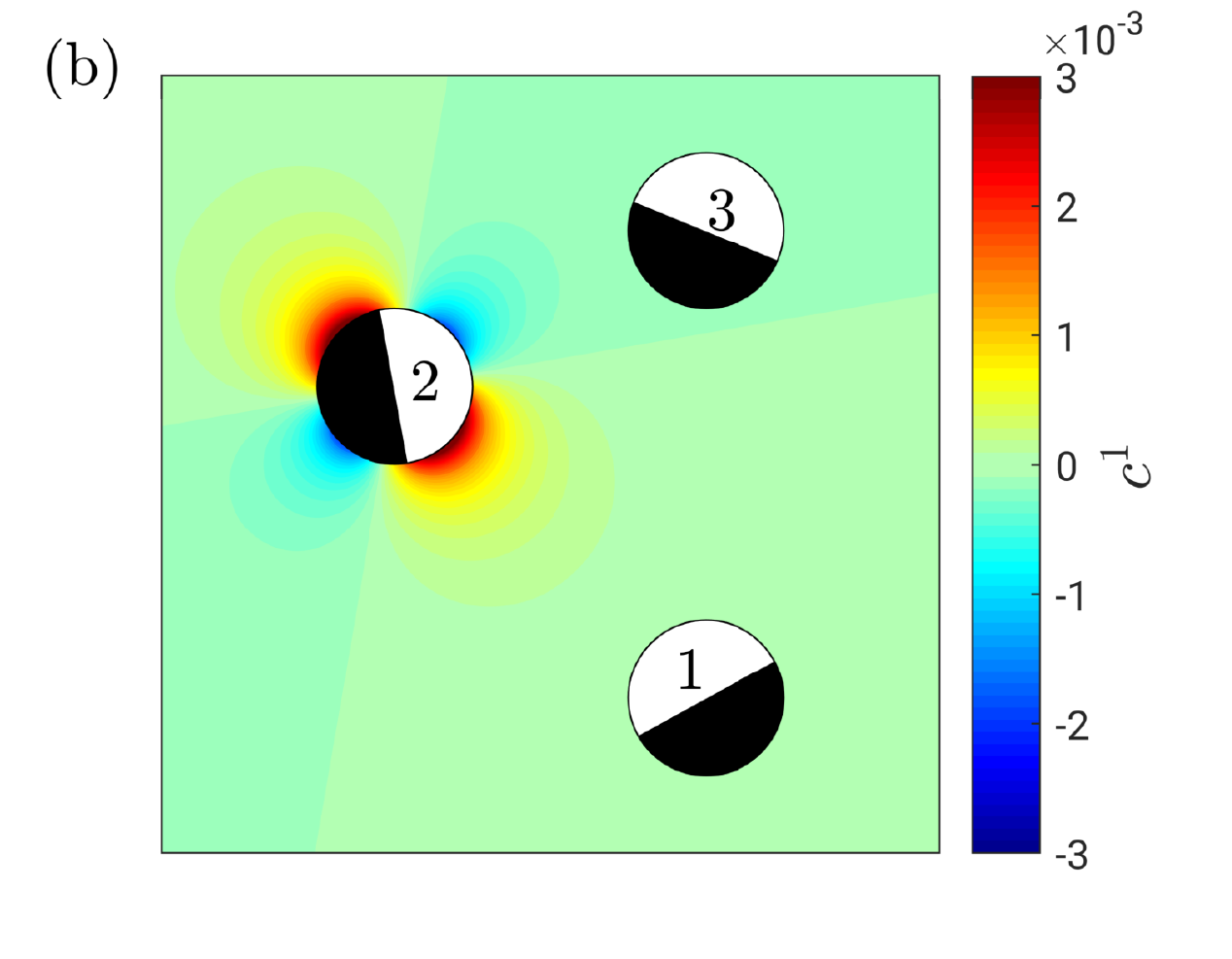} &
\includegraphics[width=.33\textwidth]{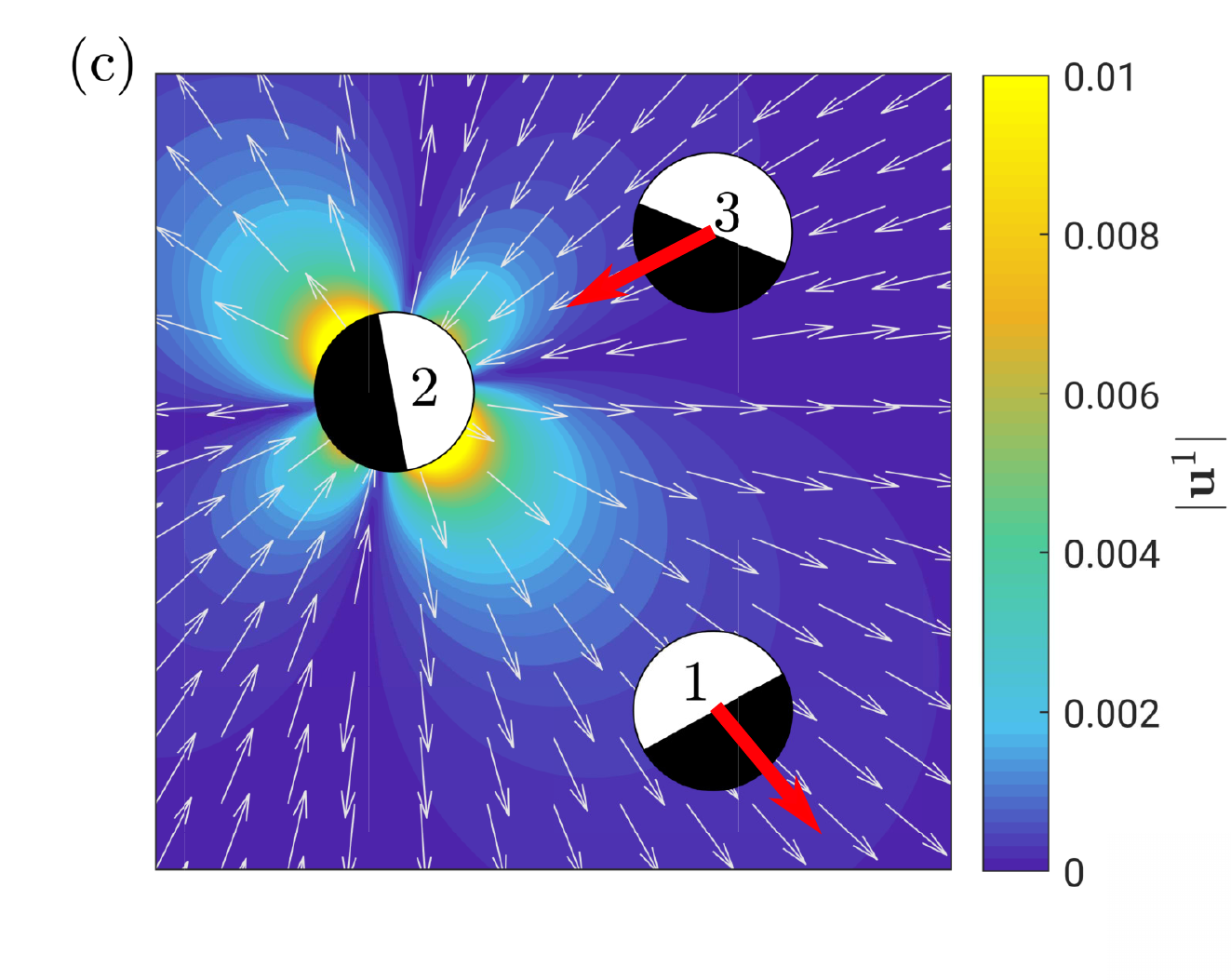} 
\end{tabular}
\caption{Illustration of one of the dominant chemo-hydrodynamic interactions resulting from a single reflection of the concentration field: (a) the chemical source from particle 1 induces a quadrupolar correction of the concentration field near particle 2 (b). This source quadrupole induces a hydrodynamic stresslet (c) which is responsible for the drift of particles 1 and 3 (red arrows). In (a) and (b), the concentration fields are shown, while (c) shows the velocity magnitude (color) and direction (white arrows).}
\label{fig:chemhydr}
\end{center}
\end{figure}

From a practical point of view, hydrodynamic reflections are initiated with Eqs.~\eqref{eq:U_zero}--\eqref{eq:P_zero} using $\Ck{r\geq 1}{q}$ (chemical reflections), and recursive relations in Eqs.~\eqref{eq:U_rec}--\eqref{eq:X_rec} are used to obtain the hydrodynamic drifts $\Ub_k^{\chi h}$ and $\Omegab_k^{\chi h}$. 
The dominant such contribution involves three particles (one chemical reflection, $r=1$, and one hydrodynamic reflection, $p=1$. A force multipole of order $q\geq 2$, $\mathbf{P}_{j,q}^0$, generated by the reflected $O(\varepsilon^{q+1})$ concentration multipole $\mathbf{C}_{j,q}^1$, Eq.~\eqref{eq:P_zero}, results in a $O(\varepsilon^{2q+1})$ drift velocity $\Ub_k^{\chi h}$ on a third particle. Similarly, a potential multipole of order $q\geq 1$, $\boldsymbol\Phi_{j,q}^0$, generated by the $O(\varepsilon^{q+1})$ reflected concentration multipole $\mathbf{C}_{j,q}^1$, Eq.~\eqref{eq:P_zero}, results in a $O(\varepsilon^{2q+3})$ drift velocity $\Ub_k^{\chi h}$. The two dominant interactions, which scale as $O(\varepsilon^5)$, therefore correspond to (i) the drift on particle $k$ induced by the potential dipole of particle $j$ created by the chemical dipole of particle $l$, and (ii) the drift on particle $k$ induced by the stresslet of particle $j$ resulting from the chemical quadrupole of particle $l$ (Figure~\ref{fig:chemhydr}), all other interactions being subdominant. Using Eqs.~\eqref{eq:U_rec}, \eqref{eq:Phi1_rec}, \eqref{eq:U_zero} and \eqref{eq:P_zero}, the dominant chemo-hydrodynamic drift is obtained as
\begin{align}
\Ub^{\chi h}_k = \sum_{j \neq k} \biggl[\boldsymbol\Phi_{j,1}^0 \cdot \boldsymbol{\mathcal{F}^1}_{jk}(1,1) + \mathbf{P}_{j,2}^0:\boldsymbol{\mathcal{F}^3}_{jk}(1,2)\biggr],\qquad \textrm{with   }\boldsymbol\Phi_{j,1}^0=M_ja_j^3\mathbf{C}_{j,1}^1,\qquad \mathbf{P}_{j,2}^0=15M_ja_j^2\mathbf{C}_{j,2}^1.
\end{align}
which is finally obtained explicitly using Eq.~\eqref{eq:conc_ref_rec}
\begin{align}
\Ub^{\chi h}_k & = \sum_{l}\sum_{j\neq(l,k)} M_jC_{l,0}^0\biggl[ \frac{a_l a_j^4}{2 d_{jk}^3 d_{lj}^2} \ssb_{lj} \cdot [3 (\ssb_{jk}\cdot\ssb_{lj})\ssb_{jk}-\ssb_{lj}] +\frac{5 a_l a_j^4}{2 d_{jk}^2 d_{lj}^3} [3 (\ssb_{lj}\cdot\ssb_{jk})^2-1]\ssb_{jk}\biggr].\label{eq:chemhydro}
\end{align}

It should be noted that any rotation induced by $3$-particle chemo-hydrodynamic interactions is at most $O(\varepsilon^6)$ and is therefore ignored. 

\vspace{.5cm}

Assembling the contributions to the interactions velocities provided in Eqs.~\eqref{eq:selfprop_final} (self-propulsion), Eqs.~\eqref{eq:chem1} and~\eqref{eq:chem2} (purely chemical interactions), Eqs.~\eqref{eq:hydro1a}, \eqref{eq:hydro1b} and \eqref{eq:hydro2} (purely hydrodynamic interations and Eq.~\eqref{eq:chemhydro} (chemo-hydrodynamic interactions) provide a consistent asymptotic approximation of the particles' dynamics with a $\varepsilon^5$ accuracy. It should be noted that a similar approach can be used to obtain velocities with a prescribed arbitrary accuracy of $\varepsilon^n$ with $n\geq 6$.

%%%%%%%%%%%%%%%%%%%%%%%%%%%%%%%%%%%%%%%%%%%

\section{Dynamics of multiple Janus phoretic particles}
\label{sec:validations}
In this section, the $\varepsilon^5$-accurate framework based on the Method of Reflections proposed in the previous section (thereafter referred to as MoR) is used to compute the dynamics of multiple active Janus particles, and its predictions are compared with the exact solution of the full interaction problem (obtained either analytically or numerically depending on the problem's symmetries) and simple far-field approximations (Section~\ref{sec:farfield}). This provides both a validatation of these results as well as the opportunity to analyze the accuracy gained in the description of the collective dynamics by accounting for higher-order interactions (in particular, $3$-particle and chemo-hydrodynamic interactions).  

Note that the present framework, in its long-range asymptotic formulation, is expected to be particularly accurate for large particle distances but does not include intrinsically a description of the lubrication interactions of particles. Further, phoretic interactions may be attractive in the near-range~\cite{yariv16}. To prevent particles' overlapping each other,  steric repulsion is accounted for by implementing an additional repulsive velocity between any pair of particles $(j,k)$, 
\begin{equation}
\mathbf{u}^{\mbox{rep}}_{jk} = -A\;\Big[1-\tanh\Big(\frac{d_{c_{jk}}}{\delta_\textrm{rep}}\Big)\Big],
\end{equation}
with $d_{c_{jk}}=d_{jk}-a_j-a_k$ the contact distance between particles $j$ and $k$. In the following, we use $A=35$ and $\delta_\textrm{rep}=0.04$, so that this repulsion velocity is sufficient to prevent the particles' overlap but is only significant when the particles' surfaces are distant by less than about a tenth of their radii~\cite{Varma18}.

\subsection{Axisymmmetric relative translation of two Janus particles}
\label{sec:axisymm}

\begin{figure}[t]
\begin{minipage}{0.45\textwidth}
\centering
\includegraphics[scale=0.6]{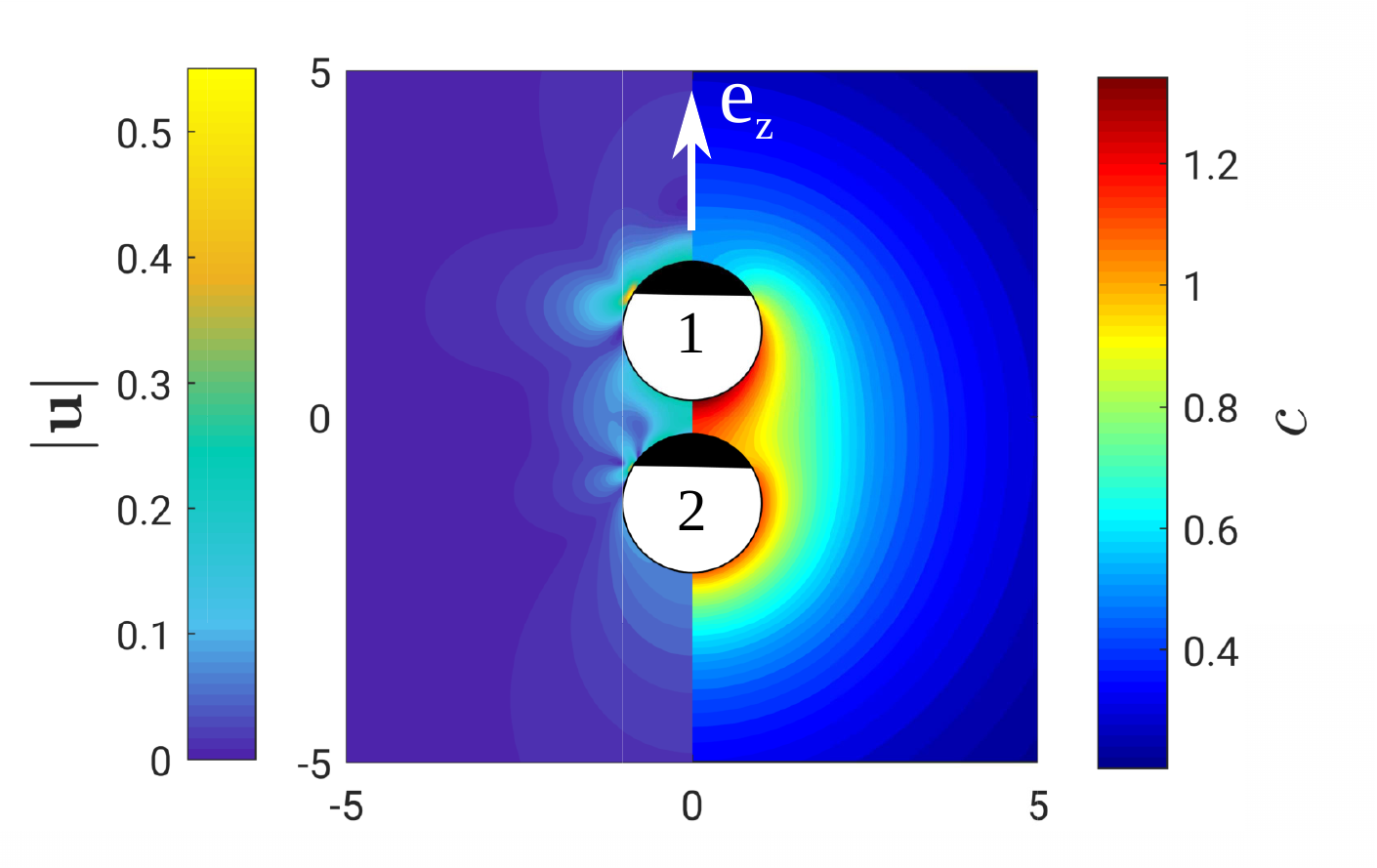}
\end{minipage}
\begin{minipage}{0.45\textwidth}
\centering
\end{minipage}
\begin{minipage}{0.25\textwidth}
\centering
\end{minipage}
\caption{Interactions of two aligned Janus particles: (left) flow velocity magnitude obtained using BEM and (right) concentration field obtained analytically (Appendix~\ref{app:bispherical}). Both Janus particles have positive mobility ($M=1$) and equal unit radius, with 3/4th of their surface releasing solute at a fixed rate ($A=1$, white region) while the rest of their surface is inert ($A=0$, black region). The particles have a contact distance $d_c=d-2=0.5$, and swim toward their inert cap when isolated (i.e. along $+\eb_z$).} 
\label{fig:axisymm_conc}
\end{figure}

 We first consider the case of two Janus particles arranged axisymmetrically, both aligned in the same direction as shown in Figure~\ref{fig:axisymm_conc}. Both  particles have unit radius ($a=1$) and uniform and positive mobility ($M=1$); $3/4$-th of their surface is active ($A=1$), the rest being inert ($A=0$). In isolation, each particle would hence swim with a velocity $\Ub^\textrm{self}= 3/16\;\eb_z$. A $3/4$-th active Janus is chosen here so as to test the framework with the most generic chemical and hydrodynamic reflections computed in Section~\ref{sec:interactions_ep5} (hemispheric Janus particles of uniform mobility have no intrinsic stresslet). 
 
In this highly-symmetric setting, the chemical and hydrodynamic fields as well as the particles' velocities can be obtained analytically for an arbitrary distance using bispherical coordinates~\cite{Michelin15,Varma18} (Appendix~\ref{app:bispherical}).  The resulting flow and concentration fields are reported on Figure~\ref{fig:axisymm_conc}. In the gap between the particles, the diffusion of the solute emitted from particle 1's active cap is limited by the confining effect of particle 2's proximity, leading to increased levels of concentration and modified slip velocity at the particles' surface in this region. The resulting hydrodynamic field is further modified by lubrication effects at close contact. 

\begin{figure}[t]
\begin{center}
\begin{tabular}{ccc}
\includegraphics[height=7cm]{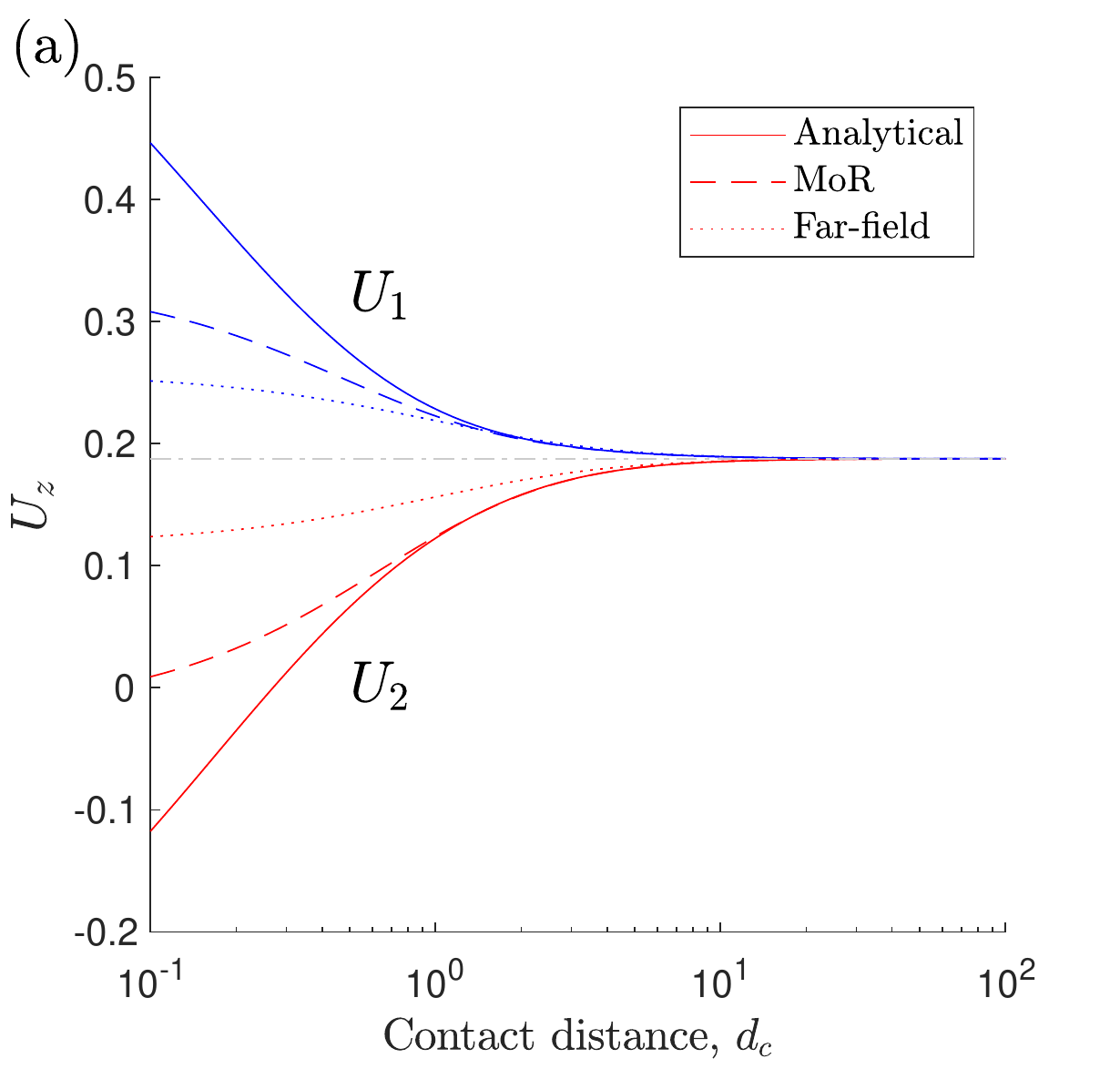} &\hspace{2cm} &
\includegraphics[height=7cm]{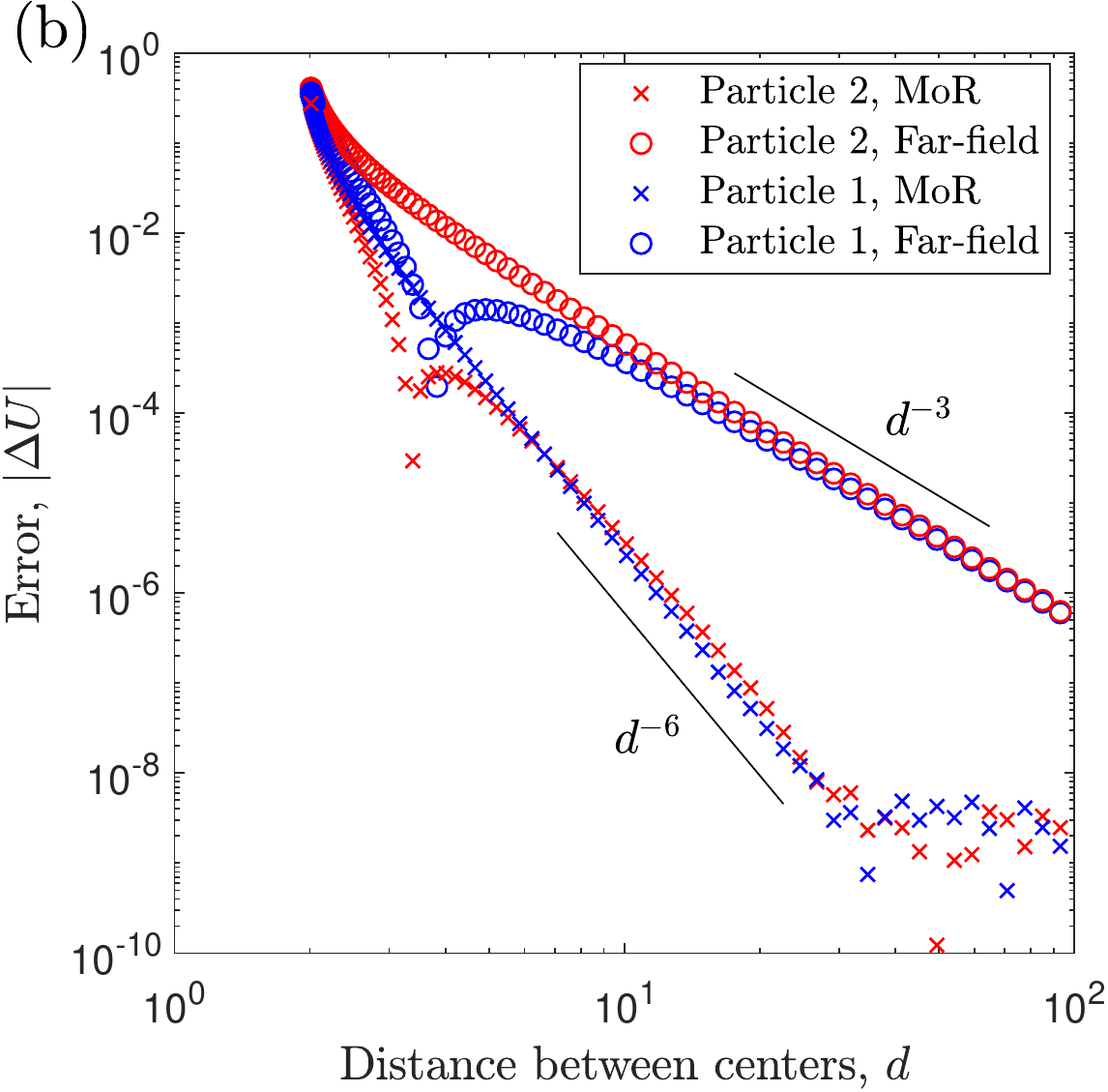}
\end{tabular}
\caption{Translation of two aligned Janus particles (see Figure~\ref{fig:axisymm_conc}): (a) Swimming velocities of particle 1 (blue) and particle 2 (red) as a function of their contact distance $d_c=d-2$, as obtained analytically (solid) or using MoR (dashed) or far-field models (dotted). The reference self-propulsion velocity of an isolated particle, $U^{\mbox{\small self}} = 0.1875 $ is also shown (dot-dashed).  (b) Error magnitude $|\Delta U|$ in the velocity prediction of far-field and MoR models with respect to the analytical solution. 
}

\label{fig:axi_U}
\end{center}
\end{figure}

Due to this confinement-induced modification of the concentration field, the contrast between the front and back of the leading particle 1 is enhanced, while it is reduced for the trailing particle 2, leading to an increased velocity of the former and a reduced velocity for the latter (see Figure~\ref{fig:axi_U}). In fact, the trailing particle is brought to rest at contact distance $d_c= 0.27$, and further reduction in contact distance leads to reversal in its swimming direction. Moreover, since $U_2\leq U_1$ for all $d_c$, particles drift away from each other. 

As seen in Figure~\ref{fig:axi_U}, the reduction (resp. enhancement) of the velocity of the trailing particle (resp. leading) particle is captured by the far-field and MoR models. Moreover, both underestimate the velocity of particle 1 and overestimate that of particle 2 when the particles are close ($d_c<1$). The propulsion velocity predicted using only far-field model deviate from analytical solution below contact distances of a few radii while that predicted using MoR provides a good estimate even for contact distances slightly smaller than a particle radius. Asymptotically, when $\varepsilon=1/d\ll 1$, the expected error scalings are observed, i.e. $O(\varepsilon^3)$ for the far-field approximation and $O(\varepsilon^6)$ for the MoR model (Figure~\ref{fig:axi_U}b).

\begin{figure}[t]
\begin{center}
\begin{tabular}{ccc}
\includegraphics[height=7cm]{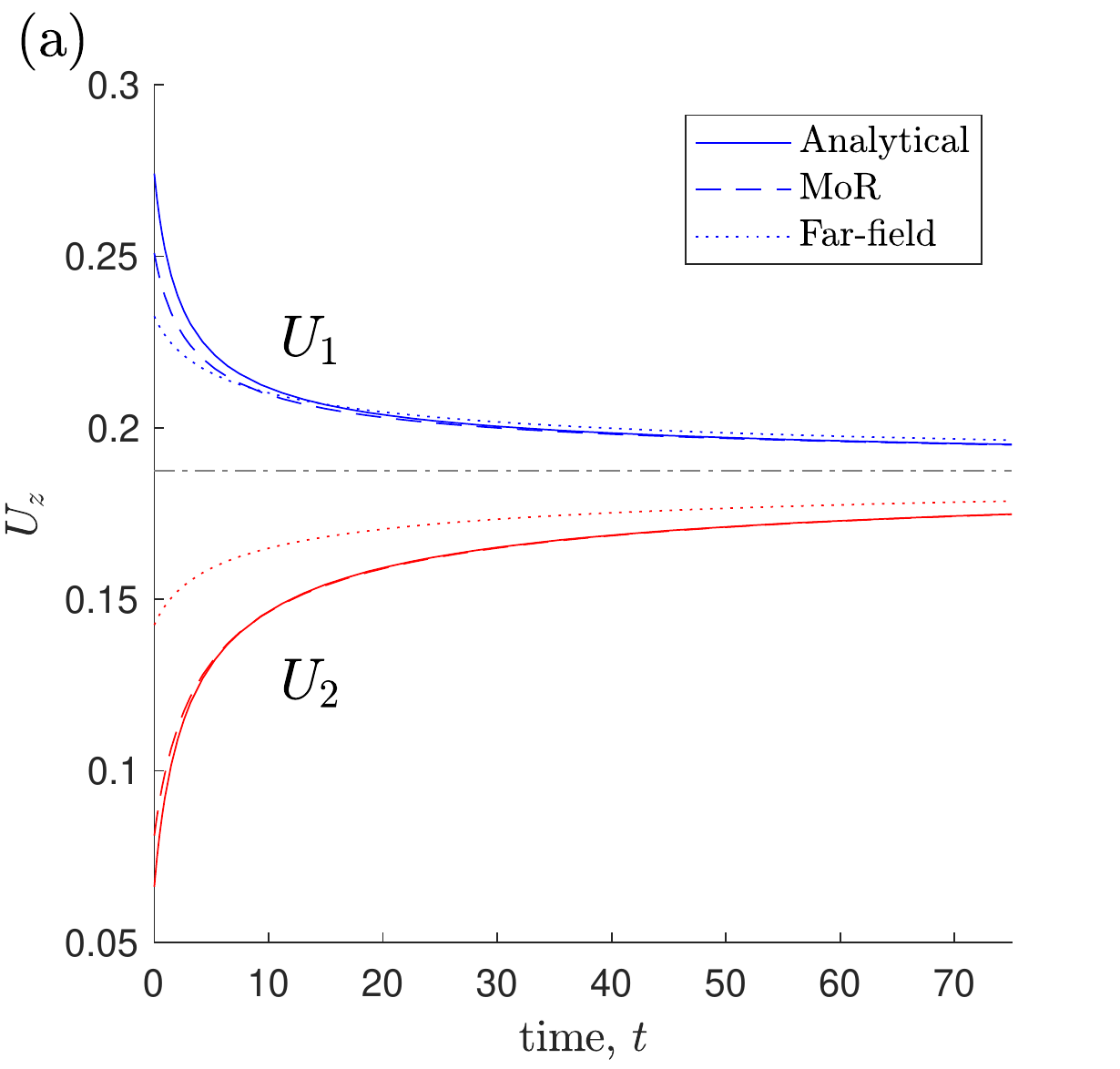} &\hspace{2cm} &
\includegraphics[height=7cm]{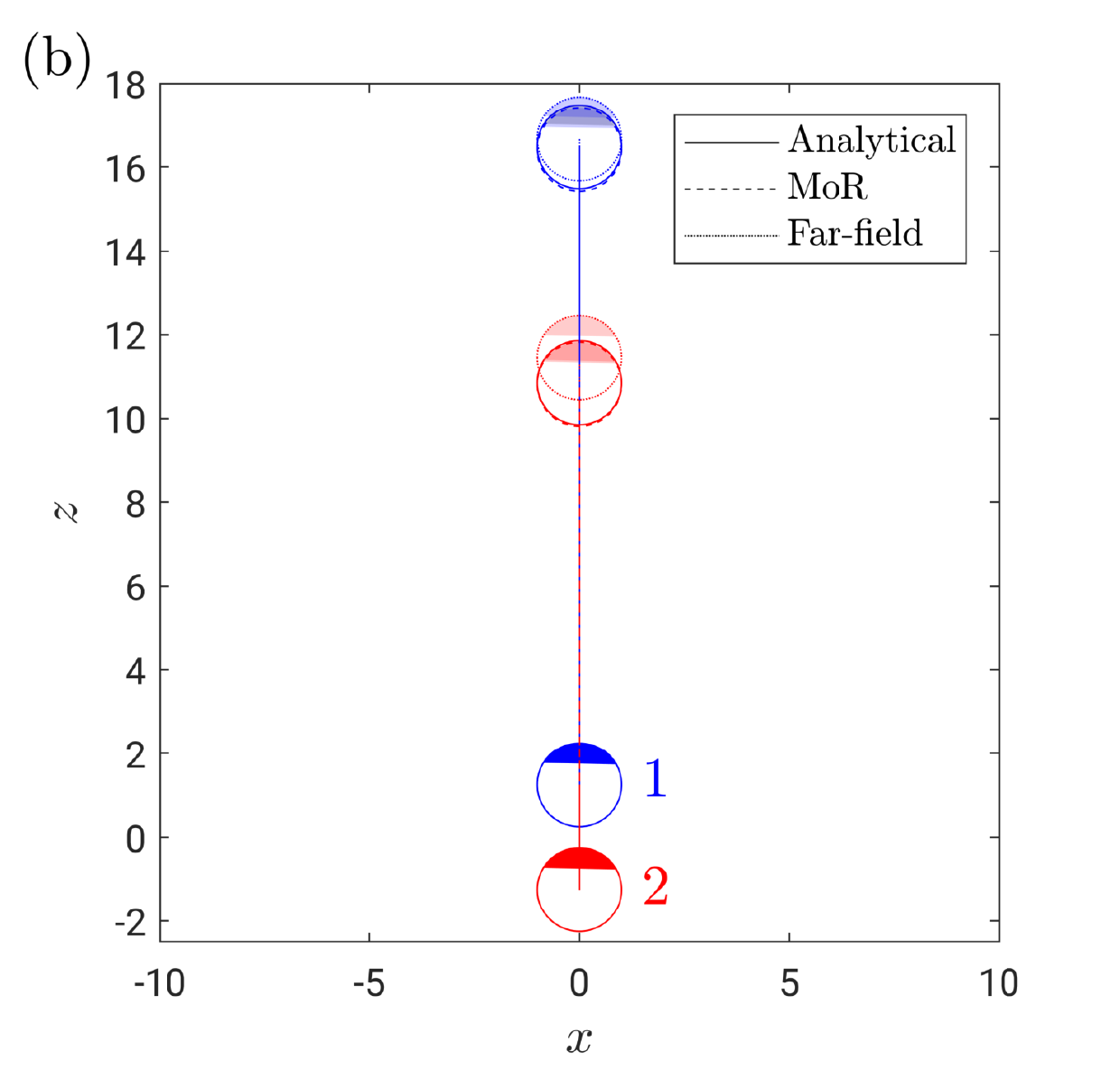}
\end{tabular}
\caption{Translation of two aligned Janus particles (see Figure~\ref{fig:axisymm_conc}): (a) Evolution in time of the particles' velocity for an initial separation distance $d_c=0.5$. The predictions for MoR and far-field modes are computed at the relative positions described analytically. (b) Trajectories of the particles. Positions of particles at $t=0$ and $t=75$ are shown. 
}

\label{fig:axi_vel}
\end{center}
\end{figure}

The previous considerations focused on instantaneous velocity predictions (for a fixed geometry). We now evaluate the far-field and MoR models performance in predicting the long-term dynamics of two particles  initially positioned at $d_c=0.5$ (Figure~\ref{fig:axi_vel}). The particles swim in the same direction but drift apart as $U_1>U_2$. As time progresses, their relative influence and resulting relative drift reduces, and both particles approach their self-propulsion velocity asymptotically (Figure~\ref{fig:axi_vel}a). Even for small separation (e.g. $d_c=0.5$), MoR-predicted propulsion velocities have a good accuracy (the error for particle 2 when $d_c=0.5$ is $\biggr|\frac{U_2^{\textrm{mor}}-U_2^{\textrm{analytical}}}{U_2^{\textrm{\scriptsize self}}-U_2^{\textrm{analytical}}}\biggr| \times 100 \approx 15 \%$), while errors introduced by the far-field model are large ($\approx 60\%$). The cumulated error in position over time (when the particles are far away from each other) is essentially negligible for MoR, while it is of the order of the particle radius for the far-field model (Figure \ref{fig:axi_vel}b).

\subsection{Co-planar translation and rotation of $2$ Janus particles}

We next focus on the coplanar and non-axisymmetric motion of two Janus particles. In contrast with the previous highly-symmetric situation, a critical element for the prediction of the particles' trajectory lies in the correct estimation of their rotation velocities (which arise from interactions with their neighbours as particles with homogeneous mobility do not rotate when isolated). 
The axisymmetry of a pair of Janus particles is lost as soon as they are not aligned with their relative position, and while a solution in bispherical coordinates remains available in principle, it becomes rapidly cumbersome~\cite{sharifimood16}. Instead, the particles' velocities are obtained here numerically using the regularized Boundary Element Methods framework for phoretic particles (regBEM), a versatile numerical technique developed by Montenegro-Johnson \emph{et. al.} \citep{TDMJ15,Varma18}.

\begin{figure}[t]
\begin{center}
\includegraphics[width=\textwidth]{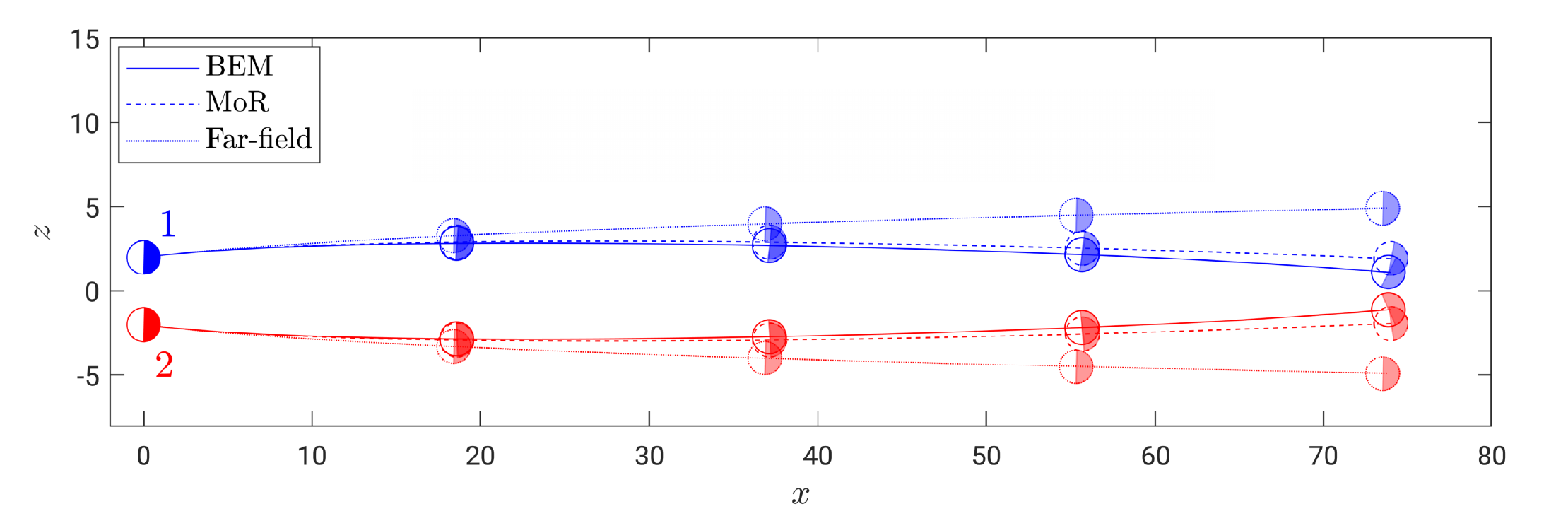}
\caption{Co-planar trajectories of two Janus particles obtained using BEM (solid), MoR (dashed) and far-field (dotted). The particles' center-to-center distance is initially $d=4a$. Particles have uniform mobility $M=1$ and hemispherical activity with $A=1$ on their active half (black) and $A=0$ on their inert cap (white). Particle locations and their orientations as obtained from BEM simulations are also shown at equal intervals of time. Note that $y$-axis is directed into the plane of the paper.}
\label{fig:planar_traj}
\end{center}
\end{figure}

The long-term dynamics of a pair of Janus particles is considered, which are initially aligned along $\eb_x$, i.e. orthogonally to their relative distance which is along $\eb_z$ (Figure~\ref{fig:planar_traj}). The particles have uniform mobility $M=1$ and hemispherical activity distribution. When isolated, these particles swim with a velocity $\Ub^\textrm{\scriptsize self}=\eb_x/4$ and are neutral squirmers (i.e. $A_2=0$, no stresslet signature).

In such an arrangement, the particle pair attract and  contact in finite time~\cite{sharifimood16}, which is indeed observed in the trajectories obtained from BEM simulations (see figure \ref{fig:planar_traj}), where the particles, exhibiting mirror symmetrical motion, first drift apart while rotating to swim toward each other at a later stage. The initial drift of the particles away from each other is easily understood by their anti-chemotactic nature: they drift and swim down the concentration gradient created by the other particle. Their rotation solely results from hydrodynamic and chemo-hydrodynamic interactions  since purely chemical interaction cannot induce rotation (for uniform mobility). 

\begin{figure}[h]
\begin{center}
\begin{tabular}{ccc}
\includegraphics[width=.33\textwidth]{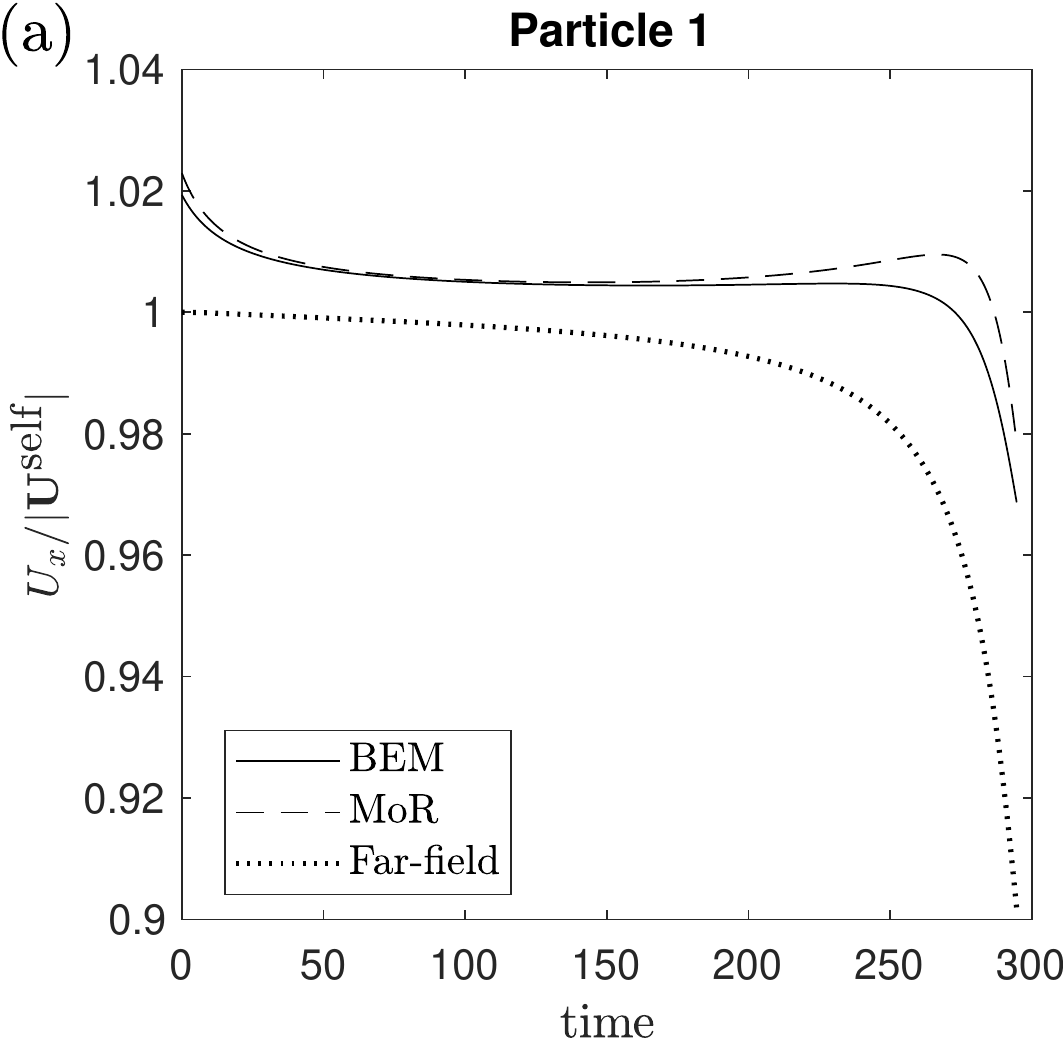} &
\includegraphics[width=.33\textwidth]{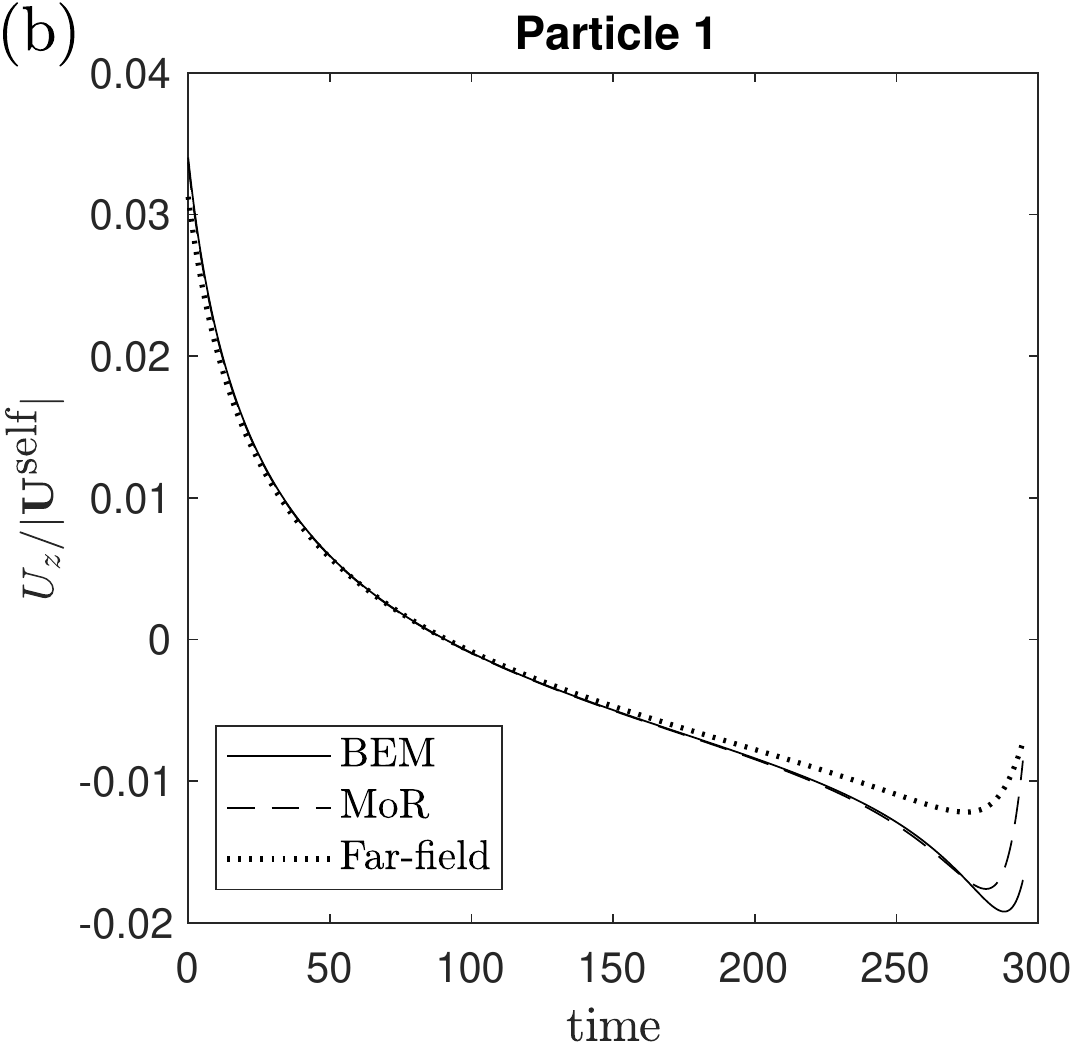} &
\includegraphics[width=.33\textwidth]{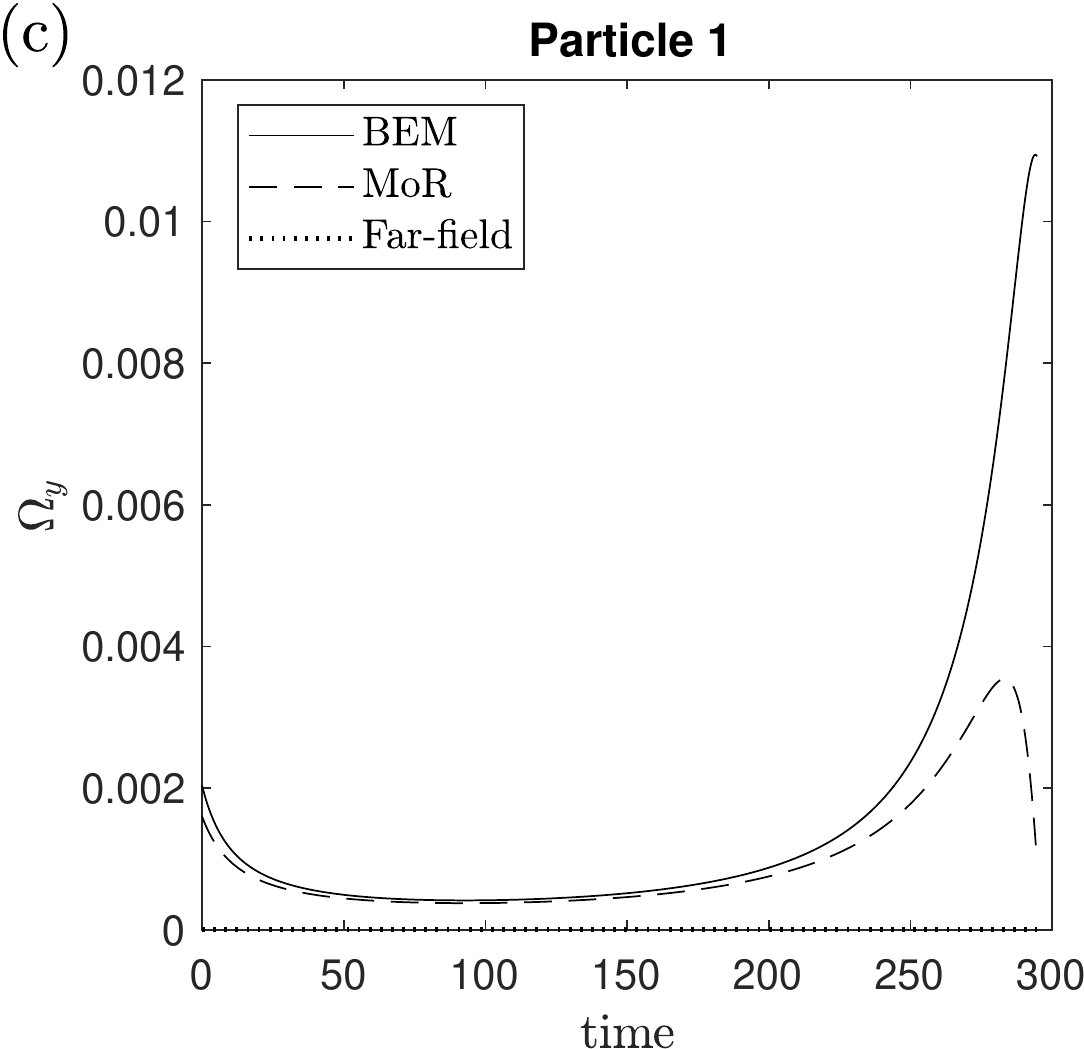} \\
\end{tabular}
\caption{(a) Horizontal, (b) vertical and (c) rotation velocity of Janus particle $1$ for two coplanar particles (see Fig~\ref{fig:planar_traj}), as obtained from BEM (solid), MoR(dashed) and the far-field model. At each time, the comparison between the prediction of the different models is performed for the same geometric configuration of the particles (i.e. that obtained from BEM simulations). The translation velocity is scaled by the self-propulsion velocity of an isolated Janus particle. Note that there is no angular velocity in the far-field model for all separations. The corresponding velocities of particle $2$ are obtained using the planar symmetry of the problem with respect to $z=0$. }
\label{fig:planar_vel}
\end{center}
\end{figure}

Instantaneous translational and angular velocities of particle 1 are shown in Figure~\ref{fig:planar_vel}. The particles' interaction results in a slight increase of their propulsion velocities (but only by a few percent). Particle 1 monotonically rotates clockwise, with a sharp increase in angular velocity arising before the particles contact. Once the particles form a cluster, they adopt a fixed tilted orientation that balances chemical, hydrodynamic and chemo-hydrodynamic interactions as well as steric repulsion; a steady co-propulsion velocity is achieved in this case.

The far-field model predicts the translational velocities reasonably well when the particles are a few radii apart but deviates strongly towards the final stages of clustering (when $d_c<1$). It however does not predict any rotation as a result of the absence of a self-generated stresslet and resulting hydrodynamic interactions for a hemispheric Janus particle of uniform mobility. It should be emphasized here that even if the particles were to have non-zero intrinsic stresslets (e.g. for non-hemispheric coverage), the angular velocities predicted using the far-field model would still be zero in this highly-symmetric setting: this is the result of the stresslet flow-field produced by each janus particle having a plane of symmetry passing through the center of the other particle, which creates no effective shear-induced rotation. Thus, in this configuration, the force-quadrupole is the leading order term responsible for the particles' reorientation. As a result, the far-field model, limited to only a force dipole, is unable to capture the qualitative trajectory (see figure \ref{fig:planar_traj}), in particular to obtain the long-term dynamics. On the other hand, Figure~\ref{fig:planar_vel} demonstrates that the  MoR model provides very accurate estimates of the translation velocities throughout the dynamics; the predicted angular velocities, accurate to $O(\varepsilon^5)$ are adequate, except for close contact where higher order corrections are necessary to fully capture lubrication effects. 

Thus, it is clearly seen from figure \ref{fig:planar_traj} that the trajectories predicted by MoR are much more accurate than far-field models both quantitatively and qualitatively. MoR further provides a good compromise between accuracy and computational performance: while BEM simulations took about 6 hours of computational time, the simulation using MoR approximation was performed in milliseconds and still captured the dynamics within an error of a particle radius.

\begin{figure}[h]
\centering
\includegraphics[scale=0.6]{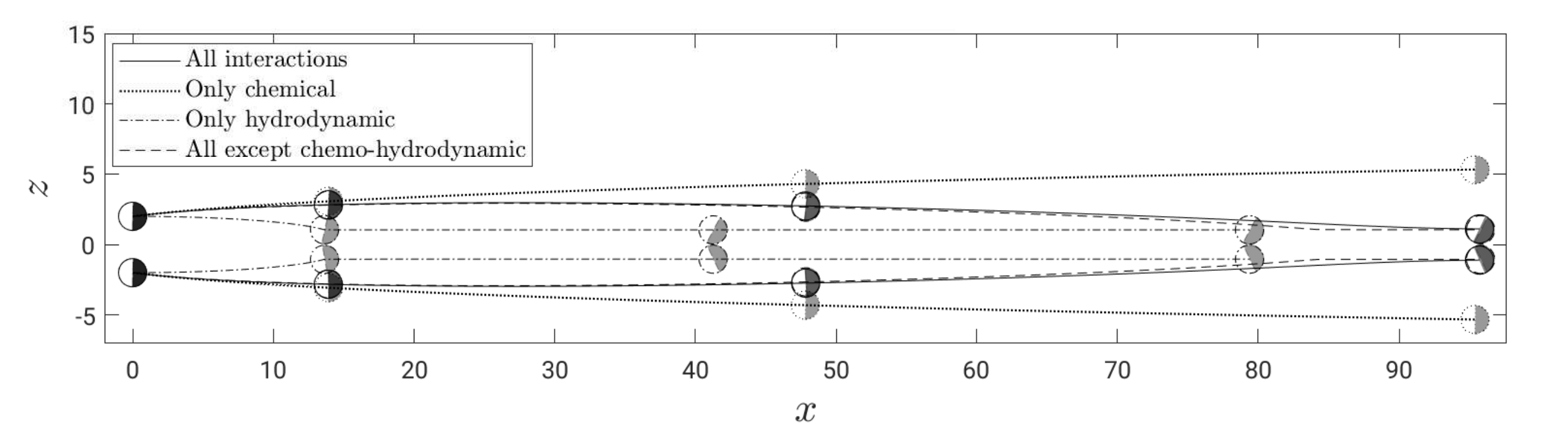}
\caption{Comparison of the effects of the different interactions on the trajectory of 2 co-planar particles using the $\varepsilon^5$-accurate MoR model. The trajectories obtained with all interactions (solid), purely chemical interactions only (dotted), purely hydrodynamic interaction only (dash-dotted) and chemical and hydrodynamic interactions (i.e. without chemohydrodynamic interactions, dashed) are shown. Particles' location and their orientation at various instances of time are shown graphically.}
\label{fig:traj_chem_vs_hydro}
\end{figure}

 Additionally, MoR clearly distinguishes chemical, hydrodynamic and chemo-hydrodynamic interactions, thus allowing us to analyse their relative and respective role in the particles' coupling by simply including or removing the appropriate interactions (Figure~\ref{fig:traj_chem_vs_hydro}). This conclusively shows that the chemical interactions are predominantly responsible for the lateral drift. As expected, purely chemical interactions do not induce any particle reorientation and the particles  drift apart laterally down the chemical gradient created by their neighbor. Hydrodynamic interactions, on the other hand, do no create any significant lateral drift but play a crucial role in reorienting the self-propelling particles toward each other, thus inducing their clustering. Chemo-hydrodynamic interactions, in the present case, are effectively repulsive but, their sharp asymptotic decay $O(\varepsilon^5)$ makes them almost non-influential in the long-term dynamics here. It is thus the competing chemical and hydrodynamic interactions that primarily gives rise to the unusual dynamics in this particular case. 

\subsection{Dynamics of randomly-arranged co-planar particles}

\begin{figure}[h]
\centering
\includegraphics[scale=0.7]{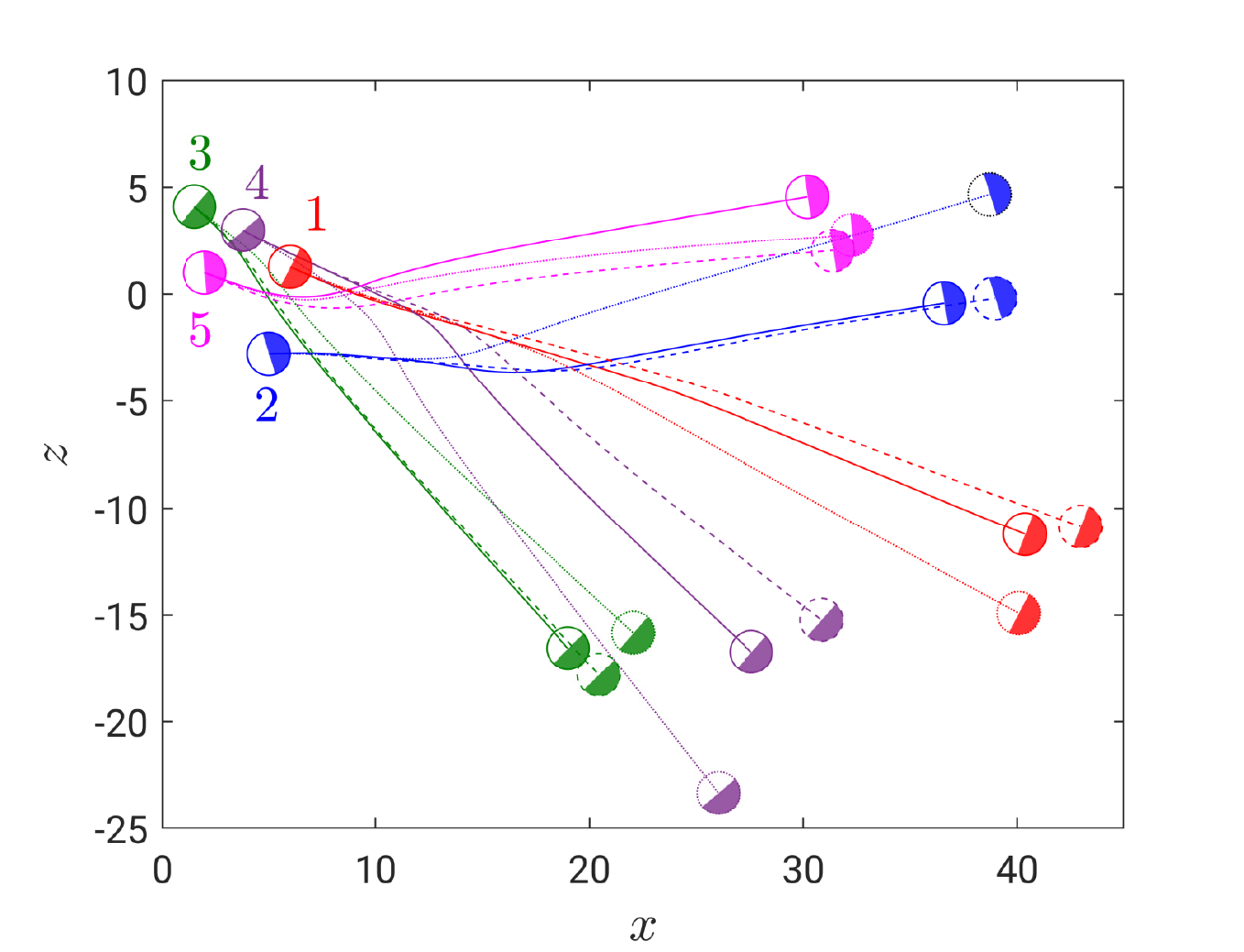}
\caption{Comparison of the trajectories of 5 Janus particles predicted by BEM (solid), MoR (dashed) and far-field models (dotted). The initial positions (at $t=0$) of the numbered particles and their predicted positions at $t=117$ are shown as well. Note that $y$-axis is directed into the plane of the paper.}
\label{fig:trajectory}
\end{figure}

In this section, we test the ability of the $O(\varepsilon^5)$-accurate MoR model to predict the dynamics of a larger number of particles ($N>2$). For simplicity of analysis and visualization, we consider here a system of 5 Janus particles initially distributed randomly in a plane (see figure \ref{fig:trajectory}), in relatively close proximity (average contact distances of the order of a few radii) . Due to the small density of particles, we restrict the choice of their random initial orientations along the plane to only within a quadrant to favour their interactions as would be expected in denser situations (i.e. with more particles). The exact dynamics are first obtained  using BEM simulations and then compared with the MoR  and far-field models (figure \ref{fig:trajectory}). Using a coarse mesh, BEM simulations required around $12$ hrs of computational time while MoR results were obtained in $5$ seconds. 

Each particle self-propels along a straight line when isolated. Any slight change of their orientation has a drastic effect on their long-term positions. Yet, Figure \ref{fig:trajectory} shows that the  MoR $O(\varepsilon^5)$-accurate model is sufficient for estimating these long-term trajectories to a reasonable accuracy and performs significantly better in that regard than the simpler far-field model. Note that the particles do not come in contact at any point in time.

\begin{figure}[h]
\begin{center}
\begin{tabular}{ccc}
\includegraphics[width=.33\textwidth]{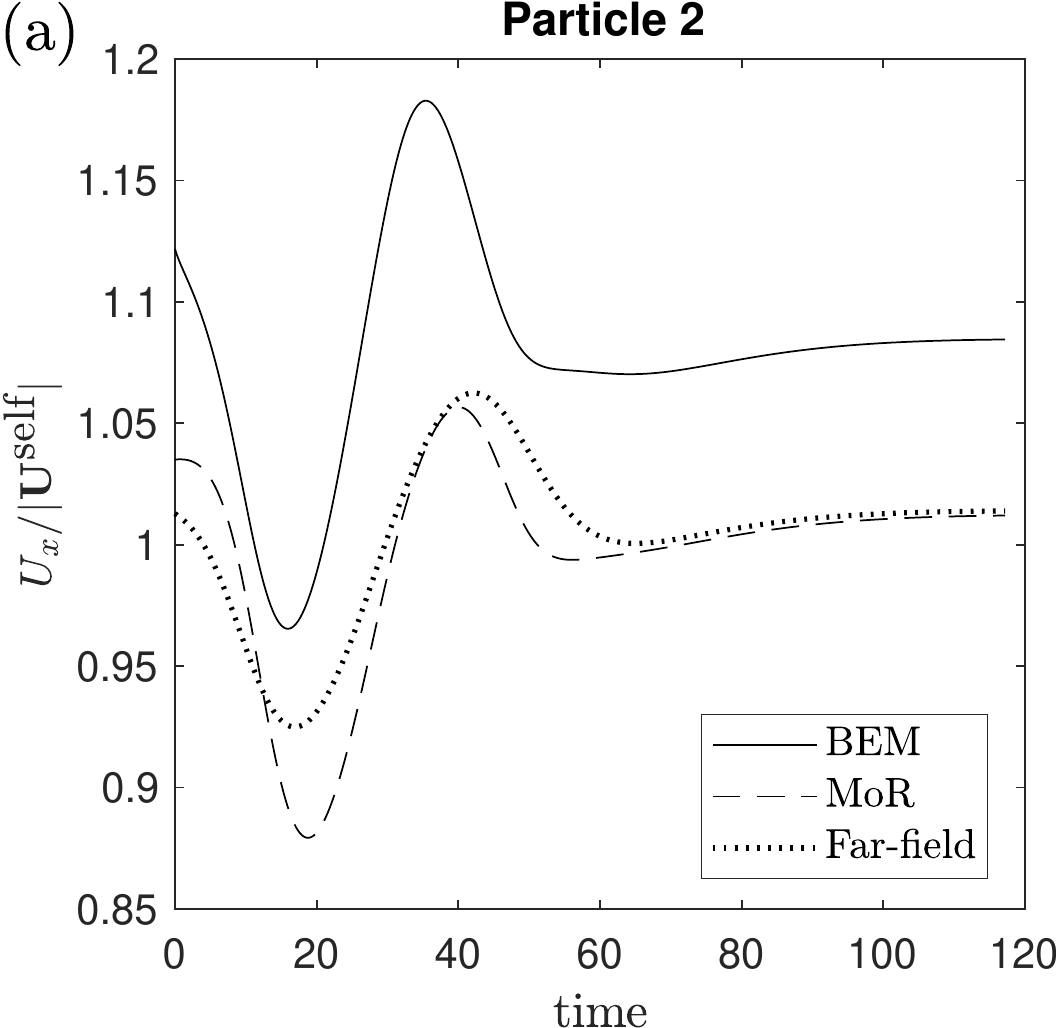}
&
\includegraphics[width=.33\textwidth]{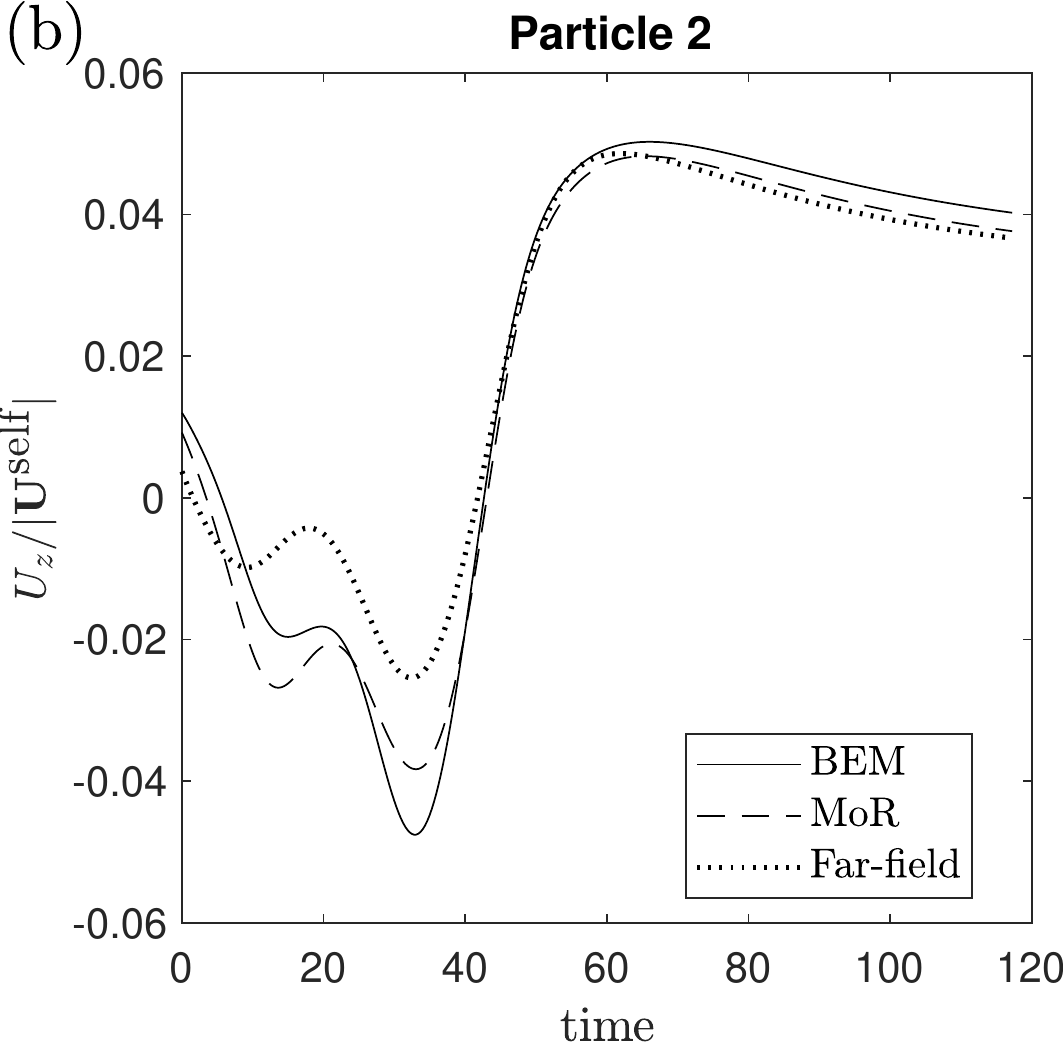}
&
\includegraphics[width=.33\textwidth]{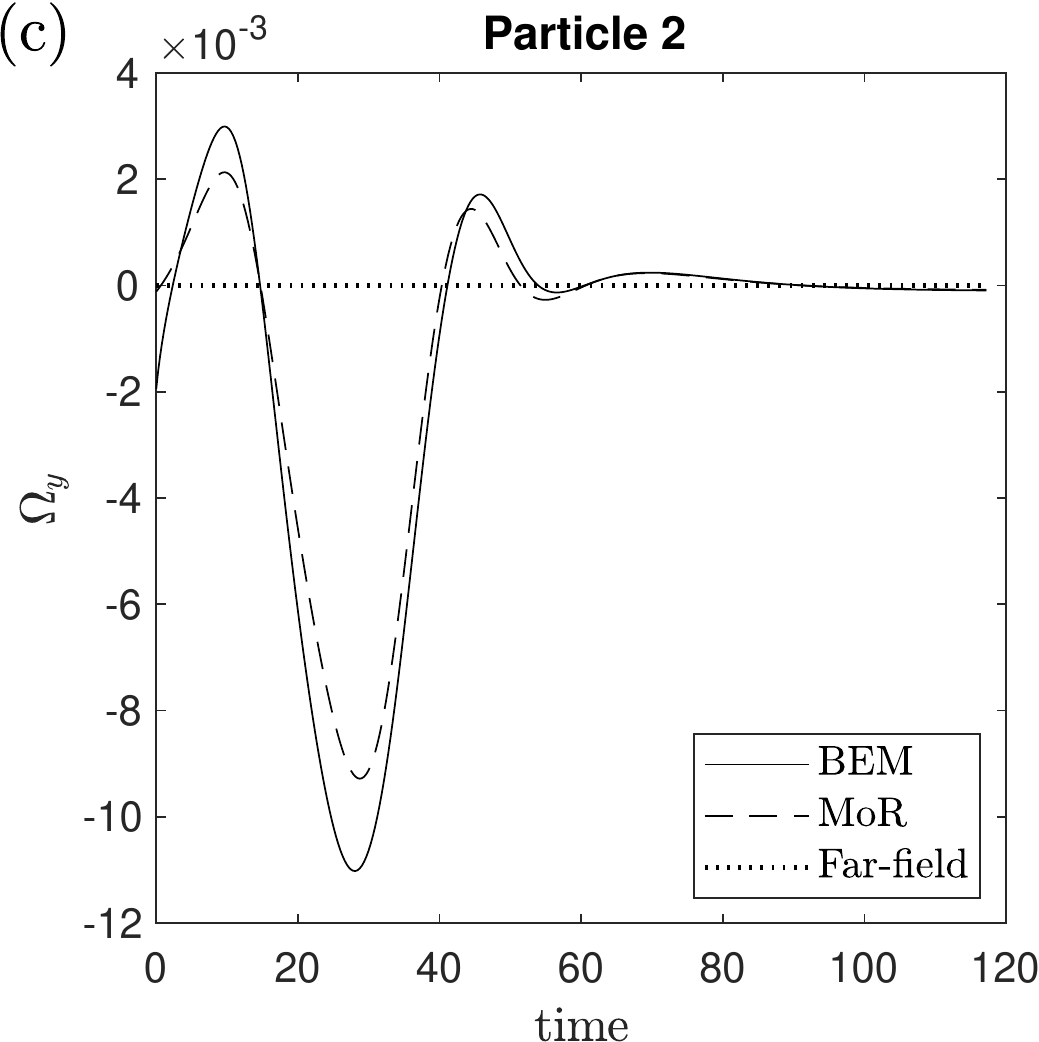}\\
\includegraphics[width=.33\textwidth]{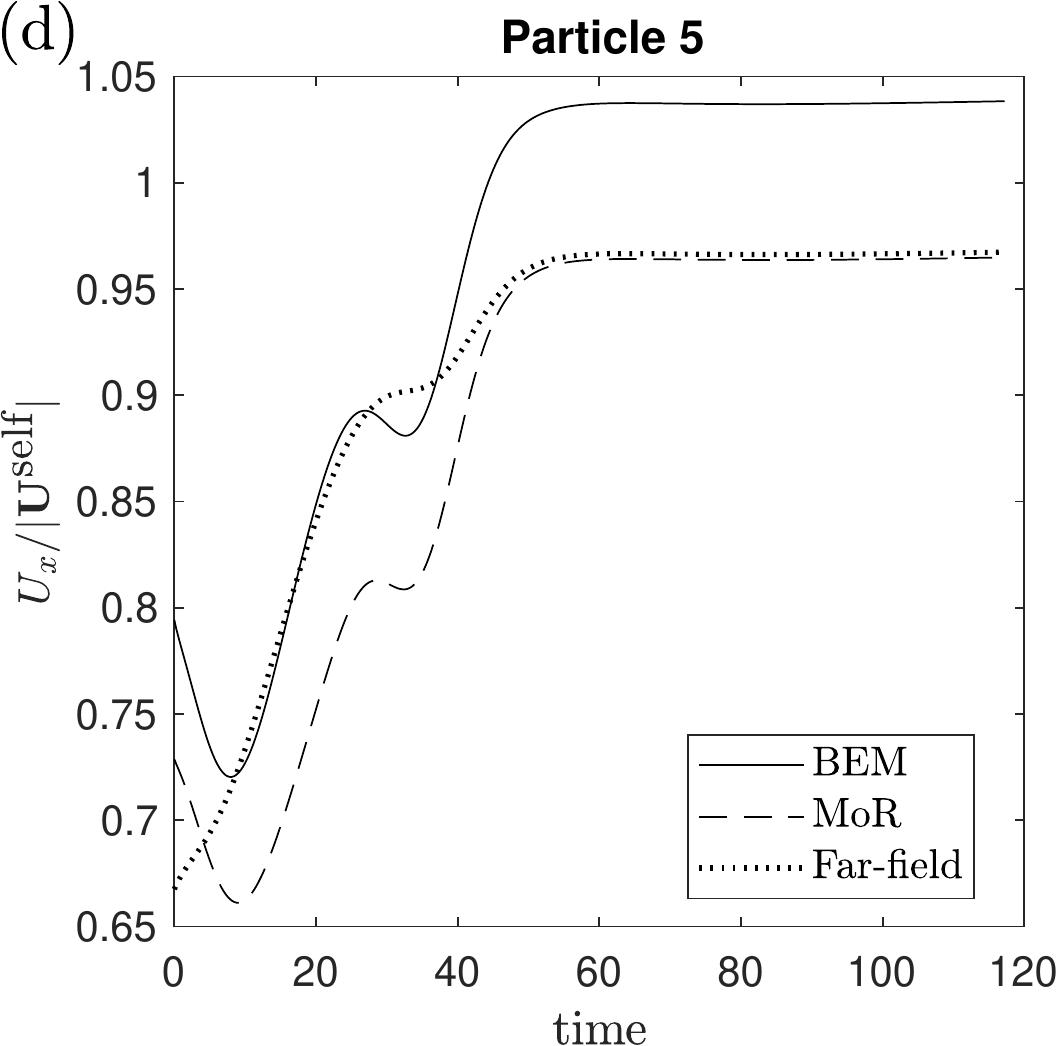}
&
\includegraphics[width=.33\textwidth]{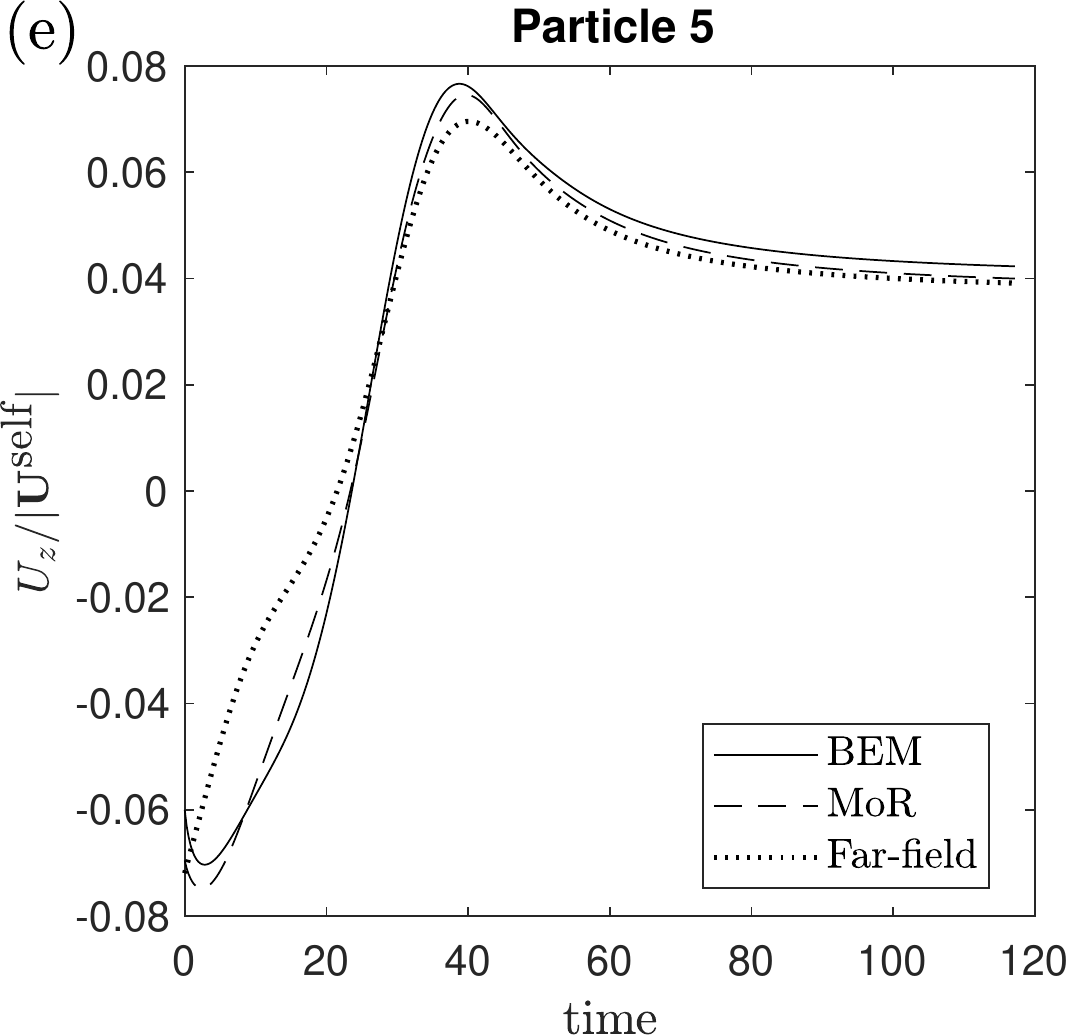}
&
\includegraphics[width=.33\textwidth]{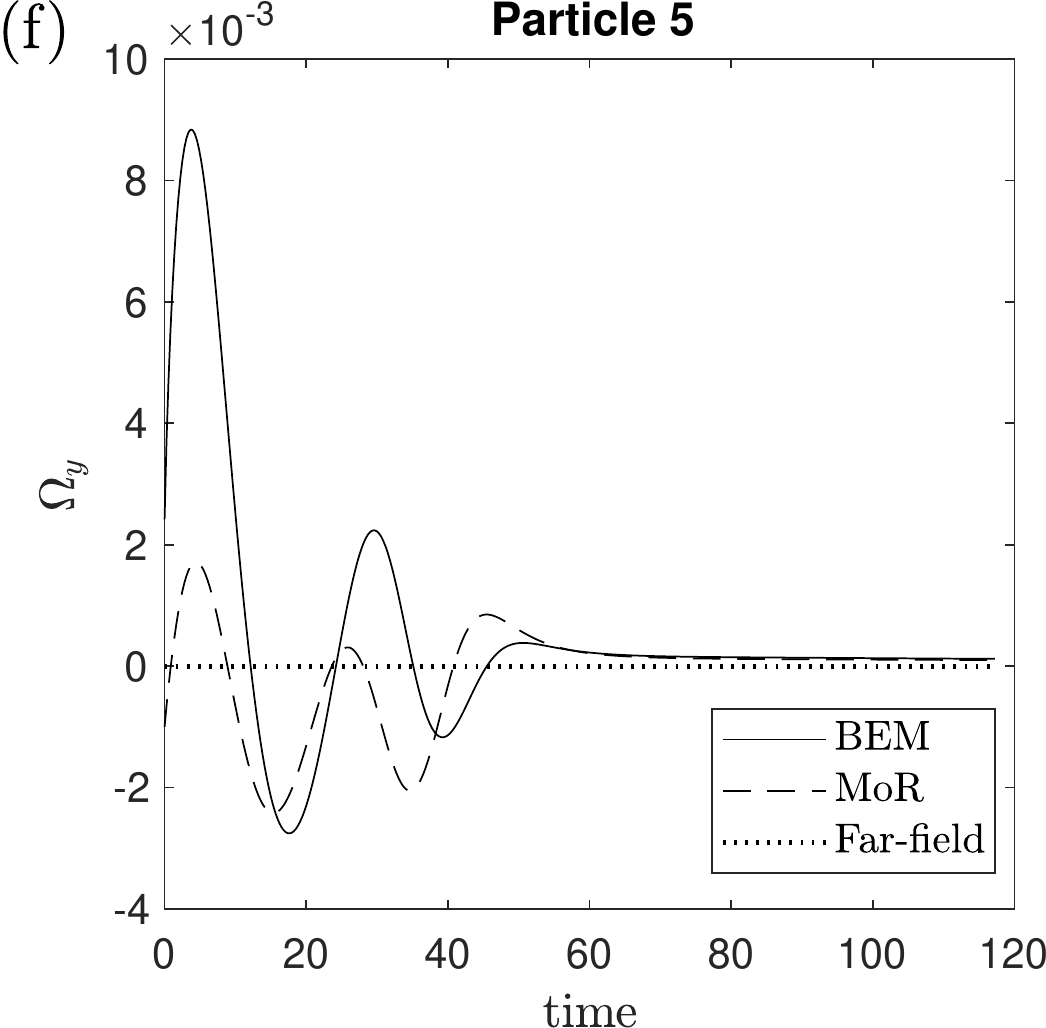}
\end{tabular}
\caption{Evolution of the particles' velocities during the planar interactions of five Janus particles (Figure~\ref{fig:trajectory}): (a,d) Horizontal, (b,e) vertical and (c,f) rotation velocties of particles 2 (top) and 5 (bottom) predicted by BEM (solid), MoR(dashed) and far-field models. The same particles' positions (obtained from the BEM simulations) are considered for all three models.
}
\label{fig:Uparticle25}
\end{center}
\end{figure}

Focusing on the instantaneous dynamics of particles $2$ and $5$, the MoR model is seen to capture the qualitative trend of the velocities much more accurately than the far-field model (Figure \ref{fig:Uparticle25}). Quantitatively, the magnitudes and the errors  in estimation of the translational velocities by MoR model are quite comparable with far-field model. Hence, the net displacement of the particles are of the same orders. However, the major advantage of the MoR model over far-field models lies is in its ability to account accurately for the particles' reorientation. Indeed, far-field models are unable to produce any change in orientation as chemical interactions do not produce any rotational effects.

MoR model correctly predicts the transient peak in angular velocity and captures the dynamics generated by its hydrodynamic interaction with other particles (Figures~\ref{fig:trajectory} and \ref{fig:Uparticle25}). This is much less the case for particle 5, for which the prediction of MoR for its angular velocity, while qualitatively correct, exhibits large errors that can be attributed to strong lubrication effects from close contact with neighbouring  particles (especially particles 3 and 4). For this particle, although the oscillatory trend in angular velocity is reproduced by MoR, the performance in terms of position predictions is significantly reduced. 

\section{Conclusions and perspectives}
\label{sec:conclusions}
In this work, we propose a general framework based on the method of reflections (MoR) to systematically determine the velocities of interacting autophoretic particles up to any order of accuracy in the particle density, under the combined influence of their chemical and hydrodynamic signatures on their environment. The explicit implementation of this framework with an $\varepsilon^5$-accuracy demonstrated its ability to capture not only the instantaneous velocity but also essential features of the long-term dynamics of phoretic particles. The performance of the predictions are significantly better, qualitatively and quantitatively, than classical far-field models which can be seen as $\varepsilon^2$-truncations of the present framework. Such far-field models are widely used due to their simplicity~\cite{Liebchen19,Zottl16,Kanso19}; yet, as they focus solely on pairwise particle interactions through the slowest-decaying hydrodynamic and chemical signatures, they fundamentally overlook more complex chemo-hydrodynamic interaction routes as well as many-body interactions. The analysis presented here demonstrate that these models become fundamentally inaccurate in not-so-dilute suspensions where particles are separated by a few radii or less. In contrast, the MoR model proposed here is observed to correctly predict the particles' velocities with a comparable computational cost,  even when the particles have contact distances of the order of a single particle radius. Further, it is able to capture quantitatively the reorientation of the self-propelled particles, an element that is critical to predict and understand their long-term trajectories and interactions. As such, the MoR model offers a promising alternative to far-field models in order to analyse  dynamics of suspensions that are not asymptotically dilute. These predictions are furthermore obtained at a computational cost that is orders of magnitude smaller than a direct numerical simulation using classical approaches such as Boundary Elements or Immersed Boundary Methods.

The complete analytical framework was presented as well as a practical application to $O(\varepsilon^5)$ accuracy for particles of uniform mobility. {Yet, with some additional tensor computations to obtain the required transfer functions, it could be extended to obtain more precise estimates, by identifying which combination of reflections (both for the Laplace and Stokes problems) lead to interactions of greater asymptotic order than the requested accuracy.} The chosen accuracy is motivated here as the smallest order at which $3$-particle interactions  become significant and combine both chemical and hydrodynamic coupling, in contrast with far-field models that simply superimpose pairwise interactions that involve solely chemical or hydrodynamic effects. The uniformity of the particles' mobility significantly simplifies the final expression of the interaction velocities as there is a direct mapping between the concentration multipole intensities and the velocity field singularities used for initializing the hydrodynamic reflections. Yet the entire framework presented here is directly applicable to particles of arbitrary mobility distribution, provided this initialization step is modified by adding a tensorial reduction process as discussed in Section~\ref{sec:hydro-ref}. 

Despite its asymptotic nature and the fact that it is inherently not designed to represent near-contact dynamics, such as lubrication effects, the method converges rapidly: an accuracy of $O(\varepsilon^5)$ in propulsion velocities were obtained using just a single reflection for the hydrodynamic problem and two reflections for the chemical field. The rapid convergence of the hydrodynamic problem is linked to the particles being force-free so that the slowest-decaying hydrodynamic singularity, the Stokeslet with an $1/r$ decay rate, is absent here. Similarly, the rapid convergence of the chemical problem is associated with the absence of a monopole in the subsequent reflections. Besides its rapid convergence, the MoR method is also surprisingly accurate as it is able to capture many of the particles' dynamics and predict their velocity even for inter-particle contact distances of the order of their radius.

The presentation of the framework followed here, for simplicity, is that of a  \emph {parallel} implementation of the method of reflections~\cite{Golusin35}, i.e. a Jacobi-type iteration where corrections near a given particle  are based on the information from all the other particles  at the previous iteration. An alternative approach is the historical \emph{sequential} approach~\cite{Smoluchowski11,Luke89}, for which the newest correction near any particle is used as soon as it becomes available in a Gauss-Seidel-type iteration (i.e. even during the same reflection near the subsequent particles). As noted in Section~\ref{sec:chem-ref}, the present framework can be straightforwardly implemented sequentially rather than in parallel (see the discussion of Eq.~\eqref{eq:conc_ref_rec}), and a similar remark holds for the hydrodynamic reflection sequence, Eqs.~\eqref{eq:U_rec}--\eqref{eq:X_rec}. Mathematically, when truncating at a given number of reflections, the sequential method is proved to converge exponentially for the \emph{mobility problem} considered here, where the forces on particles are prescribed \cite{Luke89} (it wouldn't be the case for a resistance problem where particles' velocities are imposed~\cite{Ichiki01}). In contrast, mathematical convergence of the parallel implementation is still an open question. However, this does not impact the implementation of the method proposed here, which is based on a truncation of the series approximation based on a fixed maximum order of the different terms in powers of $\varepsilon=a/d$ rather than a fixed number of reflections:  with this physically-based approach, both the sequential and parallel methods then lead to retaining the same contributions. 

 \begin{figure}[h]
\begin{center}
\begin{tabular}{ccc}
\includegraphics[width=.33\textwidth]{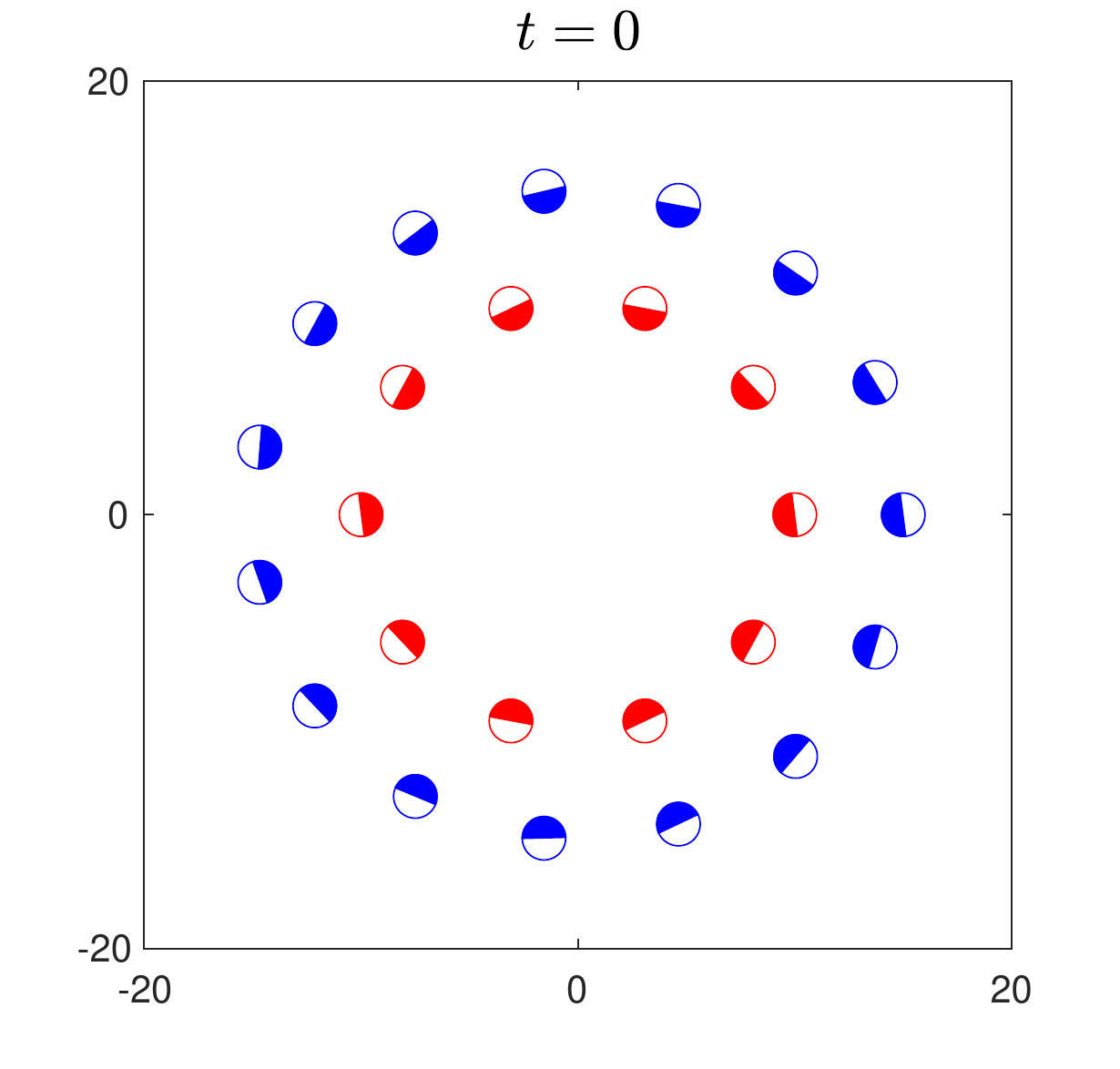}&
\includegraphics[width=.33\textwidth]{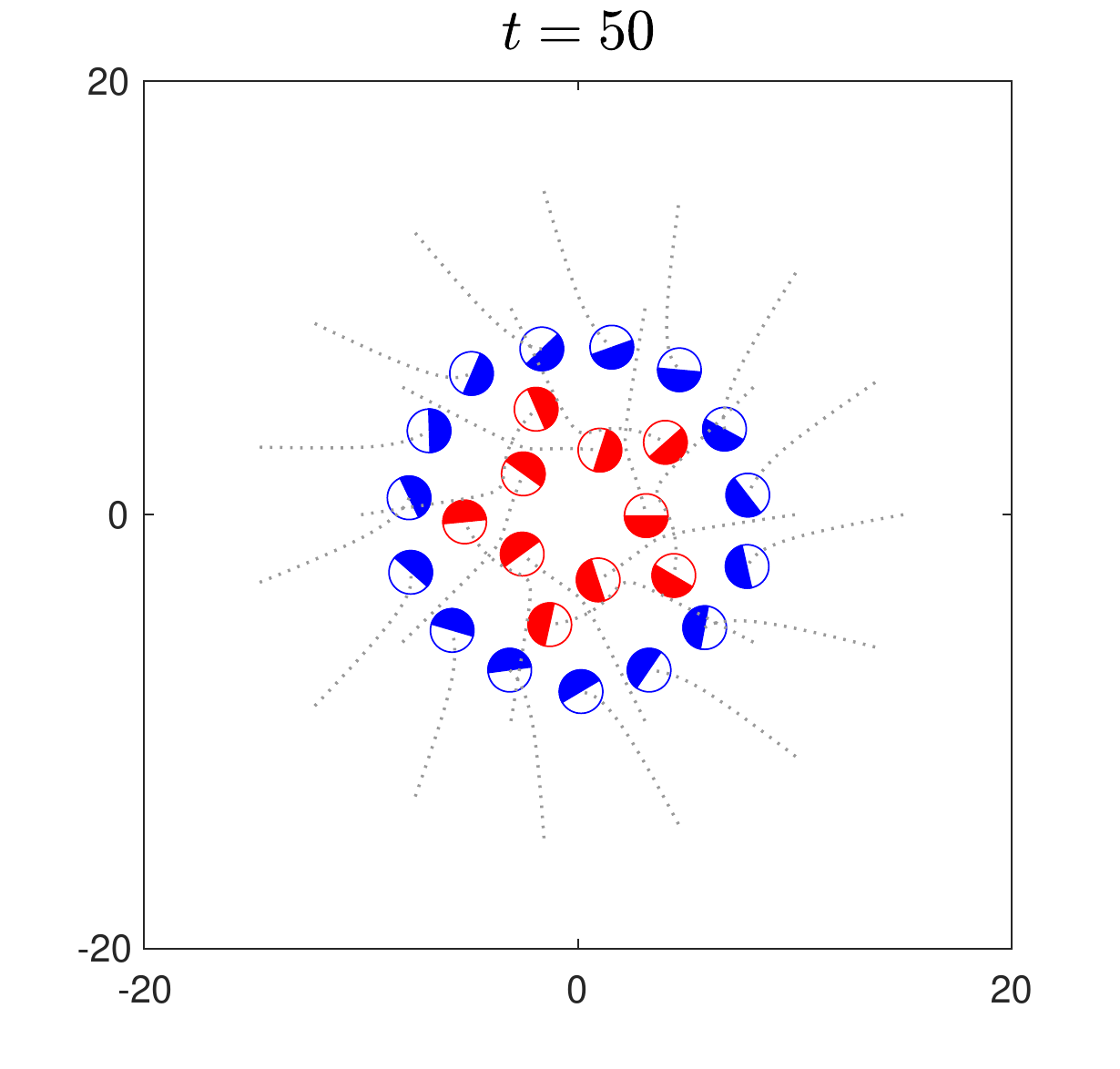}&
\includegraphics[width=.33\textwidth]{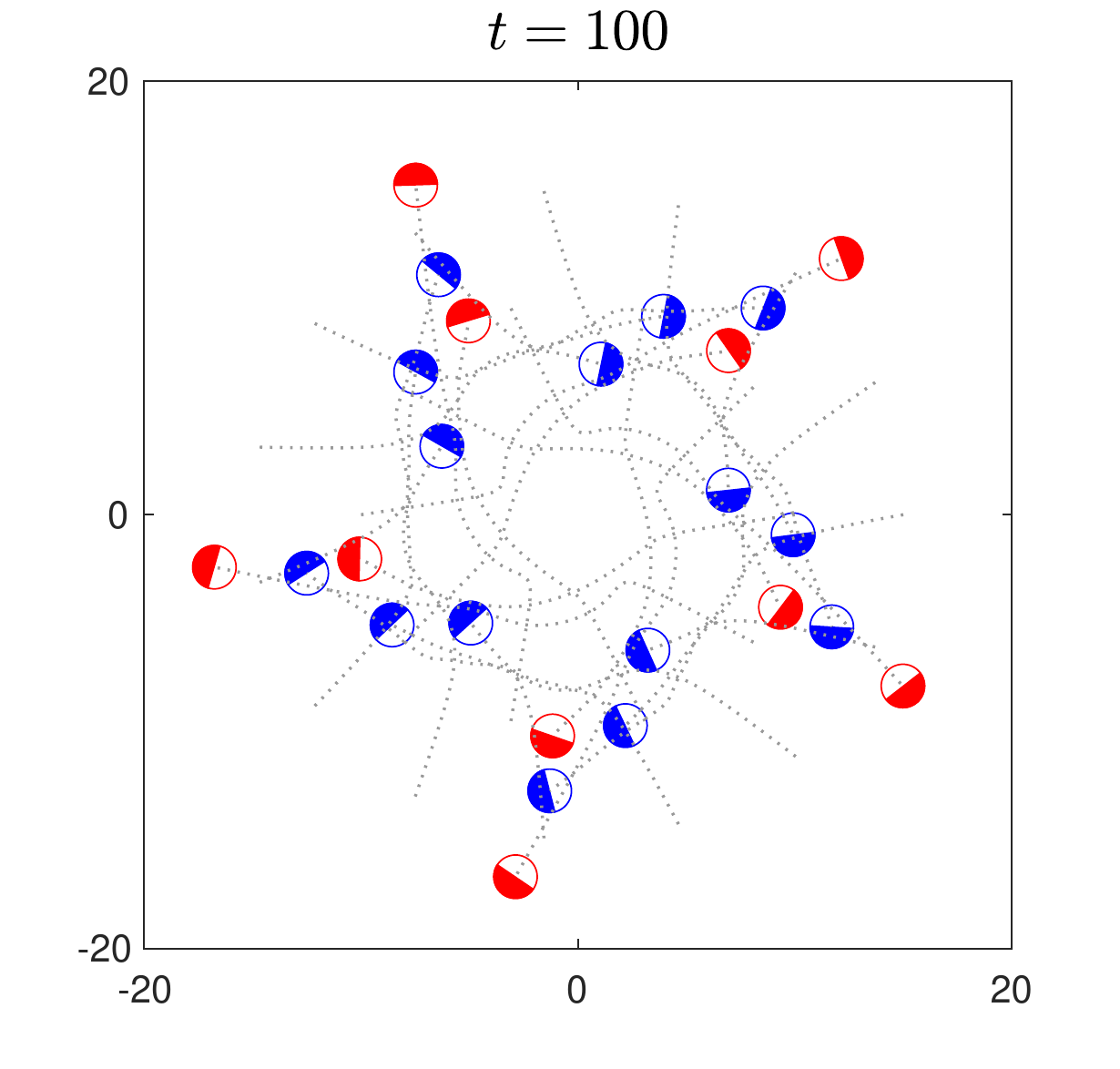}
\end{tabular}
\caption{Collective dynamics of 25 Janus particles with uniform positive mobility and hemispherical activity. A set of 10 particles (in red) are arranged on a circle of radius 10 units, aligned offset from the radial direction by an angle $0.05 \pi$ and 15 particles (in blue) are arranged on a circle of radius 15 units with the same angular offset from the radial direction. Particles' position computed using the MoR framework of Section~\ref{sec:interactions_ep5} are shown for various instances of time, $t$.
}
\label{fig:circle_janus}
\end{center}
\end{figure}

An important feature and fundamental interest of this approach, from a physical point of view, is to clearly identify the physical mechanisms resulting in the different components of the particles' interaction velocities, as demonstrated in Section~\ref{sec:interactions_ep5}. The interaction of phoretic particles are indeed commonly and perhaps short-sightedly considered as the juxtaposition of two independent and fundamentally different physical mechanisms, namely the effect of their non-uniform chemical signature and the hydrodynamic flow they create while swimming, and the question of their relative weight is attracting much debate~\cite{Soto14,Liebchen19,Kanso19,Zottl16}. This picture, inherited implicitly from far-field models is misleading: in essence, the only physical mechanism leading to the particles' displacement is hydrodynamics as particles do not have any direct chemical or physical interactions (i.e. so-called chemical interactions are in fact due to the hydrodynamic slip generated by the neighboring particles' chemical effect). It further overlooks the intricate coupling of the hydrodynamic and chemical problems, and the most generic interactions are  in fact chemo-hydrodynamic and involve many particles, rather than being simply pairwise. The present framework in fact provides a unique opportunity to analyse rigorously the relative weight of different interaction routes, as each interaction type can be turned on or off easily in the model (a feature that is much more difficult to implement on a full numerical simulation for example).

The MoR model was implemented and tested here in the limit of a small number of particles, to enable quantitative comparisons with direct numerical simulations. However, it can  straightforwardly be applied to analyse complex dynamics of larger systems. As an illustration, Figure~\ref{fig:circle_janus} shows the interactions and scattering dynamics of 25 Janus particles initially distributed regularly. Its low computational cost  makes this method particularly well-suited for analysing the dynamics of a very large number of particles and of suspensions. An important  element influencing the cost of the method is obviously the degree of the highest order multipoles considered, which is directly linked to the desired degree of accuracy. Nevertheless, the successive computations of chemical and hydrodynamic moments as linear combinations of the same moments evaluated independently around each of the other particles at the previous order of reflection confer interesting scalability properties to this method with the number of particles $N$, with a $O(N^2)$-computational cost for large numbers of particles, which makes it a very compelling candidate to obtain quantitative insights in the behavior of large active suspensions.

\section*{Acknowledgments}
This work was supported by the European Research Council (ERC) under the European Union's Horizon 2020 research and innovation program (Grant Agreement No. 714027 to S.M.).

\appendix
\section{Solution of the hydrodynamic reflection problem}
\label{sec:app_hydro}
\subsection{Spherical harmonics decomposition}

%%%------------
{The first step in solving the hydrodynamic reflection problem (i.e. finding the flow field $\ub_k^p$ for $p\geq 1$) is to determine the intensity of the flow singularities involved in Eq.~\eqref{eq:sol_ukp} as a function of the velocity gradients generated near particle $k$ at the previous reflection. The $(q-1)$-th gradient of the flow field can first be decomposed by isolating its symmetric part with respect to all indices:
\begin{equation}
\overset{q-1}{\grad}\ub_j^{p-1}=\frac{1}{q}\sum_{s=1}^q(\overset{q-1}{\grad}\ub_j^{p-1})^{T_{1s}}+\Big[\overset{q-1}{\grad}\ub_j^{p-1}-\frac{1}{q}\sum_{s=1}^q(\overset{q-1}{\grad}\ub_j^{p-1})^{T_{1s}}\Big],
\end{equation}
where the terms in bracket do not contribute to Eq.~\eqref{eq:dec_vn}, where it is contracted with a fully-symmetric tensor, $\nb_k\overset{q}{\otimes}\nb_k$. Here, ${\bf A}^{T_{1s}}$ corresponds to the transpose of ${\bf A}$ with respect to indices $1$ and $s$. When $q\geq 3$, the first part (i.e. the symmetric part) is not necessarily trace-free with respect to any pair of the last $q-1$ indices and must therefore be further decomposed as 
\begin{align}
\frac{1}{q}\sum_{s=1}^q(\overset{q-1}{\grad}\ub_j^{p-1})^{T_{1s}}=&\overbrace{\frac{1}{q}\sum_{s=1}^q(\overset{q-1}{\grad}\ub_j^{p-1})^{T_{1s}}-\frac{q-2}{q(2q-1)}\sum_{1\leq l<m\leq q}\left[\overset{q-3}{\grad}(\mathbf{I}\otimes\grad^2\ub_j^{p-1})\right]^{T_{1l},T_{2m}}}^{\overbracket{\overset{q-1}{\grad}\ub_j^{p-1}}}\nonumber\\
&+\frac{q-2}{q(2q-1)}\sum_{1\leq l<m\leq q}\left[\mathbf{I}\otimes\overset{q-3}{\grad}(\grad^2\ub_j^{p-1})\right]^{T_{1l},T_{2m}},
\end{align}
and the first part ($\overbracket{\overset{q-1}{\grad}\ub_j^{p-1}}$) denotes the fully symmetric and deviatoric part of the $(q-1)$-th velocity gradient. Then, noting that the last terms includes $q(q-1)/2$ different terms contributing identically once contracted with $\nb_k\overset{q}{\otimes}\nb_k$,
\begin{align}
\sum_{q\geq 1}\left[\frac{a_k^{q-1}}{(q-1)!}\overset{q-1}{\grad}\ub_j^{p-1}\right]\overset{q}{\odot}[\nb_k\overset{q}{\otimes}\nb_k]&=\sum_{q\geq 1}\left[\frac{a_k^{q-1}}{(q-1)!}\overbracket{\overset{q-1}{\grad}\ub_j^{p-1}}\right]\overset{q}{\odot}[\nb_k\overset{q}{\otimes}\nb_k]+\sum_{q\geq 3}\frac{1}{2(2q-1)}\frac{a_k^{q-1}}{(q-3)!}\overset{q-3}{\grad}\Big(\grad^2\ub_j^{p-1}\Big)\overset{q-2}{\odot}[\nb_k\overset{q-2}{\otimes}\nb_k]\nonumber\\
&=\sum_{q\geq 1}\left[\frac{a_k^{q-1}}{(q-1)!}\Big(1+\frac{a_k^2}{2(2q+3)}\nabla^2\Big)\overbracket{\overset{q-1}{\grad}\ub_j^{p-1}}\right]\overset{q}{\odot}[\nb_k\overset{q}{\otimes}\nb_k]
\end{align}
since $\overset{q-1}{\grad}\left(\nabla^2\ub_j^{p-1}\right)$ is traceless with respect to any pair of its indices.
Eqs.~\eqref{eq:dec_divsurf} and \eqref{eq:dec_rotsurf} can be decomposed similarly, noting that $\overset{q-1}{\grad}\omegab_j^{p-1}$ is already fully deviatoric, and lead to Eqs.~\eqref{eq:dec_vn}--\eqref{eq:dec_rotsurf}. 
Using the expression for the surface velocity $\vb_k^p$, Eq.~\eqref{eq:surf_vel_exp}, the normal velocity, surface divergence and  surface vorticity are thus obtained as
\begin{align}
\vb_k^p\cdot\nb_k&=-\sum_{q\geq 1}\left[\sum_{j\neq k}\frac{a_k^{q-1}}{(q-1)!}\left(1+\frac{a_k^2}{2(2q+3)}\nabla^2\right)\overbracket{\overset{q-1}{\grad}\ub_j^{p-1}}\biggr|_{r_k=0}\right]\overset{q}{\odot}[\nb_k\overset{q}{\otimes}\nb_k],\label{eq:dec_vn}\\
-a_k\nabla_s\cdot\vb_k^p&=-\sum_{q\geq 1}\left[\sum_{j\neq k}\frac{a_k^{q-1}}{(q-1)!}\left(q-1+\frac{(q+1)a_k^2}{2(2q+3)}\nabla^2\right)\overbracket{\overset{q-1}{\grad}\ub_j^{p-1}}\biggr|_{r_k=0}\right]\overset{q}{\odot}[\nb_k\overset{q}{\otimes}\nb_k],\label{eq:dec_divsurf}\\
a_k\nb_k\cdot[\nabla_s\times\vb_k^p]&=-\sum_{q\geq 1}\left[\sum_{j\neq k}\frac{a_k^{q}}{q!}\overbracket{\overset{q-1}{\grad}\boldsymbol\omega_j^{p-1}}\biggr|_{r_k=0}\right]\overset{q}{\odot}[\nb_k\overset{q}{\otimes}\nb_k].\label{eq:dec_rotsurf}
\end{align}
Identifying the singularities' intensity using Eqs.~\eqref{eq:Phi_kq_def}--\eqref{eq:X_kq_def} requires decomposing these three functions into spherical harmonics along the particle's surface as in Eq.~\eqref{eq:sphere_harm_dec}. 
Comparing Eqs~\eqref{eq:dec_vn}--\eqref{eq:dec_rotsurf} with Eqs.~\eqref{eq:Phi_kq_def}--\eqref{eq:X_kq_def}, the fundamental singularities at reflection $p$ can be identified readily in terms of the gradients of the velocity fields introduced at the previous reflections. For force- and torque-free particles, $\mathbf{P}^p_{k,1}=\mathbf{X}_{k,1}^p=0$, 
\begin{align}
\Ub_k^p&=\sum_{j\neq k}\Big(1+\frac{a_k^2}{6}\nabla^2\Big)\ub_j^{p-1}\biggr|_{r_k=0},\qquad
\Omegab_k^p=\frac{1}{2}\sum_{j\neq k}\omegab_j^{p-1}\biggr|_{r_k=0},\qquad
\boldsymbol\Phi_{k,1}^p=-\frac{a_k^5}{30}\sum_{j\neq k}\nabla^2\ub_j^{p-1}\biggr|_{r_k=0},
\end{align}
which recovers Faxen's laws exactly, and for $q\geq 2$,
\begin{align}
\boldsymbol\Phi^p_{k,q}&=-\frac{2q-1}{2(q+1)}\frac{a_k^{2q+1}}{(q-1)!}\sum_{j\neq k}\Big(1+\frac{(2q+1)a_k^2}{2(2q-1)(2q+3)}\nabla^2\Big)\overbracket{\overset{q-1}{\grad}\ub_j^{p-1}}\biggr|_{r_k=0},\label{eq:Phi_p_calc}\\
\mathbf{P}^p_{k,q}&=-\frac{2q+1}{2(q+1)}\frac{a_k^{2q-1}}{(q-1)!}\sum_{j\neq k}\Big(1+\frac{a_k^2}{2(2q+1)}\nabla^2\Big)\overbracket{\overset{q-1}{\grad}\ub_j^{p-1}}\biggr|_{r_k=0},\label{eq:P_p_calc}\\
\mathbf{X}^p_{k,q}&=-\frac{1}{ q(q+1)}\frac{a_k^{2q+1}}{q!}\sum_{j\neq k}\overbracket{\overset{q-1}{\grad}\omegab_j^{p-1}}\biggr|_{r_k=0}.\label{eq:X_p_calc}
\end{align}
}

%%%%%----------

\subsection{Recursive relations}
From Eqs.~\eqref{eq:Phi_p_calc}--\eqref{eq:X_p_calc}, obtaining recursive relations in $p$ between the three sets of tensors $\boldsymbol\Phi_{k,q}^p$, $\mathbf{P}_{k,q}^p$ and $\mathbf{X}_{k,q}^p$ therefore requires determining $\overset{q-1}{\grad}\ub_j^{p-1}$, $\overset{q-1}{\grad}\omegab_j^{p-1}$ and $\nabla^2(\overset{q-1}{\grad}\ub_j^{p-1})=\overset{q}{\grad}p_j^{p-1}$ associated with each singularity at reflection $p-1$ at the center of particle $k$. Rewriting Eq.~\eqref{eq:sol_ukp} in terms of the set of tensors $\boldsymbol\Phi_{k,q}^p$, $\mathbf{P}_{k,q}^p$ and $\mathbf{X}_{k,q}^p$:
\begin{align}
\ub_j^{p-1}=\sum_{s=1}^\infty&\left\{\boldsymbol{\Phi}_{j,s}^{p-1}\overset{s}{\odot}\grad\left(\frac{\rb_j\overset{s}{\otimes}\rb_j}{r_j^{2s+1}}\right)-\mathbf{X}_{j,s}^{p-1}\overset{s}{\odot}\left[s\left(\frac{\rb_j\overset{s-1}{\otimes}\rb_j}{r_j^{2s+1}}\right)\otimes(\boldsymbol\varepsilon\cdot\rb_j)\right]\right.\nonumber\\
&\left.+\frac{\mathbf{P}_{j,s}^{p-1}}{2(2s-1)}\overset{s}{\odot}\left[\frac{\rb_j\overset{s-1}{\otimes}\rb_j}{r_j^{2s-1}}\otimes\left((2s-1)\frac{\rb_j\rb_j}{r_j^2}-(s-2)\mathbf{I}\right)\right]\right\},
\end{align}
the required gradients are computed as
\begin{align}
\overset{q-1}{\grad}\ub_j^{p-1}=\sum_{s=1}^\infty&\left\{\boldsymbol{\Phi}_{j,s}^{p-1}\overset{s}{\odot}\overset{q}{\grad}\left(\frac{\rb_j\overset{s}{\otimes}\rb_j}{r_j^{2s+1}}\right)-\mathbf{X}_{j,s}^{p-1}\overset{s}{\odot}\overset{q-1}{\grad}\left[s\left(\frac{\rb_j\overset{s-1}{\otimes}\rb_j}{r_j^{2s+1}}\right)\otimes(\boldsymbol\varepsilon\cdot\rb_j)\right]\right.\nonumber\\
&\left.+\frac{\mathbf{P}_{j,s}^{p-1}}{2(2s-1)}\overset{s}{\odot}\overset{q-1}{\grad}\left[\frac{\rb_j\overset{s-1}{\otimes}\rb_j}{r_j^{2s-1}}\otimes\left((2s-1)\frac{\rb_j\rb_j}{r_j^2}-(s-2)\mathbf{I}\right)\right]\right\},\\
\overset{q-1}{\grad}\omegab_j^{p-1}=\sum_{s=1}^\infty&\overset{q-1}{\grad}\left(-s\grad\chi_{j,s}^{p-1}+\frac{1}{s}\grad p_{j,s}^{p-1}\times\rb_j\right)\nonumber\\
=\sum_{s=1}^\infty&\left\{-s\mathbf{X}_{j,s}^{p-1}\overset{s}{\odot}\overset{q}{\grad}\left[\frac{\rb_j\overset{s}{\otimes}\rb_j}{r_j^{2s+1}}\right]-\mathbf{P}_{j,s}^{p-1}\overset{s}{\odot}\overset{q-1}{\grad}\left[\frac{\rb_j\overset{s-1}{\otimes}\rb_j}{r_j^{2s+1}}\otimes(\boldsymbol\varepsilon\cdot\rb_j)\right]\right\}\\
\nabla^2(\overset{q-1}{\grad}\ub_j^{p-1})=\sum_{s=1}^\infty& \mathbf{P}_{j,s}^{p-1}\overset{s}{\odot}\overset{q}{\grad}\left[\frac{\rb_j\overset{s}{\otimes}\rb_j}{r_j^{2s+1}}\right]
\end{align}
Using these results, the transfer functions between two successive orders of reflections are obtained as
\begin{align}
\Ub_k^p&=\sum_{j\neq k}\sum_{s\geq 1}\left[\boldsymbol\Phi_{j,s}^{p-1}\overset{s}{\odot}\boldsymbol{\mathcal{F}^1}_{jk}(1,s)-\mathbf{X}_{j,s}^{p-1}\overset{s}{\odot}\boldsymbol{\mathcal{F}^2}_{jk}(1,s)+\mathbf{P}_{j,s}^{p-1}\overset{s}{\odot}\left(\boldsymbol{\mathcal{F}^3}_{jk}(1,s)+\frac{a_k^2}{6}\boldsymbol{\mathcal{F}^1}_{jk}(1,s)\right)\right],\\
\Omegab_k^p&=-\frac{1}{2}\sum_{j\neq k}\sum_{s\geq 1}\left[\mathbf{P}_{j,s}^{p-1}\overset{s}{\odot}\boldsymbol{\mathcal{F}^2}_{jk}(1,s)+s\mathbf{X}_{j,s}^{p-1}\overset{s}{\odot}\boldsymbol{\mathcal{F}^1}_{jk}(1,s)\right],\\
\boldsymbol\Phi_{j,1}^{p}&=-\frac{a_k^5}{30}\sum_{j\neq k}\sum_{s\geq 1}\left[\mathbf{P}_{j,s}^{p-1}\overset{s}{\odot}\boldsymbol{\mathcal{F}^1}_{jk}(1,s)\right]
\end{align}
and for $q\geq 2$
\begin{align}
\boldsymbol\Phi^p_{k,q}&=\sum_{j\neq k}\sum_{s\geq 1}\left[\boldsymbol\Phi^{p-1}_{j,s}\overset{s}{\odot}\boldsymbol{\mathcal{F}^{\Phi\rightarrow\Phi}}_{jk}(q,s)+\mathbf{P}^{p-1}_{j,s}\overset{s}{\odot}\boldsymbol{\mathcal{F}^{P\rightarrow\Phi }}_{jk}(q,s)+\mathbf{X}^{p-1}_{j,s}\overset{s}{\odot}\boldsymbol{\mathcal{F}^{X\rightarrow\Phi}}_{jk}(q,s)\right]\\
\mathbf{P}^p_{k,q}&=\sum_{j\neq k}\sum_{s\geq 1}\left[\boldsymbol\Phi^{p-1}_{j,s}\overset{s}{\odot}\boldsymbol{\mathcal{F}^{\Phi\rightarrow P}}_{jk}(q,s)+\mathbf{P}^{p-1}_{j,s}\overset{s}{\odot}\boldsymbol{\mathcal{F}^{P\rightarrow P}}_{jk}(q,s)+\mathbf{X}^{p-1}_{j,s}\overset{s}{\odot}\boldsymbol{\mathcal{F}^{X\rightarrow P}}_{jk}(q,s)\right]\\
\mathbf{X}^p_{k,q}&=\sum_{j\neq k}\sum_{s\geq 1}\left[\boldsymbol\Phi^{p-1}_{j,s}\overset{s}{\odot}\boldsymbol{\mathcal{F}^{\Phi\rightarrow X}}_{jk}(q,s)+\mathbf{P}^{p-1}_{j,s}\overset{s}{\odot}\boldsymbol{\mathcal{F}^{P\rightarrow X}}_{jk}(q,s)+\mathbf{X}^{p-1}_{j,s}\overset{s}{\odot}\boldsymbol{\mathcal{F}^{X\rightarrow X}}_{jk}(q,s)\right],
\end{align}
with
\begin{align}
\boldsymbol{\mathcal{F}^{\Phi\rightarrow \Phi}}_{jk}(q,s)&=-\frac{(2q-1)a_k^{2q+1}}{2(q+1)(q-1)!}\boldsymbol{\mathcal{F}^1}_{jk}(q,s),\hspace{3.5cm}
\boldsymbol{\mathcal{F}^{X\rightarrow \Phi}}_{jk}(q,s)=\frac{(2q-1)a_k^{2q+1}}{2(q+1)(q-1)!}\boldsymbol{\mathcal{F}^2}_{jk}(q,s)\label{eq:transfer1}\\
\boldsymbol{\mathcal{F}^{P\rightarrow \Phi}}_{jk}(q,s)&=-\frac{(2q-1)a_k^{2q+1}}{2(q+1)(q-1)!}\left[\boldsymbol{\mathcal{F}^3}_{jk}(q,s)+\frac{(2q+1)a_k^2}{2(2q-1)(2q+3)}\boldsymbol{\mathcal{F}^1}_{jk}(q,s)\right]\label{eq:transfer2}\\
\boldsymbol{\mathcal{F}^{\Phi\rightarrow P}}_{jk}(q,s)&=-\frac{(2q+1)a_k^{2q-1}}{2(q+1)(q-1)!}\boldsymbol{\mathcal{F}^1}_{jk}(q,s),\hspace{3.5cm}
\boldsymbol{\mathcal{F}^{X\rightarrow P}}_{jk}(q,s)=\frac{(2q+1)a_k^{2q-1}s}{2(q+1)(q-1)!}\boldsymbol{\mathcal{F}^2}_{jk}(q,s)\label{eq:transfer3}\\
\boldsymbol{\mathcal{F}^{P\rightarrow P}}_{jk}(q,s)&=-\frac{(2q+1)a_k^{2q-1}}{2(q+1)(q-1)!}\left[\boldsymbol{\mathcal{F}^3}_{jk}(q,s)+\frac{a_k^2}{2(2q+1)}\boldsymbol{\mathcal{F}^1}_{jk}(q,s)\right] \label{eq:transfer4}\\
\boldsymbol{\mathcal{F}^{\Phi\rightarrow X}}_{jk}(q,s)&=0,\qquad 
\boldsymbol{\mathcal{F}^{P\rightarrow X}}_{jk}(q,s)=\frac{a_k^{2q+1}}{q(q+1)\times q!}\boldsymbol{\mathcal{F}^2}_{jk}(q,s),\qquad 
\boldsymbol{\mathcal{F}^{X\rightarrow X}}_{jk}(q,s)=\frac{a_k^{2q+1}s}{q(q+1)\times q!}\boldsymbol{\mathcal{F}^1}_{jk}(q,s)\label{eq:transfer5}
\end{align}
where the following $(q+s)$-order tensors, which are fully-symmetric and deviatoric with respect to their last $q$ indices, have been defined (with their respective order in $\varepsilon=a/d$ shown):
\begin{align}
\boldsymbol{\mathcal{F}^1}_{jk}(q,s)&=\left[\overset{q}{\grad}\left(\frac{\rb_j\overset{s}{\otimes}\rb_j}{r_j^{2s+1}}\right)\right]_{r_k=0}=O(\varepsilon^{s+q+1}),\quad \boldsymbol{\mathcal{F}^2}_{jk}(q,s)=s \left[\overbracket{\overset{q-1}{\grad}\left(\frac{\rb_j\overset{s-1}{\otimes}\rb_j}{r_j^{2s+1}}\otimes(\boldsymbol\varepsilon\cdot\rb_j)\right)}^{q}\right]_{r_k=0}=O(\varepsilon^{s+q}),\label{eq:transfer6}\\
\boldsymbol{\mathcal{F}^3}_{jk}(q,s)&=\frac{1}{2(2s-1)}\left[\overbracket{\overset{q-1}{\grad}\left(\frac{\rb_j\overset{s-1}{\otimes}\rb_j}{r_j^{2s-1}}\otimes\left((2s-1)\frac{\rb_j\rb_j}{r_j^2}-(s-2)\mathbf{I}\right)\right)}^q\right]_{r_k=0}=O(\varepsilon^{s+q-1})\label{eq:transfer7}
\end{align}

\section{Axisymmetric motion of two Janus particles}
\label{app:bispherical}
Here, we use bispherical coordinates to compute the  velocities of two axisymmetric Janus particles aligned along their common axis of symmetry (see Figure~\ref{fig:axisymm_conc}). 
 The analysis is presented here for two Janus particles of identical radius $a$. The case of two particles with different radii can be obtained following a similar approach (e.g.~\cite{Michelin15}).

 In this coordinate system, the orthogonal coordinates $(\tau, \xi, \phi)$ are related to the cylindrical coordinates $(\rho, \phi, z)$ through
\begin{equation}
\rho = \frac{\kappa \sqrt{1-\xi^2}}{\cosh \tau - \xi}, \qquad z=\frac{\kappa \sinh \tau}{\cosh \tau - \xi}\cdot
\end{equation}
The surface of two identical spheres are represented by  $\tau = \pm \tau_0$ (which defines particles 1 and 2 respectively). The spheres have a radius of $a=\kappa/|\sinh \tau_0|$ and their centers are at a distance $d=2 \kappa \coth{\tau_0}$ (this defines $\kappa$ and $\tau_0$ uniquely). On the surface of each sphere, $\xi$ varies monotonically from $\xi=-1$ (at the pole facing the other particle) to $\xi=1$ (at the pole facing away from the other particle).

Let $\xi=\xi^c_{i}$ demarcate the region of activity on the surface of a Janus particle $i$ (i.e. the active regions are $[-1,\xi^c_{1}]$ for particle 1, and $[\xi^c_{2},1]$ for particle 2). Noting ${\cal S}_i^c$ the fraction of the particle surface that is chemically-active (e.g. coated with a catalyst):
\begin{equation}
\xi^c_{i=1,2} = \frac{1\pm (2 {\cal S}^c_i-1)\cosh(\pm \tau_0)}{\cosh (\pm\tau_0) \pm (2 {\cal S}^c_i-1)}\cdot
\end{equation}
Because of the particle structure of the bispherical coordinate system, $\xi_i^c$ is also a function of the instantaneous distance bewteen the particles. Hemispheric Janus particles correspond to ${\cal S}_1^c={\cal S}_2^c=1/2$ while Section \ref{sec:axisymm} focuses on ${\cal S}^c_1={\cal S}^c_2=3/4$. 

The solute concentration field produced by the particles obeys the diffusion equation, Eq.~\eqref{eq:laplace}, whose general far-field decaying solution is given by \cite{stimson26,Michelin15}:
\begin{align}
c(\tau,\xi) & = \sqrt{\cosh \tau - \xi}\; \sum_{n = 0}^\infty c_n(\tau) L_n(\xi) \quad \mbox{with,} \quad c_n (\tau) = a_n \exp^{(n+1/2)(\tau-\tau_0)}+  b_n \exp^{-(n+1/2)(\tau-\tau_0)}.
\label{eq:bisphc_n}
\end{align}
The normal flux boundary condition on the surface two particles,
\begin{align}
\frac{\cosh \tau - \xi}{\kappa} \frac{\partial c}{\partial \tau} \biggr|_{\tau= \pm \tau_0} = \pm A \;H(\xi,\xi^c_{i=1,2})\quad \mbox{where,} \quad
H(\xi,\xi^c_{i=1,2}) &= 
\begin{cases} 
      1 & l_1<\xi \leq l_2 \\
      0 & \textrm{otherwise}
   \end{cases},
   \label{eq:bc_analy}
\end{align}
with, $l_1=-1$ and $l_2=\xi^c_{1}$ for particle 1 and $l_1=\xi^c_{2}$ and $l_2=1$ for particle 2. 
Eqs.~\eqref{eq:bisphc_n}--\eqref{eq:bc_analy} provide after projection along $L_n(\xi)$:
\begin{equation}
\frac{c_n (\pm \tau_0) \sinh(\tau_0)}{2(2n+1)} + \frac{c_n'(\pm \tau_0)\;\cosh(\pm \tau_0)}{2n+1} - \frac{(n+1)c_{n+1}'(\pm \tau_0)}{(2n+1)(2n+3)} -\frac{n c_{n-1}'(\pm \tau_0)}{(2n+1)(2n-1)} = \pm\int_{l_1}^{l_2} \frac{ A |\sinh \tau_0| L_n(\xi) d\xi}{2 \sqrt{\cosh \tau_0-\xi}}\cdot
\label{eq:cn_boundary}
\end{equation}

The integral in Eq.~\eqref{eq:cn_boundary} is computed numerically and Eqs.~\eqref{eq:bisphc_n} and  \eqref{eq:cn_boundary} together provide a linear system for $(a_n,b_n)$ whose solution determines the concentration field. The surface concentration gradients induce an effective slip velocity along $\eb_\xi$,
\begin{equation}
\tilde{u}_\xi(\pm \tau_0,\xi) = \frac{M \sqrt{1-\xi^2}}{\kappa}(\cosh \tau_0 - \xi) \frac{\partial c}{\partial \xi}\biggr|_{\tau=\pm\tau_0}\cdot
\end{equation}

To obtain the particles' velocities, the common strategy employed in low Reynolds hydrodynamics is to develop an auxiliary problem whose solution is known or can be computed easily (e.g. rigid body dynamics) and thereafter, use Lorentz reciprocal theorem to obtain velocity or forces of the original problem~\cite{stone96}. We consider here an auxiliary problem  $(\ub^*,\sigmab^*)$ corresponding to the flow field around the same particles considered here, with particle $i$ translating rigidly with velocity $\Ub_i=U_i\eb_z$ with a net hydrodynamic force $\Fb_i=F_i\eb_z$. It satisfies
\begin{align}
\nabla^2 \ub^* &= \nabla p^*,\qquad \nabla\cdot\ub^*=0,\qquad\ub^* (\rb \to \infty) \to 0,
\end{align}
and $\ub=\Ub^*_i$ and $\int_{\mathcal{S}_i}\sigmab^*\cdot\nb\dd S=\Fb_i^*$ on particle $i$. Applying Lorentz reciprocal theorem to this auxiliary problem and to the dynamics of the two Janus particles provide that for any $(F_1^*,F_2^*)$
\begin{equation}
F_1^*U_1+F^*_2U_2=-\int_{\mathcal{S}_1,\mathcal{S}_2}\tilde\ub\cdot\sigmab^*\cdot\nb\,\dd S.
\end{equation}
Applying this result for the particular choice of auxiliary problem with $F_1^*=F_2^*$ (resp. $F_2^*=-F_1^*$) provides the global velocity $U_1+U_2$ (resp. relative velocity $U_1-U_2$) and hence reconstruct the individual velocities of the particles.

In each case, the relation between the translation velocity of each sphere $U_i^*$, the total hydrodynamic force $F_i^*$ and corresponding fluid stress tensor $\sigmab^*$ is well known \cite{stimson26}, and we therefore only briefly summarize the main results. The auxiliary problem is axisymmetric and can be formulated in terms of a streamfunction $\psi^*$ 
\begin{align}
\psi^*(\tau,\xi) & = (\cosh \tau - \xi)^{-3/2} \sum_{n = 1}^\infty (1-\xi^2) L_n'(\xi) \;V_n(\tau) \qquad \textrm{where} \\
V_n (\tau) & = \alpha_n \cosh \left(n+\frac{3}{2}\right) \tau + \beta_n \sinh \left(n+\frac{3}{2}\right) \tau + \gamma_n \cosh \left(n-\frac{1}{2}\right) \tau + \delta_n \sinh \left(n-\frac{1}{2}\right) \tau.
\end{align}
The coefficients $\alpha_n$, $\beta_n$, $\gamma_n$, and $\delta_n$ are computed from the no-slip boundary condition on the spheres, i.e $\ub^* =U_i^* \eb_z$ on particle $i$ (i.e. $\tau=\pm\tau_0$) \cite{stimson26}.  
Once the coefficients are determined, one can evaluate the surface shear stress,
\begin{align}
& \sigma_{\tau,\xi}^*(\pm \tau_0,\xi) = \frac{\sqrt{1-\xi^2}}{\kappa} \left[\sum_{n \geq 1} L_n'(\xi) S_n - \cosh \tau_0 +\frac{\sinh^2 \tau_0}{2(\cosh \tau_0 - \xi)} \right] \quad \textrm{with,} \label{eq:analy_stress} \\
& S_n = -(\cosh \tau_0-\xi)^{3/2} V_n''(\pm \tau_0) + \frac{\sqrt{\cosh \tau_0 - \xi}}{2} (\pm V_n'(\pm \tau_0) \sinh \tau_0 + 3 V_n(\pm \tau_0) \cosh \tau_0),
\end{align}
and the total hydrodynamic force on each sphere is then obtained as \cite{stimson26},
\begin{equation}
F_1^* = \frac{2\pi \sqrt{2}}{\kappa} \sum_{n \geq 1} n(n+1)(\alpha_n +\beta_n+ \gamma_n+\delta_n) \quad \mbox{and} \quad F_2^* = \frac{2\pi \sqrt{2}}{\kappa} \sum_{n \geq 1} n(n+1)(\alpha_n-\beta_n + \gamma_n-\delta_n).
\label{eq:analy_forces}
\end{equation}

%\bibliography{Biblio}
%\bibliographystyle{unsrt}

\end{document}